\documentclass[aps,prd,twocolumn,10pt,longbibliography,superscriptaddress,floatfix]{revtex4-1}

\usepackage{amsmath}
\usepackage{amsfonts}
\usepackage{amssymb}
\usepackage{mathtools}
\usepackage{times}
\usepackage{hyperref}
\usepackage{bm}
\usepackage{graphicx}
\usepackage{xcolor}
\usepackage{enumerate}
\usepackage{braket}

\begin{document}

\preprint{APS/123-QED}

\title{A quantum Monte Carlo method on asymptotic Lefschetz thimbles for quantum spin systems: \\ An application to the Kitaev model in a
magnetic field}

\author{Petr A. Mishchenko}
\affiliation{Department of Applied Physics, University of Tokyo, Tokyo 113-8656, Japan}
\affiliation{NTT Secure Platform Laboratories, Tokyo 180-8585, Japan}

\author{Yasuyuki Kato}
\affiliation{Department of Applied Physics, University of Tokyo, Tokyo 113-8656, Japan}

\author{Yukitoshi Motome}
\affiliation{Department of Applied Physics, University of Tokyo, Tokyo 113-8656, Japan}

\date{\today}

\begin{abstract}
The quantum Monte Carlo method on asymptotic Lefschetz thimbles is a numerical algorithm devised specifically for alleviation of the sign
problem appearing in the simulations of quantum many-body systems. In this method, the sign problem is alleviated by shifting the
integration domain for the auxiliary fields, appearing for example in the conventional determinant quantum Monte Carlo method, from real
space to an appropriate manifold in complex space. Here we extend this method to quantum spin models with generic two-spin interactions, by
using the Hubbard-Stratonovich transformation to decouple the exchange interactions and the Popov-Fedotov transformation to map the quantum
spins to complex fermions. As a demonstration, we apply the method to the Kitaev model in a magnetic field whose ground state is predicted
to deliver a topological quantum spin liquid with non-Abelian anyonic excitations. To illustrate how the sign problem is alleviated in this
method, we visualize the asymptotic Lefschetz thimbles in complex space, together with the saddle points and the zeros of the fermion
determinant. We benchmark our method in the low-temperature region in a magnetic field and show that the sign of the action is recovered
considerably and unbiased numerical results are obtained with sufficient precision.
\end{abstract}

\maketitle

\section{
	\label{sec:intro}
	Introduction
}

The quantum Monte Carlo (QMC) simulation based on the path integral formalism is one of the most widely used methods to study quantum
many-body problems. In this method, the action of the system in $d$ dimensions is written by a functional integral in terms of the
auxiliary fields in $(d + 1)$ dimensions, and the integration is performed by means of the importance sampling on the configurations of the
auxiliary fields with the use of the action as the Monte Carlo (MC) weight for each configuration. The QMC method is a versatile tool as it
provides numerically exact results within the statistical errors in principle, but it often encounters with a serious obstacle called the
sign problem. It originates from the fact that, except in a few limited cases, the action is complex, rather than positive definite real,
which makes hard to use it as the MC weight. In the conventional QMC method, the complex action has been dealt with by using the
reweighting technique. In this technique, the MC sampling is done by the real part of the action $\mathrm{Re}\mathcal{S}(\bm{\varphi})$,
where $\bm{\varphi}$ represents the auxiliary fields, and the imaginary part $\mathrm{Im}\mathcal{S}(\bm{\varphi})$ is measured together
with the observable $O(\bm{\varphi})$; namely, the thermal average of $O(\bm{\varphi})$ is computed as
\begin{align}
\braket{O} =
\frac{\int d \bm{\varphi} O(\bm{\varphi}) e^{ -\mathcal{S}(\bm{\varphi}) }}
     {\int d \bm{\varphi}                 e^{ -\mathcal{S}(\bm{\varphi}) }}
           =
\frac{\big \langle O(\bm{\varphi}) e^{ -i\mathrm{Im}\mathcal{S}(\bm{\varphi}) } \big \rangle_{\mathrm{Re}\mathcal{S}(\bm{\varphi})}}
     {\big \langle                 e^{ -i\mathrm{Im}\mathcal{S}(\bm{\varphi}) } \big \rangle_{\mathrm{Re}\mathcal{S}(\bm{\varphi})}},
\label{eq:observable_real}
\end{align}
where $\langle A \rangle_{\mathrm{Re}\mathcal{S}(\bm{\varphi})}$ denotes the thermal average of $A$ obtained with the weight of
$e^{ -\mathrm{Re}\mathcal{S}(\bm{\varphi}) }$. This reweighting is used in a broad range of the research fields; for instance, it has been
used in the so-called determinant QMC (D-QMC) method for many fermionic models in solid state physics~\cite{blankenbecler_prd_24_1981,
scalapino_prb_24_1981,linden_pys_rep_220_1992,loh_gubernatis_1992,santos_bjp_33_2003,bercx_scipost_phys_3_2017,assaad_evertz,
sato_arxiv_2012.12283}. The problem is that the practical evaluation of Eq.~(\ref{eq:observable_real}) is exponentially difficult as
follows. By definition, the evaluation is feasible when the ensemble sampled according to $\mathrm{Re}\mathcal{S}(\bm{\varphi})$ has a
sufficient overlap with that by the full action $\mathcal{S}(\bm{\varphi})$. Consequently, the method is successful only as long as the
average sign, which is defined by
\begin{align}
S_\mathrm{D\textnormal{-}QMC} =
\left \lvert \big \langle e^{ -i\mathrm{Im}\mathcal{S}(\bm{\varphi}) } \big \rangle_{\mathrm{Re}\mathcal{S}(\bm{\varphi})} \right \lvert,
\label{eq:action_sign_real}
\end{align}
retains a sufficiently large value. In general, however, $S_\mathrm{D\textnormal{-}QMC}$ becomes exponentially small with respect to the
inverse temperature and the system size. Thus, simulations using the reweighting method become exponentially harder at lower temperatures
and for larger system sizes. This is the notorious sign problem, which has hampered full understanding of many interesting quantum
many-body phenomena, such as exotic phases in quantum chromodynamics~\cite{philipsen_2009,aarts_jpcf_706_2016}, physics of Feshbach
resonances in cold atomic Fermi gases~\cite{inguscio_pispef_164_2007,bloch_rmp_80_2008,giorgini_rmp_80_2008}, possible superconductivity in
doped Mott insulators~\cite{baeriswyl_1995,lee_rmp_78_2006}, and quantum spin liquids in frustrated quantum spin systems~\cite{wen_2007,
balents_nature_464_2010,lacroix_2011,diep_2013,savary_rpp_80_2017}. We note that the sign problem is shown to be an NP-hard problem for
Ising spin-glass systems~\cite{troyer_prl_94_2005}, which suggests that its fundamental solution is unlikely to be available.

Under this circumstance, however, there have been a lot of efforts to avoid or alleviate the sign problem. One of such efforts is to extend
the auxiliary fields from real to complex and shift the integration domain form real space to an appropriate manifold in complex space. An
approach based on this scheme is employing the idea of the Lefschetz thimbles~\cite{written_arxiv_1001.2933,written_arxiv_1009.6032}. This
method also demonstrated its power in a variety of applications in the field of high-energy physics~\cite{cristoforetti_prd_86_2012,
cristoforetti_prd_88_2013,fujii_jhep_10_2013,mukherjee_prd_88_2013,cristoforetti_lattice_2013,cristoforetti_prd_89_2014,aarts_jhep_10_2014,
tanizaki_prd_91_2015,renzo_prd_92_2015,fujii_jhep_11_2015,kanazawa_jhep_03_2015,fukushima_ptep_2015_2015,fujii_jhep_12_2015,
tanizaki_njp_18_2016,tsutsui_prd_94_2016,alexandru_prd_93_2016,alexandru_jhep_05_2016,alexandru_prd_93_2017,alexandru_prd_96_2017_094505,
tanizaki_jhep_10_2017,alexandru_prd_96_2017,fukuma_ptep_2017_2017,renzo_prd_97_2018,bluecher_scipost_5_2018,alexandru_prd_97_2018,
alexandru_prl_121_2018,alexandru_prd_98_2018} and solid state physics~\cite{mukherjee_prb_90_2014,ulybyshev_arxiv_1712.02188,
ulybyshev_arxiv_1906.02726,fukuma_prd_100_2019,fukuma_arxiv_1912.13303,ulybyshev_prd_101_2020,ulybyshev_ppn_51_2020,
wynen_arxiv_2006.11221}. It is, however, important to note that in the case of fermionic systems, straightforward application of the
Lefschetz thimble method faces difficulties in practice; while the efficient calculations require prior knowledge of the structure of the
Lefschetz thimbles in complex space, it is in general unknown a priori~\cite{alexandru_prd_93_2016}. To overcome the difficulty, the
possibility of integrating on an asymptotic form of the Lefscehtz thimbles was proposed, which in principle enables one to take into
account all the relevant thimbles without prior knowledge of their structure~\cite{alexandru_jhep_05_2016}. One of the most important
improvements along this approach is a technique similar to the parallel tempering which was demonstrated to give unbiased results with
great efficiency~\cite{alexandru_prd_96_2017,fukuma_ptep_2017_2017,fukuma_prd_100_2019,fukuma_arxiv_1912.13303}. Despite all these
achievements, however, to the best of our knowledge, the previous studies were all limited to the models with interacting mobile fermions
like the Hubbard model, and there are no applications to quantum spin models where the fermions are immobile and only their spin degrees of
freedom remain active.

In this paper, we develop the QMC method based on the asymptotic Lefscehtz thimbles, which we call the ALT-QMC method, into a form
applicable to quantum spin models with generic two-spin interactions. In our framework, we obtain the functional integral of the action by
two steps. The first step is the Hubbard-Stratonovich transformation~\cite{stratonovich_spd_2_1958,hubbard_prl_3_1959} in which the
exchange interactions are decoupled into one-body terms by introducing the auxiliary fields. The second step is the Popov-Fedotov
transformation~\cite{popov_jetp_67_1988,*popov_psim_184_1991} by which the quantum spins are mapped to complex fermions. These end up with
the functional integral for the generic quantum spin Hamiltonian to which the ALT-QMC method can be applied. We demonstrate the efficiency
of the method for the Kitaev model in a magnetic field, which has recently attracted much attention since it delivers a topological quantum
spin liquid with non-Abelian anyonic excitations. We visualize the asymptotic Lefschetz thimbles in complex space, together with the saddle
points and the zeros of the fermion determinant. By careful comparison with the results obtained by the D-QMC method, we show that our
ALT-QMC method indeed alleviates the sign problem and potentially extends the accessible parameter regions to lower temperatures and larger
system sizes.

The paper is structured as follows. In Sec.~\ref{sec:lt_method}, we briefly review the Lefschetz thimble method. We introduce the general
formalism of the Lefschetz thimbles in Sec.~\ref{subsec:lt_formalism} and the asymptotic Lefschetz thimbles in
Sec.~\ref{subsec:a_l_thimbles}. In Sec.~\ref{sec:qmc}, we construct the framework for the ALT-QMC method applicable to generic spin models.
After introducing a class of the models to which the method can be applied in Sec.~\ref{subsec:generic_model}, we introduce the
Hubbard-Stratonovich transformation in Sec.~\ref{subsec:hs_transform}, the Popov-Fedotov transformation in Sec.~\ref{subsec:pf_transform},
and derive the functional integral of the action in Sec.~\ref{subsec:path_integral}. In Sec.~\ref{subsec:remarks}, we make some remarks on
the implementation of the simulation and the definitions of the metrics to estimate the severity of the sign problem. In
Sec.~\ref{sec:kh_model}, we present the results by the ALT-QMC method for the Kitaev model in a magnetic field. After introducing the model
in Sec.~\ref{subsec:Kitaev_model}, we visualize the structure of the Lefschetz thimbles in complex space and present the benchmark of the
ALT-QMC technique for a small system in Secs.~\ref{subsec:visualization} and \ref{subsec:alleviation_sign}, respectively. In
Sec.~\ref{subsec:benchmark}, we show the results for the Kitaev model in a magnetic field for a larger system size. Finally,
Sec.~\ref{sec:summary_and_outlook} is devoted to the summary and outlook.

\section{
	\label{sec:lt_method}
	Lefschetz thimble method
}

In this section, we briefly review the Lefschetz thimble method. In Sec.~\ref{subsec:lt_formalism}, we describe the fundamentals of the
Lefschetz thimbles, and in Sec.~\ref{subsec:a_l_thimbles}, we present the framework of asymptotic Lefschetz thimbles.

\subsection{
	\label{subsec:lt_formalism}
	Lefschetz thimbles
}

\begin{figure}[b]
	\includegraphics[width=\columnwidth,clip]{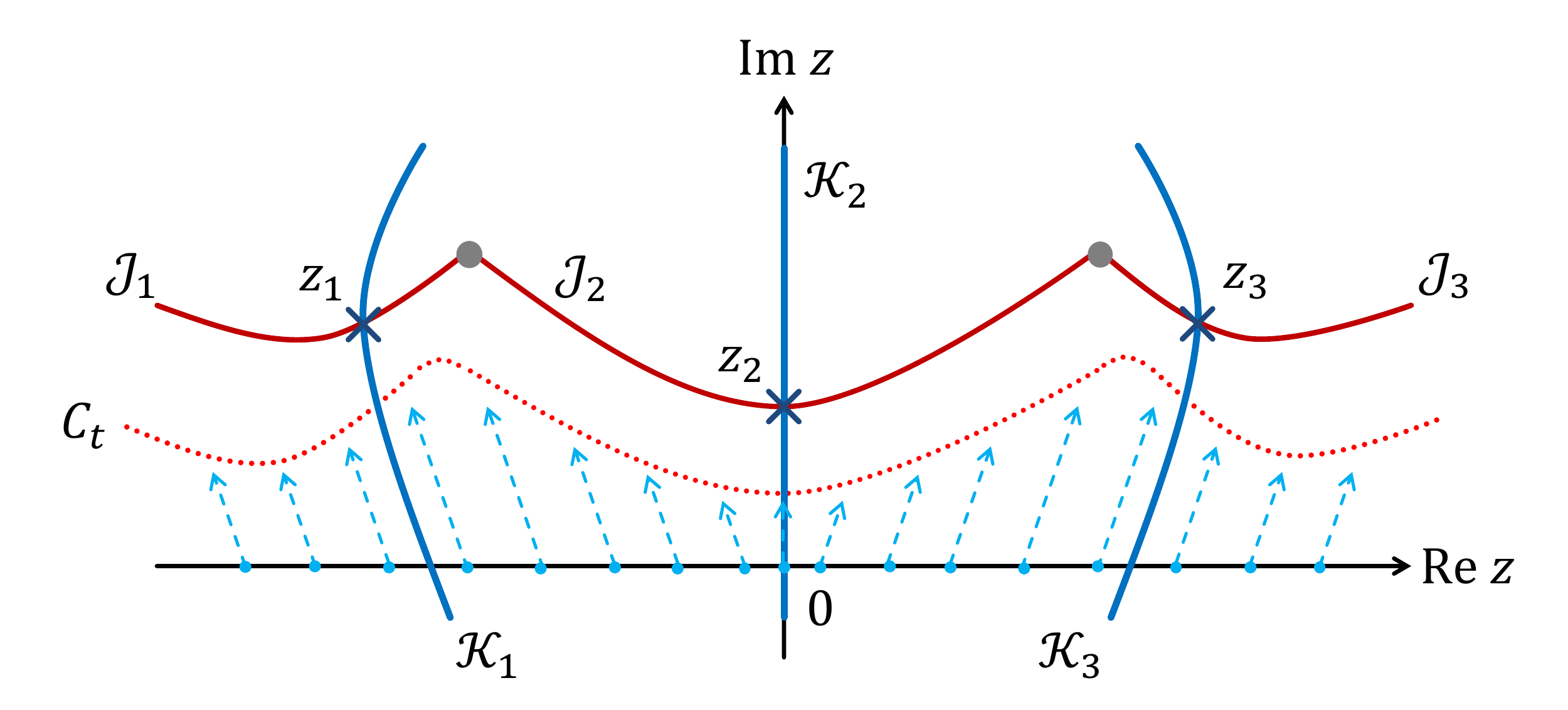}
	\caption{
		Schematic illustration of the Lefschetz thimbles and the asymptotic ones for a single auxiliary field $z$. The solid red lines
		represent the true Lefschetz thimbles $\mathcal{J}_\nu$ associated with the saddle points $z_\nu$ (dark blue crosses) and the zeros of
		the fermion determinant separating different thimbles (gray dots). The solid blue lines represent the antithimbles $\mathcal{K}_\nu$.
		The dotted red line represents the asymptotic Lefschetz thimble $\mathcal{C}_t$ obtained by the time evolution from the real axis which
		is represented by the dashed blue arrows.
	}
	\label{fig:alt_illustration}
\end{figure}

In general, one can estimate the action $\mathcal{S}(\bm{\varphi})$ by extending the functional integral to complex space by analytic
continuation from the real auxiliary fields $\bm{\varphi}$ to the complex ones $\bm{z}$. In fact, one can define an integration path in the
complex space along which the following two properties are satisfied: (i) Along the integration path, the real part of the action,
$\mathrm{Re}\mathcal{S}(\bm{z})$, decreases the fastest when moving towards the saddle point, and (ii) the imaginary part of the action,
$\mathrm{Im}\mathcal{S}(\bm{z})$, is constant on the integration path which takes the same value at the saddle point. Such a path is called
the Lefschetz thimble~\cite{written_arxiv_1001.2933,written_arxiv_1009.6032}. In general, the action $\mathcal{S}(\bm{z})$ may have
multiple saddle points $\bm{z}_\nu$ with multiple Lefschetz thimbles $\mathcal{J}_\nu$ associated to them (hereafter we use $\nu$ for
labeling the saddle points and associated Lefschetz thimbles), and therefore, the original integral in the real space is equivalent to the
sum of the integrals over the Lefschetz thimbles in the complex space. It is important to note that in fermionic systems, different
Lefschetz thimbles are separated from each other by the points where the fermion determinant vanishes and the effective action diverges.
See the schematic in Fig.~\ref{fig:alt_illustration}.

One can identify the saddle points $\bm{z}_\nu$ of the action $\mathcal{S}(\bm{z})$ by solving the set of equations
\begin{align}
\frac{\partial \mathcal{S}(\bm{z})}{\partial z_l}\bigg|_{\bm{z} = \bm{z}_\nu} = 0,
\label{eq:saddle}
\end{align}
and the Lefschetz thimble $\mathcal{J}_\nu$ attached to the saddle point $\bm{z}_\nu$ by the flow equation
\begin{align}
\frac{d z_l}{d t} = - \left[ \frac{\partial \mathcal{S}(\bm{z})}{\partial z_l} \right]^*,
\label{eq:neg_flow}
\end{align}
where $l$ is the index of the auxiliary variable $\bm{z}$, and $d z_l/d t$ means the ``time'' derivative of the auxiliary variable $z_l$.
Indeed the Lefschetz thimbles determined by Eq.~(\ref{eq:neg_flow}) satisfy the two properties described above.

Besides the thimbles $\mathcal{J}_\nu$, the so-called antithimbles $\mathcal{K}_\nu$ attached to the saddle point $\bm{z}_\nu$ are also
important in the following calculations. The antithimble for the saddle point $\bm{z}_\nu$ is defined by
\begin{align}
\frac{d z_l}{d t} = + \left[ \frac{\partial \mathcal{S}(\bm{z})}{\partial z_l} \right]^*.
\label{eq:pos_flow}
\end{align}

Once all the thimbles $\mathcal{J}_\nu$ and the antithimbles $\mathcal{K}_\nu$ are identified, one can calculate the statistical average of
an observable $O(\bm{z})$ by using the sum of the integrals over $\mathcal{J}_\nu$~\cite{written_arxiv_1001.2933,written_arxiv_1009.6032}:
\begin{align}
\braket{O} =
\frac{
\sum_\nu n_\nu e^{-i\mathrm{Im}\mathcal{S}(\bm{z}_\nu)} \int_{\mathcal{J}_\nu} d \bm{z} O(\bm{z}) e^{-\mathrm{Re}\mathcal{S}(\bm{z})}
}
{
\sum_\nu n_\nu e^{-i\mathrm{Im}\mathcal{S}(\bm{z}_\nu)} \int_{\mathcal{J}_\nu} d \bm{z}           e^{-\mathrm{Re}\mathcal{S}(\bm{z})}
},
\label{eq:thimble_decomposition}
\end{align}
where $n_\nu$ is an integer given by the number of the intersections between the corresponding antithimble $\mathcal{K}_\nu$ and the
original integration domain in the real space. Thus, one can in principle compute the original integral in the real space by the integrals
on the Lefschetz thimbles in the complex space. The important point here is that the sign problem is alleviated since
$\mathrm{Im}\mathcal{S}(\bm{z}_\nu)$ is constant on each Lefschetz thimble and put outside the integrals in
Eq.~(\ref{eq:thimble_decomposition}).

\subsection{
	\label{subsec:a_l_thimbles}
	Asymptotic Lefschetz thimbles
}

Although the framework of the Lefschetz thimbles is exact and suppresses the sign problem, the practical application is not straightforward
as it is difficult to identify all the thimbles $\mathcal{J}_\nu$ with all the corresponding coefficients $n_\nu$. For this reason, the
Lefschetz thimble technique has been applied only to some limited cases. For instance, the method was applied when the models are
sufficiently simple and all the Lefschetz thimbles can be identified, for the bosonic~\cite{
aarts_jhep_10_2014,
tanizaki_prd_91_2015,
fukushima_ptep_2015_2015
} and fermionic cases~\cite{
kanazawa_jhep_03_2015,
tanizaki_njp_18_2016
}. Also, QMC simulations were performed on dominant thimbles for both bosonic~\cite{
renzo_prd_92_2015,
tsutsui_prd_94_2016,
cristoforetti_prd_88_2013,
fujii_jhep_10_2013,
mukherjee_prd_88_2013,
cristoforetti_lattice_2013,
cristoforetti_prd_89_2014
} and fermionic systems~\cite{
fujii_jhep_11_2015,
fujii_jhep_12_2015,
alexandru_prd_93_2016
}. However, especially in the fermionic systems, the integration over the dominant thimbles is often insufficient to obtain the precise
estimate of the observables; it is not obvious how to take into account all the relevant contributions.

To overcome the difficulty, an alternative method was proposed~\cite{alexandru_jhep_05_2016}, in which the integral is taken on an
asymptotic form of the Lefschetz thimbles rather than the true Lefschetz thimbles. This technique enables one to take into account the
contributions from all the relevant thimbles without knowing all the true thimbles a priori~\cite{alexandru_jhep_05_2016}. It is achieved
by starting the process of finding the thimbles not from the saddle points $\bm{z}_\nu$ by solving Eqs.~(\ref{eq:saddle}) and
(\ref{eq:neg_flow}) but rather from the original integration domain with time evolution by using Eq.~(\ref{eq:pos_flow}). Such time
evolution gradually deforms the integration domain to a manifold in the complex space, as schematically shown in
Fig.~\ref{fig:alt_illustration}. The manifold approaches the true Lefschetz thimbles asymptotically, and hence the time evolved manifold is
called the asymptotic Lefschetz thimbles, which we denote as $\mathcal{C}_t$ at time $t$. In this time evolution, some special points will
flow to the saddle points $\bm{z}_\nu$ and other points close to them will flow closely to $\mathcal{J}_\nu$ attached to $\bm{z}_\nu$. All
the points except the ones flowing to the saddle points will eventually flow to the singularities of the action in the long time limit,
which are zeros of the fermion determinant in the fermionic problems. As such, the flow from the real space can collect the contributions
from all the relevant thimbles once the sampling and the time evolution are performed appropriately.

Using the above method, one can perform MC sampling on the asymptotic Lefschetz thimbles $\mathcal{C}_t$ obtained by time evolution of the
samples proposed in the real space. The regions with large MC weights in the real space become even larger in the complex space because the
time evolution according to Eq.~(\ref{eq:pos_flow}) always increases the real part of the action. On the other hand, the imaginary
part of the action on $\mathcal{C}_t$ remains as the original one in the real space because the time evolution keeps the imaginary part
of the action unchanged. Thus, in the regions of $\mathcal{C}_t$ with large MC weights, the fluctuation of the phase
$e^{ -i\mathrm{Im}\mathcal{S}(\bm{z}) }$ tends to be suppressed, and hence, the sign problem is alleviated for the samples on the complex
manifold $\mathcal{C}_t$ compared to the original ones in the real space.

Mathematically, the deformation from the real space to the complex manifold $\mathcal{C}_t$ by Eq.~(\ref{eq:pos_flow}) corresponds to a
change of variables from real $\bm{\varphi}$ to complex $\bm{z}$ as
\begin{align}
\braket{O} = ~
&\frac{\int_{\mathcal{C}_t} d \bm{z} O(\bm{z}) e^{ -\mathcal{S}(\bm{z}) }}
      {\int_{\mathcal{C}_t} d \bm{z}           e^{ -\mathcal{S}(\bm{z}) }}& \nonumber \\
           = ~
&\frac{\int d \bm{\varphi} O\bm{(}\bm{z}(\bm{\varphi})\bm{)} e^{ -\mathcal{S}\bm{(}\bm{z}(\bm{\varphi})\bm{)} } \mathrm{det} J }
      {\int d \bm{\varphi}                                   e^{ -\mathcal{S}\bm{(}\bm{z}(\bm{\varphi})\bm{)} } \mathrm{det} J },&
\label{eq:int_complex}
\end{align}
where $J$ is the Jacobian given by $J_{l,m} = \partial z_l / \partial \varphi_m$; the integrals in the first line are taken on the
asymptotic Leschetz thimbls $\mathcal{C}_t$ in the complex space, while those in the second line are taken on the original domain in the
real space. Equation~(\ref{eq:int_complex}) indicates that $\bm{\varphi}$ parametrize the asymptotic Lefschetz thimbles $\mathcal{C}_t$ in
the complex space after a flow time $t$. This allows one to estimate the statistical average by proposing $\bm{\varphi}$ in the real
space by the Markov chain MC sampling and evolving them to the asymptotic Lefschetz thimbles $\mathcal{C}_t$ by Eq.~(\ref{eq:pos_flow}).
During the time evolution, the Jacobian along the flows of the samples can be calculated by
\begin{align}
\frac{d J_{l,m}}{d t} = \left[ \sum_n \frac{\partial^2 \mathcal{S}\left( \bm{z} \right)}{\partial z_l \partial z_n}J_{n,m} \right]^*,
\label{eq:jacobian_flow}
\end{align}
with the initial condition $J = \mathbb{I}$ (identity matrix).

This technique allows one to perform MC sampling without prior knowledge of relevant Lefschetz thimbles $\mathcal{J}_\nu$ and the
corresponding saddle points $\bm{z}_\nu$. In the simulation, the MC sampling is performed for the configurations of the real auxiliary
variables $\bm{\varphi}$ by using $e^{ -\mathrm{Re}\mathcal{S}\bm{(}\bm{z}(\bm{\varphi})\bm{)} }$ as the MC weight, and the integral in
Eq.~(\ref{eq:int_complex}) is computed after the time evolution by measuring the phase
$e^{ -i\mathrm{Im}\mathcal{S}\bm{(}\bm{z}(\bm{\varphi})\bm{)} }$ and $\mathrm{det} J $ together with the observable
$O\bm{(}\bm{z}(\bm{\varphi})\bm{)}$ as
\begin{align}
\braket{O} =
\frac{
\big \langle
O\bm{(}\bm{z}(\bm{\varphi})\bm{)} e^{ -i\mathrm{Im}\mathcal{S}\bm{(}\bm{z}(\bm{\varphi})\bm{)} } \mathrm{det} J
\big \rangle_{\mathrm{Re}\mathcal{S}\bm{(}\bm{z}(\bm{\varphi})\bm{)}}
}
{
\big \langle                      e^{ -i\mathrm{Im}\mathcal{S}\bm{(}\bm{z}(\bm{\varphi})\bm{)} } \mathrm{det} J
\big \rangle_{\mathrm{Re}\mathcal{S}\bm{(}\bm{z}(\bm{\varphi})\bm{)}}
}.
\label{eq:alt_qmc_formula}
\end{align}
Here, in order to save the computational time during the calculation of Eq.~(\ref{eq:alt_qmc_formula}), the determinant of the Jacobian,
$\mathrm{det} J$, is included into the observable $O\bm{(}\bm{z}(\bm{\varphi})\bm{)}$ and is calculated only once per MC sweep, following
the previous studies~\cite{ulybyshev_arxiv_1906.02726,ulybyshev_prd_101_2020,ulybyshev_ppn_51_2020}.

Let us remark on the dimension of the original integration domain, the true and asymptotic Lefschetz thimbles, and the zeros of the fermion
determinant. The dimension of the original integration domain in the real space is defined by the number of the auxiliary fields
$\bm{\varphi}$, $N_{\bm{\varphi}}$. The domain $\mathbb{R}^{N_{\bm{\varphi}}}$ is evolved by Eq.~(\ref{eq:pos_flow}) into a manifold
$\mathbb{R}^{N_{\bm{\varphi}}}$ embedded in the complex domain $\mathbb{C}^{N_{\bm{\varphi}}}$. Therefore, the asymptotic Lefschetz
thimbles are $\mathbb{R}^{N_{\bm{\varphi}}}$-dimensional manifolds. The true Lefschetz thimbles are also
$\mathbb{R}^{N_{\bm{\varphi}}}$-dimensional manifolds. On the other hand, the zeros of the fermion determinant constitute
$\mathbb{C}^{N_{\bm{\varphi}} - 1}$-dimensional manifolds embedded in $\mathbb{C}^{N_{\bm{\varphi}}}$ in the generic
case~\cite{alexandru_jhep_05_2016}.

\section{
	\label{sec:qmc}
	Quantum Monte Carlo method for generic quantum spin models
}

In this section, we show a QMC method based on the asymptotic Lefschetz thimbles, which we call the ALT-QMC method, applicable to a generic
quantum spin model with arbitrary two-spin interactions and the Zeeman coupling. While this technique can be applied to generic spin
magnitudes $S$, for the sake of simplicity, we limit our description to the $S = 1/2$ case. In Sec.~\ref{subsec:generic_model}, we
introduce the generic model to which the ALT-QMC method can be applied. In Sec.~\ref{subsec:hs_transform}, we introduce the
Hubbard-Stratonovich transformation to decouple the two-spin interactions. In Sec.~\ref{subsec:pf_transform}, we describe an exact mapping
from the quantum spins to complex fermions by means of the Popov-Fedotov transformation. In Sec.~\ref{subsec:path_integral}, we discretize
the partition function via the Suzuki-Trotter decomposition and derive the functional integral of the action. In Sec.~\ref{subsec:remarks},
we make a remark on the details of the implementation and the estimate of the sign problem.

\subsection{
	\label{subsec:generic_model}
	Model
}

We consider a generic $S = 1/2$ model with arbitrary two-spin interactions, whose Hamiltonian is given by 
\begin{align}
\mathcal{H} = - \sum_{p,q}\sum_{\alpha,\beta} K_{p,q}^{\alpha,\beta} \sigma_p^\alpha \sigma_q^\beta
              - \sum_p \sum_\alpha h_p^\alpha \sigma_p^\alpha,
\label{eq:gqs_model}
\end{align}
where the spin degree of freedom at site $p$ is described by the Pauli matrices $\sigma_p^\alpha$, and $K_{p,q}^{\alpha,\beta}$ is the
coupling constant for the two-spin interaction ($\alpha$, $\beta = x$, $y$, $z$); the second term describes the Zeeman coupling to a
magnetic field whose $\alpha$ component at site $p$ is denoted as $h_p^\alpha$. Both $K_{p,q}^{\alpha,\beta}$ and $h_p^\alpha$ can be
spatially inhomogeneous. The following formulation is applicable to the model in Eq.~(\ref{eq:gqs_model}) on any lattice geometry with any
boundary conditions.

\subsection{
	\label{subsec:hs_transform}
	Hubbard-Stratonovich transformation
}

For constructing the QMC method based on the path integral formalism, we decompose the two-body interactions by using the
Hubbard-Stratonovich transformation~\cite{stratonovich_spd_2_1958,hubbard_prl_3_1959}. Among several choices, we use the transformation
with continuous auxiliary variables which is suitable for the present purpose to develop the ALT-QMC method. Specifically, rewriting the
interaction term in Eq.~(\ref{eq:gqs_model}) as
\begin{align}
-K_{p,q}^{\alpha,\beta} \sigma_p^\alpha \sigma_q^\beta
= -\frac{1}{2} K_{p,q}^{\alpha,\beta} \left[ \left( \sigma_p^\alpha + \sigma_q^\beta \right)^2 - 2 \right],
\label{eq:interactions_to_powers}
\end{align}
we decompose it by using the Hubbard-Stratonovich transformation as
\begin{align}
&\mathrm{exp}\left[ \frac{\Delta}{2} K_{p,q}^{\alpha,\beta} \left( \sigma_p^\alpha + \sigma_q^\beta \right)^2 \right] =
\sqrt{\frac{\Delta}{2\pi}} \int d \varphi_{p,q}^{\alpha,\beta} \nonumber \\
&\times \mathrm{exp} \left[
-\frac{\Delta}{2} \left( \varphi_{p,q}^{\alpha,\beta} \right)^2 - \Delta \varphi_{p,q}^{\alpha,\beta}
\sqrt{K_{p,q}^{\alpha,\beta}} \left( \sigma_p^\alpha + \sigma_q^\beta \right)
\right],
\label{eq:hubbard_stratonovich}
\end{align}
where $\Delta$ is a positive constant that will be introduced in the Suzuki-Trotter decomposition in Sec.~\ref{subsec:path_integral}. Note
that the sign of $K_{p,q}^{\alpha,\beta}$ can be both positive and negative; in case of $K_{p,q}^{\alpha,\beta} < 0$,
$\sqrt{K_{p,q}^{\alpha,\beta}}$ in Eq.~(\ref{eq:hubbard_stratonovich}) becomes a pure imaginary. Note that, in this formulation, we
introduce one continuous real auxiliary variable of Gaussian type, $\varphi_{p,q}^{\alpha,\beta}$, for each interaction term
$-K_{p,q}^{\alpha,\beta} \sigma_p^\alpha \sigma_q^\beta$, which will save the computational cost; see Sec.~\ref{subsec:remarks}.

\subsection{
	\label{subsec:pf_transform}
	Popov-Fedotov transformation
}

To utilize the framework based on the asymptotic Lefschetz thimble technique for the fermionic systems, we proceed with mapping from the
quantum spins to complex fermions via the Popov-Fedotov transformation~\cite{popov_jetp_67_1988,*popov_psim_184_1991}. To begin with, let
us express the spin degree of freedom described in terms of the Pauli matrices by complex fermions by using
\begin{align}
\sigma_p^\alpha \rightarrow \sum_{\gamma,\gamma^\prime} f_{p,\gamma}^\dag \sigma_{\gamma,\gamma^\prime}^\alpha f_{p,\gamma^\prime},
\label{eq:spin_fermion}
\end{align}
where $f_{p,\gamma}^\dag$ and $f_{p,\gamma}$ are the creation and annihilation operators of a complex fermion, respectively, at site $p$
with spin $\gamma = \uparrow$ or $\downarrow$. It is important to note that the relation in Eq.~(\ref{eq:spin_fermion}) enlarges the size
of the Hilbert space per site from two for the original spin to four for the fermion as $f_{p,\uparrow}^\dag \ket{0}$,
$f_{p,\downarrow}^\dag \ket{0}$, $\ket{0}$, $f_{p,\uparrow}^\dag f_{p,\downarrow}^\dag \ket{0}$ ($\ket{0}$ is the vacuum). The first two
states with fermion occupation number unity are physical, whereas the last two are unphysical. One can eliminate the unphysical states by
adding a term to the fermion Hamiltonian, which gives zero when acting on the physical Hilbert space and keeps the partition function
intact. The explicit form of such a term is given by~\cite{popov_jetp_67_1988,*popov_psim_184_1991}
\begin{align}
\mathcal{H}_\mu = \frac{ i \pi}{2\beta}\sum_p \left(\sum_\gamma f_{p,\gamma}^\dag f_{p,\gamma} - 1 \right),
\label{eq:complex_chemical}
\end{align}
which corresponds to the introduction of an imaginary chemical potential $-i \pi / 2\beta$ depending on the inverse temperature
$\beta = 1/T$ (we set the Boltzmann constant $k_{\rm B} = 1$). It is straightforward to verify that the trace over the unphysical states
vanishes for the total Hamiltonian including Eq.~(\ref{eq:complex_chemical}) because the contributions from $\ket{0}$ and
$f_{p,\uparrow}^\dag f_{p,\downarrow}^\dag \ket{0}$ eliminate each other. Hence, the following relation for the partition function holds
exactly:
\begin{align}
\mathcal{Z} = \mathrm{Tr} \left[ \mathrm{exp} \left( -\beta\mathcal{H} \right) \right] =
\mathrm{Tr} \left\{ \mathrm{exp} \left[-\beta\left(\mathcal{H}_f + \mathcal{H}_\mu \right)\right] \right\},
\label{eq:popof_fedotov_equality}
\end{align}
where $\mathcal{H}_f$ is the fermion Hamiltonian obtained from Eq.~(\ref{eq:gqs_model}) via the transformation in
Eq.~(\ref{eq:spin_fermion}).

This method provides an exact mapping from quantum spins to complex fermions without introducing any unphysical states. It is applicable to
any $S = 1/2$ models with any two-spin interactions. Furthermore, since the mapping is defined by the local (onsite) transformation in
Eq.~(\ref{eq:spin_fermion}), it can be applied to any lattice geometry with any boundary conditions. Further details and applications of
this method are found, for example, in Refs.~\cite{popov_jetp_67_1988,gros_physb_165-166_1990,stain_zp_83_1991,veits_jsjp_4_1994,
bouis_physb_259-261_1999,azakov_ijmpb_14_2000,kiselev_prl_85_2000,kiselev_jetp_lett_71_2000,kiselev_epjb_22_2001,kiselev_prb_65_2002,
kiselev_prb_68_2003,coleman_jpcm_16_2004,dillenschneider_epjb_49_2006,prokofev_prb_84_2011}.

\subsection{
	\label{subsec:path_integral}
	Path integral of the action
}

By combining the two transformations above, we obtain the expression of the action to which the ALT-QMC method is applicable. First, let us
decompose the entire fermionized Hamiltonian into the two-body parts and the one-body part as
\begin{align}
\mathcal{H}_f + \mathcal{H}_\mu
= \sum_{m = 1}^{N_I - 1} \mathcal{H}_m + \left( \mathcal{H}_{h} + \mathcal{H}_{\mu} \right)
= \sum_{m = 1}^{N_I} \mathcal{H}_m,
\label{eq:interaction_type}
\end{align}
where $\mathcal{H}_m$ for $1 \leq m \leq N_I - 1$ is the Hamiltonian obtained by fermionizing
$- \sum_{p,q} K_{p,q}^{\alpha,\beta} \sigma_p^\alpha \sigma_q^\beta$ with a particular set of $\alpha$ and $\beta$, and $\mathcal{H}_h$
stands for the Zeeman coupling term in Eq.~(\ref{eq:gqs_model}); $\mathcal{H}_{N_I}$ denotes the one-body part,
$\mathcal{H}_{h} + \mathcal{H}_{\mu}$. With this form, we introduce the Suzuki-Trotter decomposition for the operator
$\mathrm{exp}[-\beta (\mathcal{H}_f + \mathcal{H}_\mu)]$ in Eq.~(\ref{eq:popof_fedotov_equality}) as
\begin{align}
\mathrm{exp}\left[ -\beta \left( \mathcal{H}_f + \mathcal{H}_\mu \right) \right]
\approx
\left[ \prod_{m = 1}^{N_I} \mathrm{exp}\left( -\Delta\mathcal{H}_m \right) \right]^{N_\beta},
\label{eq:suzuki_trotter}
\end{align}
where $N_\beta$ is the number of the Suzuki-Trotter discretization in the imaginary-time direction; $\Delta = \beta/N_\beta$. Note that the
approximation in Eq.~(\ref{eq:suzuki_trotter}) is valid up to $\mathcal{O}(\Delta^2)$. Then, the partition function in
Eq.~(\ref{eq:popof_fedotov_equality}) is expressed as
\begin{align}
\mathcal{Z}
\approx \mathrm{Tr} \left\{ \left[ \prod_{m = 1}^{N_I} \mathrm{exp}\left( -\Delta\mathcal{H}_m \right) \right]^{N_\beta} \right\}.
\label{eq:suzuki_trotter_fermion}
\end{align}
The exact partition function will be obtained in the limit of $\Delta \rightarrow 0$ and $N_\beta \to \infty$. Note that the systematic
error in the discretized partition function $\mathcal{Z}$ is again of $\mathcal{O}(\Delta^2)$ owing to the property of the trace
operation~\cite{sandvik_aip_1297_2010}.

The partition function $\mathcal{Z}$ in Eq.~(\ref{eq:suzuki_trotter_fermion}), with the constant contribution in
Eq.~(\ref{eq:complex_chemical}) placed outside the integral, can be expressed by using anticommuting Grassmann variables $\psi_{p,\gamma}$
and $\bar{\psi}_{p,\gamma}$ as
\begin{align}
\mathcal{Z} = e^{\frac{i \pi N_S}{2}} \int
& \prod_{p,\gamma} d\bar{\psi}_{p,\gamma} d\psi_{p,\gamma}
\mathrm{exp} \left( -\sum_{p,\gamma} \bar{\psi}_{p,\gamma} \psi_{p,\gamma} \right) \nonumber \\
& \times \braket{-\bar{\psi}|\left[ \prod_{m = 1}^{N_I} \mathrm{exp}\left( -\Delta\mathcal{H}_m \right) \right]^{N_\beta}|\psi},
\label{eq:grassmann_preliminary}
\end{align}
where $\ket{\psi} = \mathrm{exp} (-\sum_{p,\gamma} \psi_{p,\gamma} f_{p,\gamma}^\dag) \ket{0}$ and
$\bra{\bar{\psi}} = \bra{0} \mathrm{exp} (-\sum_{p,\gamma} f_{p,\gamma} \bar{\psi}_{p,\gamma})$ are the fermionic coherent states, and
$N_S$ is the system size; here, $\mathcal{H}_{N_I}$ does not include the complex constant $-i \pi N_S / 2\beta$. Then, by inserting the
identity relation for the fermionic coherent states we obtain
\begin{align}
\mathcal{Z} = e^{\frac{i \pi N_S}{2}} \int
& \prod_{l = 1}^{N_\beta} \prod_{m = 1}^{N_I} \prod_{p,\gamma} d\bar{\psi}_{p,\gamma}^{l,m} d\psi_{p,\gamma}^{l,m} \nonumber \\
& \mathrm{exp}
\left( -\sum_{l = 1}^{N_\beta} \sum_{m = 1}^{N_I} \sum_{p,\gamma} \bar{\psi}_{p,\gamma}^{l,m + 1} \psi_{p,\gamma}^{l,m} \right)
\nonumber \\
& \times \prod_{l = 1}^{N_\beta} \prod_{m = 1}^{N_I}
\braket{\bar{\psi}^{l,m}| \mathrm{exp} \left( -\Delta\mathcal{H}_{l,m} \right) |\psi^{l,m}}.
\label{eq:grassmann_suzuki_trotter}
\end{align}
where $l$ is the label for the Suzuki-Trotter slice, each containing in total $N_I$ slices numbered by $m$, and for all of the slices, the
aniticommuting Grassmann variables $\psi_{p,\gamma}^{l,m}$ and $\bar{\psi}_{p,\gamma}^{l,m}$ are defined. Here, the Grassmann variables
satisfy the boundary conditions:
\begin{align}
&-\bar{\psi}_{p,\gamma} = \bar{\psi}_{p,\gamma}^{1,1} = \bar{\psi}_{p,\gamma}^{N_\beta,N_I + 1}, \nonumber \\
&\psi_{p,\gamma}^{N_\beta,N_I} = \psi_{p,\gamma}, \quad \bar{\psi}_{p,\gamma}^{l,N_I + 1} = \bar{\psi}_{p,\gamma}^{l + 1,1}.
\label{eq:grassmann_boundary}
\end{align}

Next, with the help of the Hubbard-Stratonovich transformation in Eq.~(\ref{eq:hubbard_stratonovich}), we obtain the following relation for
each matrix element in Eq.~(\ref{eq:grassmann_suzuki_trotter}) with the two-body terms:
\begin{align}
&\mathrm{exp} \left( -\Delta\mathcal{H}_{l,m} \right) \nonumber \\
& = \mathrm{exp} \left[ \frac{\Delta}{2} \sum_n K_n^m \left(\mathcal{P}_n^m\right)^2 - \Delta \sum_n K_n^m \right] \nonumber \\
& = \left( \frac{\Delta}{2\pi} \right)^{\frac{N_m}{2}} \mathrm{exp} \left( -\Delta \sum_n K_n^m \right)
\int \prod_n d \varphi_n^m \nonumber \\
& \qquad \mathrm{exp}
\left[ -\frac{\Delta}{2} \sum_n \left(\varphi_n^m\right)^2 - \Delta \sum_n \varphi_n^m \sqrt{K_n^m} \mathcal{P}_n^m \right],
\end{align}
where $n$ denotes a pair of $p$ and $q$ for the two-body terms ($1 \leq m \leq N_I - 1$), $N_m$ stands for the total number of interaction
terms in a particular $\mathcal{H}_{l,m}$, and $\mathcal{P}_n^m$ is the bilinear fermionic operator obtained by replacing quantum spins in
Eq.~(\ref{eq:hubbard_stratonovich}) with the complex fermions as in Eq.~(\ref{eq:spin_fermion}):
\begin{align}
\mathcal{P}_n^m = \sum_{\gamma,\gamma^\prime}
( f_{p,\gamma}^\dag \sigma_{\gamma,\gamma^\prime}^\alpha f_{p,\gamma^\prime}
+ f_{q,\gamma}^\dag \sigma_{\gamma,\gamma^\prime}^\beta  f_{q,\gamma^\prime} ).
\end{align}
By using the relation for the Grassmann variables for a matrix $M$~\cite{ulybyshev_prl_111_2013,smith_prb_89_2014,buividovich_prb_86_2012}
\begin{align}
&\braket{ \bar{\psi}^{l,m} | \mathrm{exp} \left(\sum_{k,k^\prime} f_{k}^\dag M_{k,k^\prime} f_{k^\prime} \right) | \psi^{l,m} }
\nonumber \\
&= \mathrm{exp}
\left[ \sum_{k,k^\prime} \bar{\psi}_{k}^{l,m} \mathrm{exp} \left( M \right)_{k,k^\prime} \psi_{k^\prime}^{l,m} \right],
\label{fermion_to_grassmann}
\end{align}
where $k = \{p,\gamma\}$ and $k^\prime = \{q,\gamma^\prime\}$, one can write each interaction term in
Eq.~(\ref{eq:grassmann_suzuki_trotter}) as
\begin{align}
&\braket{ \bar{\psi}^{l,m} | \mathrm{exp} \left( -\Delta\mathcal{H}_{l,m} \right) | \psi^{l,m} } \nonumber \\
& = \left( \frac{\Delta}{2\pi} \right)^{\frac{N_m}{2}}
\mathrm{exp} \left( -\Delta \sum_n K_n^m \right)
\int \prod_n d \varphi_n^m \nonumber \\
& \mathrm{exp}
\left[
-\frac{\Delta}{2} \sum_n \left(\varphi_n^m\right)^2
+ \sum_{k,k^\prime} \bar{\psi}_{k}^{l,m} \mathrm{exp} \left( -\Delta h_{l,m} \right)_{k,k^\prime} \psi_{k^\prime}^{l,m}
\right].
\end{align}
Here, $h_{l,m}$ stands for the matrix element of the bilinear fermionic operator $\sum_n \varphi_n^m \sqrt{K_n^m} \mathcal{P}_n^m$
($1 \leq m \leq N_I - 1$). By using Eq.~(\ref{fermion_to_grassmann}), one can also transform the one-body part
$\braket{ \bar{\psi}^{l,N_I} | \mathrm{exp} \left( -\Delta\mathcal{H}_{l,N_I} \right) | \psi^{l,N_I} }$ which does not contain the
auxiliary fields introduced via the Hubbard-Stratonovich transformation. 

Finally, by integrating out the Grassmann variables, we obtain
\begin{align}
\mathcal{Z} = & 
e^{\frac{i \pi N_S}{2}} \mathrm{exp} \left( -\beta \sum_{m = 1}^{N_I - 1} \sum_n K_n^m \right)
\left( \frac{\beta}{2 \pi N_\beta} \right)^{\frac{N_\beta (N_I - 1)}{2}} \nonumber \\
& \int \prod_{l = 1}^{N_\beta} \prod_{m = 1}^{N_I - 1} \prod_n d \varphi_n^m
\mathrm{exp} \left[ -\frac{\Delta}{2}\sum_{l = 1}^{N_\beta} \sum_{m = 1}^{N_I - 1} \sum_n \left(\varphi_n^m\right)^2 \right] \nonumber \\
& \quad \times \mathrm{det}
\left[
\mathbb{I} - i \prod_{l = 1}^{N_\beta} \prod_{m = 1}^{N_I} \mathrm{exp} \left( -\Delta h_{l,m} \right)
\right].
\label{auxiliary_fields_partition_function}
\end{align}
Therefore, the action of this system can be obtained as
\begin{align}
\mathcal{S}\left( \bm{\varphi} \right)
= &~\frac{\Delta}{2}\sum_{l = 1}^{N_\beta} \sum_{m = 1}^{N_I - 1} \sum_n \left(\varphi_n^m\right)^2 \nonumber \\
&-\ln \mathrm{det}
\left[ \mathbb{I} - i \prod_{l = 1}^{N_\beta} \prod_{m = 1}^{N_I}
\mathrm{exp} \left( -\Delta h_{l,m} \right)
\right].
\label{action}
\end{align}
Note that, after integrating out the Grassmann variables introduced for the imaginary chemical potential term in
Eq.~(\ref{eq:complex_chemical}), the factor of $i$ appears inside the determinant. For further details of the derivation, one can refer for
example to Ref.~\cite{altland}. Given the action as a function of the auxiliary variables $\bm{\varphi}$ in Eq.~(\ref{action}), we can
perform the ALT-QMC simulations by plugging it in Eq.~(\ref{eq:alt_qmc_formula}).

\subsection{
	\label{subsec:remarks}
  Remarks
}

It is important to note that, if one tries to apply the Hubbard-Stratonovich transformation after the fermionization of the Hamiltonian
into $\mathcal{H}_f$, one needs to introduce four auxiliary variables for each interaction term. In our approach, we need only a single
auxiliary variable for each interaction term, as mentioned in Sec.~\ref{subsec:hs_transform}. This is why we apply the Hubbard-Stratonovich
transformation to the original spin Hamiltonian $\mathcal{H}$ before the fermionization. The computational cost of the present method is
$\mathcal{O}(N_\beta^3 N_\varphi^3)$, since the bottleneck is in the calculation of $\mathrm{det} J$ in Eq.~(\ref{eq:alt_qmc_formula}).
This means that our approach with $N_\varphi = N_S$ is $64$ times faster than the alternative one with $N_\varphi = 4 N_S$.

In the simulations below, we solve Eqs.~(\ref{eq:pos_flow}) and (\ref{eq:jacobian_flow}) by the Runge-Kutta-Fehlberg
algorithm~\cite{cash_tms_16_1990,numerical_recipes_in_c}. In the actual computation of $\partial \mathcal{S} (\bm{z}) / \partial z_l$ and
$\partial^2 \mathcal{S} (\bm{z}) / \partial z_l \partial z_m$, we use the analytical expressions of the derivatives obtained from
Eq.~(\ref{action}). It is important to note that although the complex logarithm in Eq.~(\ref{action}) is ambiguous up to integer multiplies
of $2 \pi i$, Eq.~(\ref{eq:pos_flow}) is well-defined because the ambiguity disappears in case of the analytical
expressions~\cite{kanazawa_jhep_03_2015}.

In order to measure the sign problem in the ALT-QMC method, we introduce three estimates~\cite{ulybyshev_arxiv_1906.02726,
ulybyshev_prd_101_2020,ulybyshev_ppn_51_2020}: the average sign of the action
\begin{align}
S_\mathrm{action} =
\left \lvert
\big \langle
e^{ -i\mathrm{Im}\mathcal{S}\bm{(}\bm{z}(\bm{\varphi})\bm{)}}
\big \rangle_{\mathrm{Re}\mathcal{S}\bm{(}\bm{z}(\bm{\varphi})\bm{)}}
\right \lvert,
\label{eq:action_sign_complex}
\end{align}
the average sign of the Jacobian
\begin{align}
S_\mathrm{Jacobian} =
\left \lvert \big \langle
e^{ i\mathrm{Im} \ln \mathrm{det} J }
\big \rangle_{\mathrm{Re}\mathcal{S}\bm{(}\bm{z}(\bm{\varphi})\bm{)}} \right \lvert,
\label{sign_jacobian}
\end{align}
and the total average sign
\begin{align}
S_\mathrm{total} = 
\left \lvert
\big \langle
e^{ -i\mathrm{Im}\mathcal{S}\bm{(}\bm{z}(\bm{\varphi})\bm{)} + i\mathrm{Im} \ln \mathrm{det} J }
\big \rangle_{\mathrm{Re}\mathcal{S}\bm{(}\bm{z}(\bm{\varphi})\bm{)}}
\right \lvert.
\label{sign_total}
\end{align}

\section{
	\label{sec:kh_model}
	Application to the Kitaev model in a magnetic field
}

In this section, we apply the ALT-QMC method developed above to a model for which the D-QMC method encounters a serious sign problem. We
here adopt the Kitaev honeycomb model in a magnetic field, for which a topological quantum spin liquid with non-Abelian anyonic excitations
is predicted by the perturbation theory in the weak field limit~\cite{kitaev_ann_phys_321_2006}. We introduce the Hamiltonian and briefly
review the fundamental properties in Sec.~\ref{subsec:Kitaev_model}. Then, applying the ALT-QMC method to this model, in
Sec.~\ref{subsec:visualization}, we visualize the time evolution of the asymptotic Lefschetz thimbles as well as the saddle points and the
zeros of the fermion determinant for a simple case with four spins and a single Suzuki-Trotter slice. In
Sec.~\ref{subsec:alleviation_sign}, analyzing the time $t$ dependence in details, we show that the ALT-QMC method needs an optimization
of $t$ to balance the gain in the sign of the MC weight and the numerical efficiency. Finally, in Sec.~\ref{subsec:benchmark}, we present
the benchmark with the detailed comparison between the D-QMC and ALT-QMC methods.

\subsection{
	\label{subsec:Kitaev_model}
	Kitaev model
}

While the scheme presented in this paper can be applied to generic quantum spin models, we here focus on the Kitaev model on a honeycomb
lattice in a uniform magnetic field~\cite{kitaev_ann_phys_321_2006}. The Hamiltonian is given by
\begin{align}
\mathcal{H} = -\sum_{\gamma = x,y,z} K^\gamma \sum_{\langle p,q \rangle_\gamma} \sigma_p^\gamma \sigma_q^\gamma 
              -\sum_{\gamma = x,y,z} h^\gamma \sum_p                            \sigma_p^\gamma,
\label{eq:kitaev_h}
\end{align}
where the sum $\langle p,q \rangle_\gamma$ is taken over all the $\gamma$ bonds corresponding to three different types of bonds on the
honeycomb lattice, and $K^\gamma$ is the exchange constant for the $\gamma$ bonds. 

In the absence of the magnetic field $h^\gamma = 0$, the ground state of the model in Eq.~(\ref{eq:kitaev_h}) can be exactly obtained by
introducing Majorana fermion operators for the spin operators~\cite{kitaev_ann_phys_321_2006}. While the ground state has gapless or gapped
Majorana excitations depending on the anisotropy in the coupling constants $K^\gamma$, it is always a quantum spin liquid with extremely
short-range spin correlations: The spin correlations $\langle \sigma_p^\gamma \sigma_q^\gamma \rangle$ are nonzero only for the
nearest-neighbor $\gamma$ bond, in addition to the onsite ones with $p = q$~\cite{baskaran_prl_98_2007}.

When $h^\gamma \neq 0$, the exact solution is no longer available. The perturbation theory in terms of $h^\gamma$, however, predicts that
the ground state becomes a topological gapped quantum spin liquid with non-Abelian anyonic excitations and a chiral Majorana edge mode in
the vicinity of the isotropic case $K^x = K^y = K^z$~\cite{kitaev_ann_phys_321_2006}. Recently, this prediction has attracted a great
attention, as the experiments for a candidate material of the Kitaev model, $\alpha$-RuCl$_3$, reported unconventional behaviors in the
field-induced paramagnetic region~\cite{baek_prl_119_2017,zheng_prl_119_2017,wang_prl_119_2017,ponomaryov_prb_96_2017,banerjee_nqm_3_2018,
jansa_nphys_14_2018,wellm_prb_98_2018,nagai_prb_101_2020,motome_jpsj_89_2020,takagi_nrp_1_2019}. Although it is still under debate whether
this field-induced paramagnetic state is the topological quantum spin liquid, the recent discovery of the half-quantized thermal Hall
conductivity provides a strong evidence of the chiral Majorana edge mode~\cite{kasahra_nature_559_2018}. Nevertheless, there are few
reliable theoretical results beyond the perturbation theory since the controlled unbiased calculations are hardly available in an applied
magnetic field. In particular, the sign-free QMC calculations are limited to zero field~\cite{nasu_prl_113_2014,nasu_prb_92_2015,
nasu_prl_115_2015,nasu_nphys_12_2016,mishchenko_prb_96_2017,eschmann_prr_1_2019,mishchenko_prb_101_2020,eschmann_prb_102_2020} and the
effective model derived by the perturbation~\cite{nasu_prl_119_2017}. In the following, we apply the ALT-QMC method to the Kitaev model in
the magnetic field in Eq.~\eqref{eq:kitaev_h} and try to extend the accessible parameter region beyond the existing methods.

\subsection{
	\label{subsec:visualization}
	Visualization of the asymptotic Lefschetz thimbles
}

\begin{figure*}[htb]
	\includegraphics[width=\textwidth,clip]{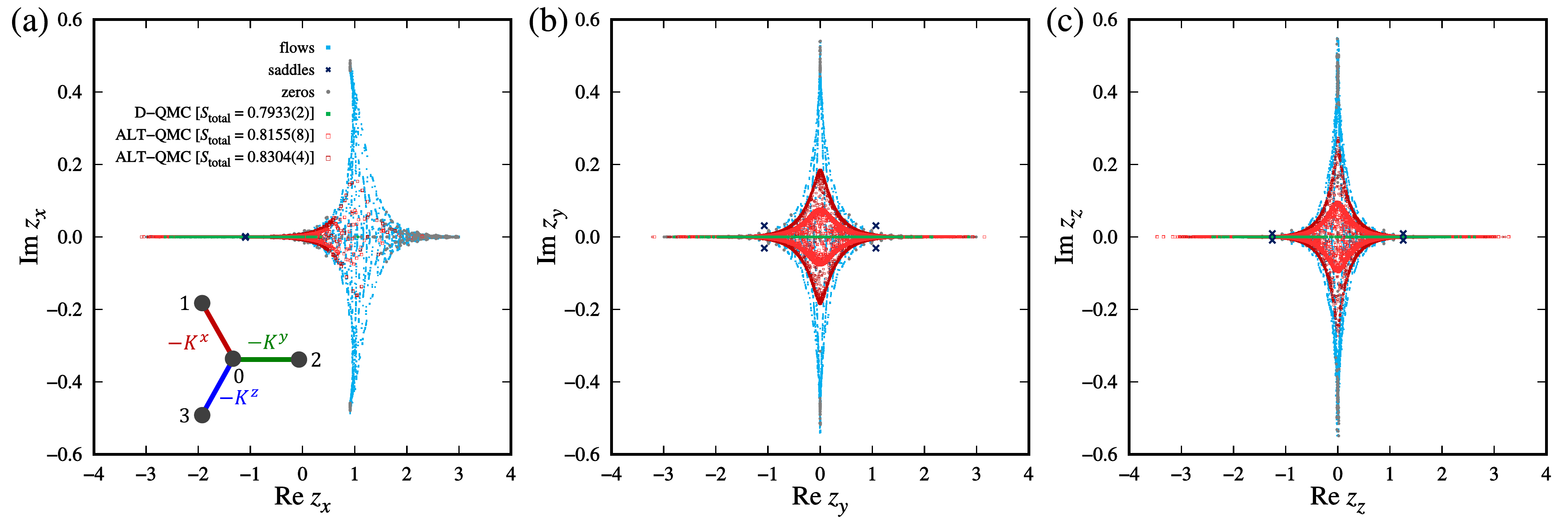}
	\caption{
		Visualization of the asymptotic Lefschetz thimbles for the four-site model in Eq.~(\ref{4_sites_kitaev_h}) with $K^x = K^y = 0.3$,
		$K^z = 0.4$, $h^x = 0.5$, and $h^y = h^z = 0$ [see the inset of (a)]. Projections onto the auxiliary variables on (a) the $x$ bond,
		$z_x$, (b) the $y$ bond, $z_y$, and (c) the $z$ bond, $z_z$ are shown. The light blue points show the time evolution of the
		discrete points on the real space up to $t = 0.16$, which represent the flows of the asymptotic Lefschetz thimbles (corresponding to
		the dashed blue arrows in Fig.~\ref{fig:alt_illustration}). The dark blue crosses and gray dots represent the saddle points and
		``zeros'' (see the text for details), respectively, similar to those in Fig.~\ref{fig:alt_illustration}. The green dots on the real
		axes represent the distribution of samples prepared by the D-QMC method in real space. The light and dark red points represent the
		samples on the asymptotic Lefschetz thimbles obtained at $t = 0.04$ and at $t = 0.08$, respectively. The average sign
		$S_\mathrm{total}$ is alleviated from $S_\mathrm{total} = 0.7933(2)$ in the D-QMC result ($t = 0$) to $S_\mathrm{total} = 0.8155(8)$
		and $S_\mathrm{total} = 0.8304(4)$ on the asymptotic Lefschetz thimbles at $t = 0.04$ and $t = 0.08$, respectively. All the data are
		obtained for $N_\beta = 1$ with $\beta = 3.125$.
	}
	\label{fig:alt_visualization}
\end{figure*}

To get some insight on how the sign problem is alleviated by the ALT-QMC method, we visualize the asymptotic Lefschetz thimbles by
following the time evolution explicitly. For this purpose, we adopt a simple model consisting of four sites of the Kitaev model, as
shown in the inset of Fig.~\ref{fig:alt_visualization}, whose Hamiltonian is given by
\begin{align}
\mathcal{H}_\mathrm{4\textnormal{-}site}
= &-K^x\sigma_0^x\sigma_1^x -K^y\sigma_0^y\sigma_2^y -K^z\sigma_0^z\sigma_3^z \nonumber \\
  &-\sum_{\gamma = x,y,z} h^\gamma (\sigma_0^\gamma + \sigma_1^\gamma + \sigma_2^\gamma + \sigma_3^\gamma).
\label{4_sites_kitaev_h}
\end{align}
We take a single Suzuki-Trotter slice, namely, $N_\beta = 1$, where we have only three auxiliary variables denoted as $\varphi_x$,
$\varphi_y$, and $\varphi_z$ on the $x$, $y$, and $z$ bonds connecting $0$-$1$, $0$-$2$, and $0$-$3$ sites, respectively. We take the
parameters as $K^x = K^y = 0.3$, $K^z = 0.4$, $h^x = 0.5$, $h^y = h^z = 0$, and $\beta = 3.125$. For this set up, we plot the projections
of the asymptotic Lefschetz thimbles onto the complex planes of $z_x$, $z_y$, and $z_z$, which are obtained via the analytic continuation
of the real variables $\varphi_x$, $\varphi_y$, and $\varphi_z$, respectively. Since in the present case the original integration domain is
$\mathbb{R}^3$, the Lefschetz thimbles will be $\mathbb{R}^3$ embedded in $\mathbb{C}^3$, and the zeros of the fermion determinant will be
$\mathbb{C}^2$ (see Sec.~\ref{subsec:a_l_thimbles}). For the explicit form of the action $\mathcal{S}(\bm{z})$ derived for the model in
Eq.~(\ref{4_sites_kitaev_h}) we refer to Appendix~\ref{sec:appendixA}.

First, we demonstrate the time flows of the asymptotic Lefschetz thimbles, which are schematically drawn by the dashed blue arrows in
Fig.~\ref{fig:alt_illustration}. For this purpose, we prepare a set of discrete points in the original integration domain $\mathbb{R}^3$ in
the parameter range where the weight $e^{ -\mathrm{Re}\mathcal{S}(\bm{\varphi}) }$ has a significant value (we confirm that the numerical
integration over the discrete points reproduce the value of the action with sufficient precision), and follow their time evolution
calculated by Eq.~(\ref{eq:pos_flow}). The results up to $t = 0.16$ are shown by the light blue points in Fig.~\ref{fig:alt_visualization}.
The flows evolve while increasing time and appear to form envelops in a different way between the three variables $z_x$, $z_y$, and $z_z$.
The envelops are expected to give the asymptotic Lefschetz thimbles $\mathcal{C}_t$, as schematically shown in
Fig.~\ref{fig:alt_illustration}, although the time flows may become numerically unstable at some point in practice (see below).

At the same time, we plot both the saddle points and the zeros of the fermion determinant in Fig.~\ref{fig:alt_visualization}. The saddle
points are obtained by solving Eq.~(\ref{eq:saddle}) directly. They are located near the real axis in all the three projections, forming
the complex conjugate pairs, as shown by the crosses in Fig.~\ref{fig:alt_visualization}. The envelops of the asymptotic Lefschetz thimbles
(light blue points) appear to approach the saddle points by the time evolution. Meanwhile, the zeros are obtained from the time
evolution by Eq.~(\ref{eq:pos_flow}) as follows. In the vicinity of zeros of the fermion determinant, the solution of
Eq.~(\ref{eq:pos_flow}) blows up, and hence, the numerical integration becomes unstable. We assume that the flow in Eq.~(\ref{eq:pos_flow})
hits a zero when the numerical value of $\mathrm{Re}\mathcal{S}(\bm{z})$ starts to decrease or when the value of
$\mathrm{Im}\mathcal{S}(\bm{z})$ starts to deviate significantly during the time evolution (we set the threshold as one percent of the
previous values in the flows with each Runge-Kutta-Fehlberg adaptive step size). We show the points obtained by this procedure by the gray
dots in Fig.~\ref{fig:alt_visualization}. Note that not all of them are true zeros of the fermion determinant, as they may include some
points where the numerical integration of Eq.~(\ref{eq:pos_flow}) simply fails by technical reasons (the points at which the
Runge-Kutta-Fehlberg adaptive step size becomes too small or the number of adaptive steps becomes too large are also included). In any
case, the results in Fig.~\ref{fig:alt_visualization} indicate the asymptotic Lefschetz thimbles appear to be terminated in the regions
where the ``zeros'' are densely distributed.

\subsection{
	\label{subsec:alleviation_sign}
	Alleviation of sign problem
}

\begin{figure}[htb]
	\includegraphics[width=\columnwidth,clip]{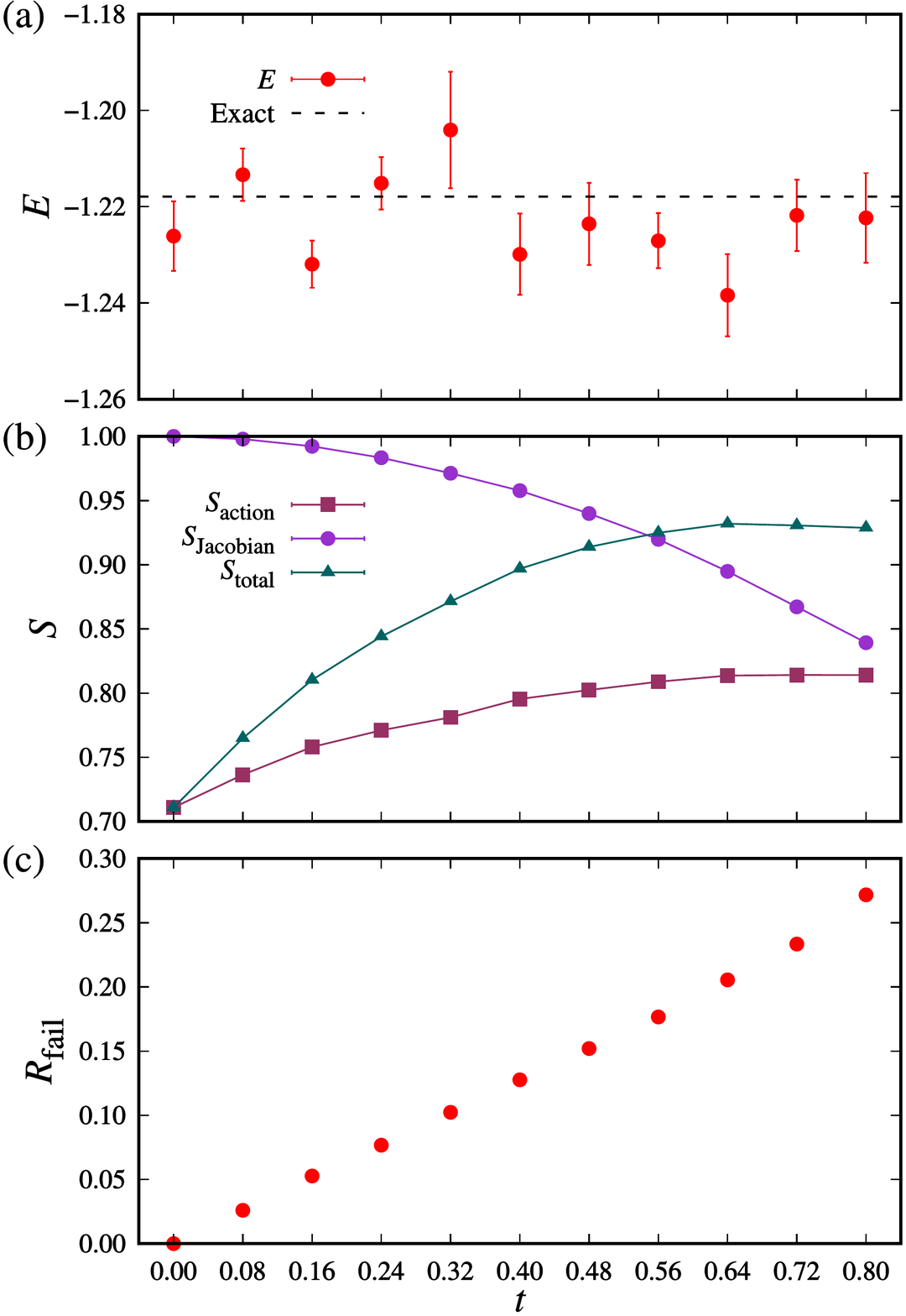}
	\caption{
		Flow time $t$ dependences of (a) the internal energy $E$, (b) the average sign of the action, $S_\mathrm{action}$ in
		Eq.~(\ref{eq:action_sign_complex}), the average sign of the Jacobian, $S_\mathrm{Jacobian}$ in Eq.~(\ref{sign_jacobian}) and the total
		average sign $S_\mathrm{total}$ in Eq.~(\ref{sign_total}), and (c) the ratio of the failed samples, $R_\mathrm{fail}$. The model is the
		four-site Kitaev model in Eq.~(\ref{4_sites_kitaev_h}) with $K^x = K^y = K^z = 1/3$, $h^x = h^y = 0$, and $h^z = 0.2$. We take
		$N_\beta = 4$ and $\beta = 5$. The statistical errorbars are calculated from eight independent samples, each consisting of
		$5 \times 10^3$ MC sweeps for thermalization and $9.5 \times 10^3$ MC sweeps for measurement.
	}
	\label{fig:alt_qmc_benchmark}
\end{figure}

\begin{table*}[t]
	\centering
	\begin{tabular}{cccccccccc}
		\hline
		\qquad    & \qquad \qquad & exact	    & \qquad \qquad & \multicolumn{2}{c}{D-QMC}                     & \qquad \qquad & \multicolumn{3}{c}{ALT-QMC}	                           \\ \hline
		$N_\beta$ & \qquad \qquad & $E$       & \qquad \qquad & $E$         & $S_\mathrm{D\textnormal{-}QMC}$ & \qquad \qquad & $E$         & $S_\mathrm{action}$ & $S_\mathrm{total}$ \\ \hline \hline
		$4$       & \qquad \qquad & $-1.2179$ & \qquad \qquad & $-1.226(7)$ & $0.711(3)$                      & \qquad \qquad & $-1.229(9)$ & $0.769(3)$          & $0.837(2)$         \\
		$6$       & \qquad \qquad & $-1.1760$ & \qquad \qquad & $-1.181(9)$ & $0.726(3)$                      & \qquad \qquad & $-1.178(5)$ & $0.799(2)$          & $0.828(2)$         \\
		$8$       & \qquad \qquad & $-1.1612$ & \qquad \qquad & $-1.16(1)$  & $0.736(4)$                      & \qquad \qquad & $-1.170(9)$ & $0.803(2)$          & $0.818(3)$         \\ \hline
	\end{tabular}
	\caption{
		Comparison of the results obtained by the exact computation of the partition function in Eq.~(\ref{eq:suzuki_trotter_fermion}), the D-QMC
		method, and the ALT-QMC method. The model, parameters, and notations are common to those in Fig.~\ref{fig:alt_qmc_benchmark}, except
		for $N_\beta$. The ALT-QMC results are obtained at $t = 0.22$. $S_\mathrm{Jacobian} \simeq 1$ for all the cases.
	}
	\label{table:flow_its}
\end{table*}

With the above visualization in mind, we demonstrate how the actual ALT-QMC simulations work. First, we present the distribution of the MC
samples obtained by the D-QMC technique. They are distributed in the original integration domain $\mathbb{R}^3$, and hence, on the real
axes in each projection, as shown by the green dots in Fig.~\ref{fig:alt_visualization}. The total average sign in Eq.~(\ref{sign_total})
is $S_\mathrm{total} = 0.7933(2)$ for the D-QMC results. Here and hereafter, the number in the parenthesis represents the statistical error
in the last digit, which is estimated by the standard error calculated from several independent MC samples, $\sigma_\mathrm{SE}$. In the
present case, we take four independent MC samples, each having $9 \times 10^5$ MC sweeps for measurement after $10^5$ MC sweeps for
thermalization. Note that not all the samples are plotted in Fig.~\ref{fig:alt_visualization}; only the values obtained every
$2.5 \times 10^3$ sweeps in one of four MC samples are shown for better visibility. The internal energy is estimated as
$E_\mathrm{D\textnormal{-}QMC} = -2.7505(9)$, which reproduces well the exact value of $E_\mathrm{exact} = -2.75032$ within the statistical
error.

Next, we show the distributions of the MC samples obtained by the ALT-QMC simulations, which correspond to the asymptotic Lefschetz
thimbles $\mathcal{C}_t$. Figure~\ref{fig:alt_visualization} displays the results with the time evolution after $t = 0.04$ and
$t = 0.08$ by the light and dark red dots, respectively. For these calculations we take four independent MC samples, each having
$9 \times 10^4$ MC sweeps for measurement after $10^4$ MC sweeps for thermalization and only the values obtained every ten sweeps in one of
four MC samples are shown in Fig.~\ref{fig:alt_visualization} for better visibility. The results indicate that the asymptotic Lefschetz
thimbles evolve from the real axis in each projection along the flows shown by the light blue points. As expected, the total average sign
increases with $t$ as $S_\mathrm{total} = 0.8155(8)$ and $S_\mathrm{total} = 0.8304(4)$ for $t = 0.04$ and $t = 0.08$, respectively. The
internal energy obtained by the ALT-QMC simulations reproduce the exact result as $E_\mathrm{ALT\textnormal{-}QMC} = -2.754(3)$ for
$t = 0.04$ and $E_\mathrm{ALT\textnormal{-}QMC} = -2.749(2)$ for $t = 0.08$.

Let us discuss the flow time $t$ dependence in more detail, by taking $\beta = 5$ and $N_\beta = 4$, for the four-site Kitaev model in
Eq.~(\ref{4_sites_kitaev_h}) with $K^x = K^y = K^z = 1/3$, $h^x = h^y = 0$, and $h^z = 0.2$. Figure~\ref{fig:alt_qmc_benchmark}(a) shows
the $t$ dependence of the internal energy $E$, indicating that $E$ reproduces well the exact value, shown by the horizontal dashed line,
for all $t$ up to $t = 0.80$ calculated here. Figure~\ref{fig:alt_qmc_benchmark}(b) shows the $t$ dependences of the total average sign
$S_\mathrm{total}$ in Eq.~(\ref{sign_total}), the average sign of the Jacobian, $S_\mathrm{Jacobian}$ in Eq.~(\ref{sign_jacobian}), and the
average sign of the action, $S_\mathrm{action}$ in Eq.~(\ref{eq:action_sign_complex}). We find that $S_\mathrm{action}$ increases with $t$
up to $t \simeq 0.64$ and then saturates to $S_\mathrm{action} \simeq 0.81$. We note that when there is only one Lefschetz thimble,
$S_\mathrm{action}$ can be arbitrarily close to $1$ while increasing $t$; however, this does not hold for the case with multiple thimbles,
since $\mathrm{Im}\mathcal{S}(\bm{z})$ can vary from one thimble to another. Hence, the above result suggests that there are more than one
relevant thimbles in the present case. On the other hand, $S_\mathrm{Jacobian}$ steadily decreases from $1$ with $t$. As a consequence,
$S_\mathrm{total}$, which includes these two contributions, steadily increases up to $t \simeq 0.64$, while it turns to decrease for larger
$t$. These results show that the total sign $S_\mathrm{total}$ does not approach $1$ and shows a maximum at some point of the time $t$.

We also measure the ratio of the failed samples, $R_\mathrm{fail}$, in the MC simulations. The failed samples are the MC samples which
collapse onto the zeros of the fermion determinant (or simply become unstable in the time evolution). In our simulations, we discard
them and do not take into account in the MC measurement. The ratio of the failed samples denoted by $R_\mathrm{fail}$ is defined as the
number of the failed samples divided by the total number of the initial MC samples. Figure~\ref{fig:alt_qmc_benchmark}(c) shows the $t$
dependence of $R_\mathrm{fail}$. We find that $R_\mathrm{fail}$ increases almost linearly with $t$, indicating that the ALT-QMC simulation
loses its efficiency while increasing $t$. 

Thus, in the practical ALT-QMC simulations, $S_\mathrm{total}$ becomes maximum at some $t$ and turns to decrease for large $t$, while
$R_\mathrm{fail}$ monotonically increases with $t$. Hence, $t$ should be optimized to retain reasonable values of $S_\mathrm{total}$ and
$R_\mathrm{fail}$ in practical simulations. The optimal value of $t$ could be determined by running test runs with small numbers of MC
samples.

Next, we examine the convergence with respect to $N_\beta$. Table~\ref{table:flow_its} summarizes the results for the same model as in
Fig.~\ref{fig:alt_qmc_benchmark} while varing $N_\beta$ from $4$ to $8$. The ALT-QMC results are obtained at $t = 0.22$, where
$R_\mathrm{fail}$ falls in the range of $0.06 \lesssim R_\mathrm{fail} \lesssim 0.07$. We note that $R_\mathrm{fail}$ becomes smaller for
larger $N_\beta$. At this relatively short flow time, $S_\mathrm{Jacobian} \simeq 1$ for all the cases. The results indicate that the
exact values of the internal energy are well reproduced in all cases within the statistical errors. In addition, we find that the ALT-QMC
method indeed improves the average sign for all $N_\beta$, compared to the D-QMC results. As a result, $S_\mathrm{total}$ in the ALT-QMC
simulation is considerably larger compared to $S_\mathrm{D\textnormal{-}QMC}$ in the D-QMC simulation, while $S_\mathrm{action}$ as well as
$S_\mathrm{total}$ is reduced gradually while increasing $N_\beta$. The results prove the efficiency of the ALT-QMC method.

\subsection{
	\label{subsec:benchmark}
	Benchmark
}

Finally, we present the benchmark of the ALT-QMC method by performing larger scale simulations. We here consider the Kitaev model in a
magnetic field on an eighteen-site cluster ($N_S = 2 \times 3^2$) with the periodic boundary conditions. In the following, we take the
parameters to retain the threefold rotational symmetry, namely, $K^x = K^y = K^z = 1/3$ and $h^x = h^y = h^z = h$ because of the following
reason. As discussed in Sec.~\ref{subsec:alleviation_sign}, the sign problem in the ALT-QMC simulations depends on the number of the
relevant thimbles and the values of $\mathrm{Im}\mathcal{S}(\bm{z})$ on each thimble; in the extreme case, when the system has only a
single thimble associated with a single saddle point, one would expect a significant alleviation of the sign problem. For this reason, we
first analyzed the structure of the saddle points by varying the model parameters $K^\gamma$ and $h^\gamma$. In order to identify the
saddle points, we performed D-QMC simulations and solve Eq.~(\ref{eq:saddle}) for each D-QMC sample in the real space. From this analysis,
we found that the system appears to have a single saddle point for the symmetric case with $K^x = K^y = K^z = 1/3$ and
$h^x = h^y = h^z = h$. We note that the values of the auxiliary fields for the saddle point are the same for all the Suzuki-Trotter slices,
while they vary with $h$ as well as temperature $T$. Hence, we take this symmetric parameter set for the following simulations.

\begin{figure}[htb]
	\includegraphics[width=\columnwidth,clip]{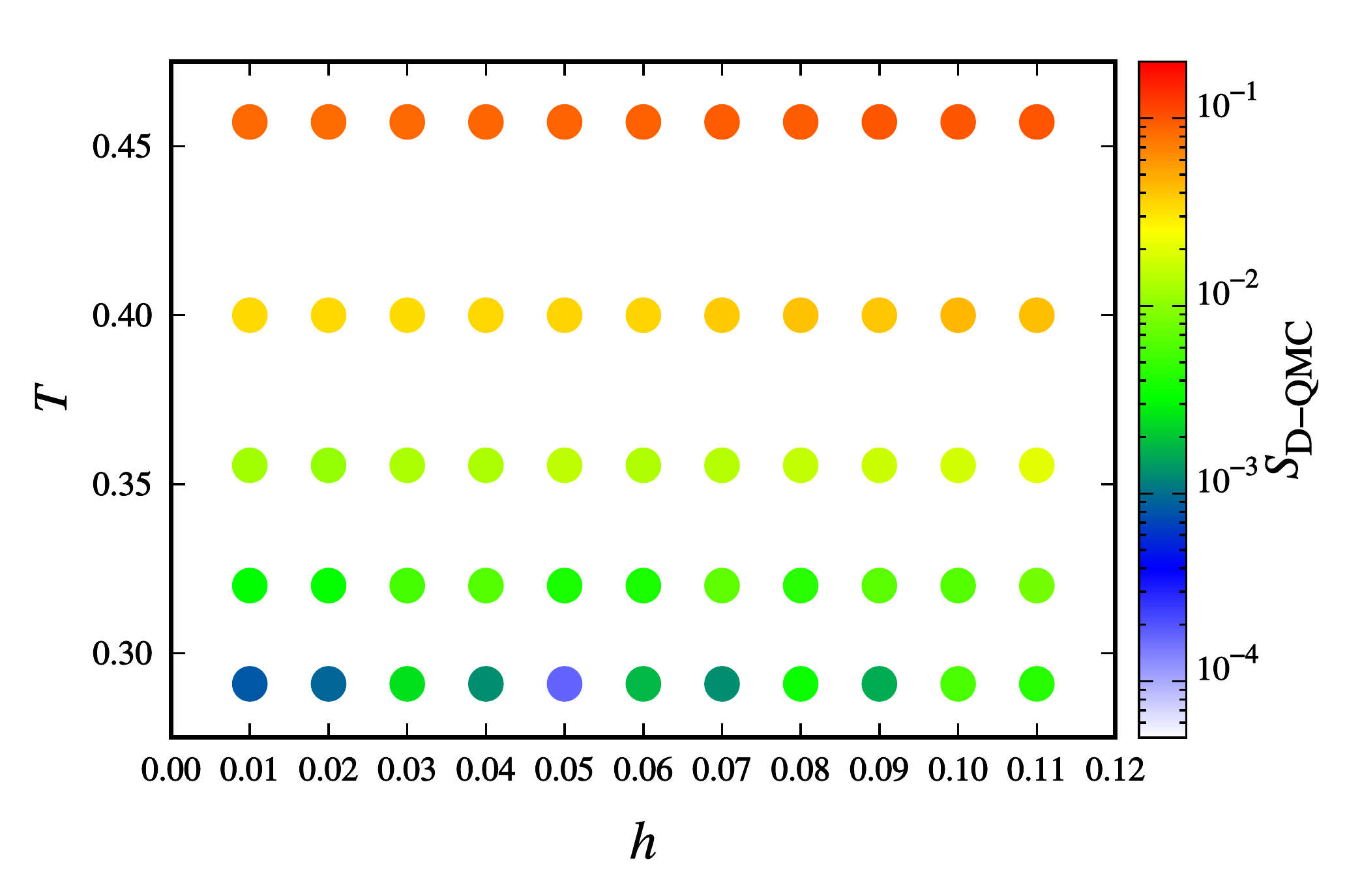}
	\caption{
		The average sign of the action, $S_\mathrm{D\textnormal{-}QMC}$ in Eq.~(\ref{eq:action_sign_real}), obtained by the D-QMC simulations
		while changing the magnetic field $h$ and temperature $T$. The results are obtained for the eighteen-site Kitaev model with
		$K^x = K^y = K^z = 1/3$ and $h^x = h^y = h^z = h$. The average values of the sign are calculated from five independent samples, each
		consisting of $10^5$ MC sweeps for thermalization and $9\times10^5$ MC sweeps for sampling. All the data are obtained for
		$N_\beta = 8$.
	}
	\label{fig:field_temperature_sign}
\end{figure}	

\begin{figure}[htb]
	\includegraphics[width=\columnwidth,clip]{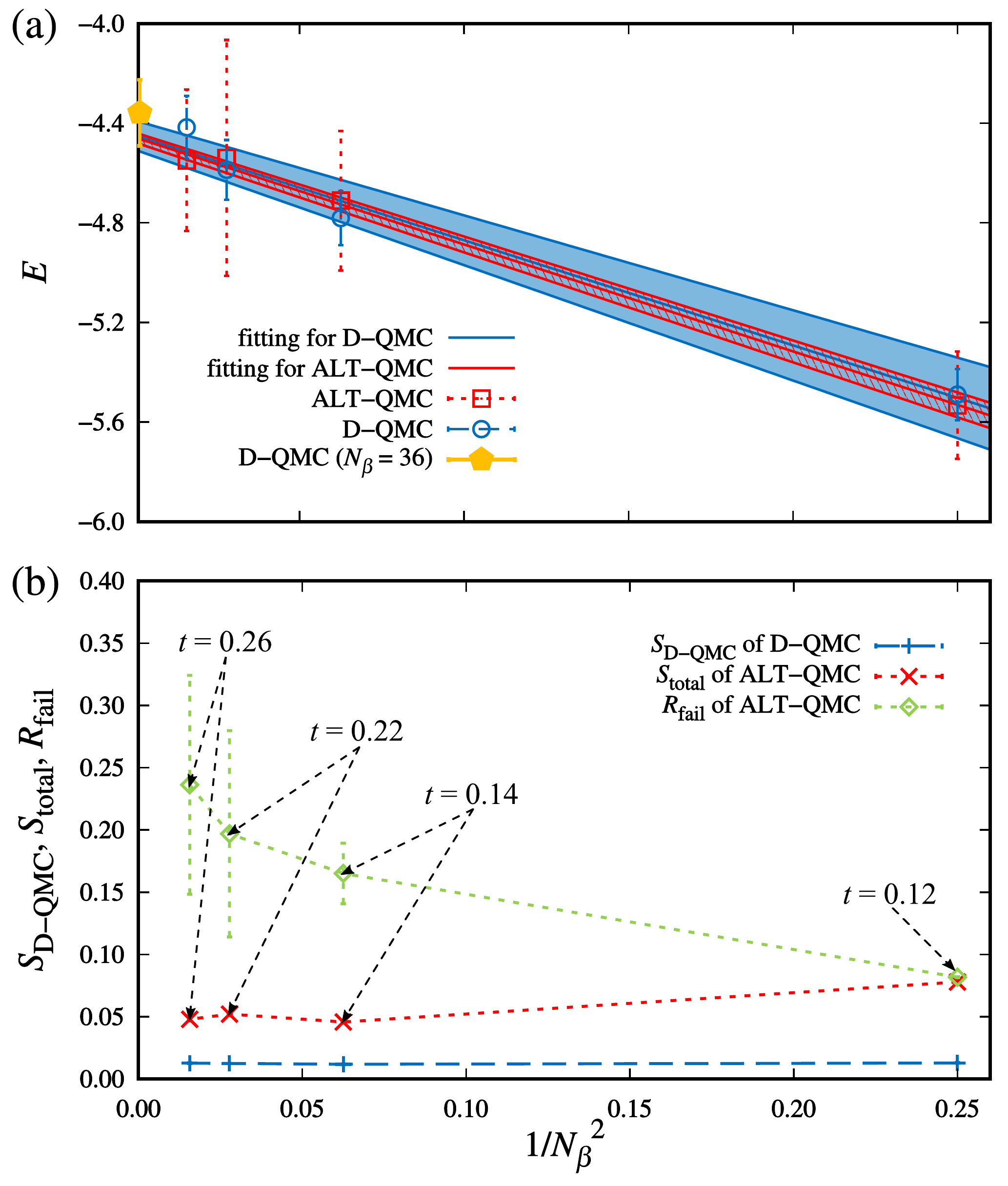}
	\caption{
		Benchmark results of the ALT-QMC simulations for the Kitaev model in the magnetic field. (a) The ALT-QMC results for the internal
		energy $E$, in comparison with the D-QMC results. The extrapolation to $N_\beta \to \infty$ is shown by the fitting with a linear
		function of $1/N_\beta^2$, plotted with the asymptotic standard errors denoted by the hatched area, for both the D-QMC and ALT-QMC
		data. Note that the D-QMC data at $N_\beta = 36$ plotted by the filled orange pentagon is not incorporated in the fitting. (b) The
		total sign $S_\mathrm{total}$ in the ALT-QMC simulations, in comparison with the sign of the action, $S_\mathrm{D\textnormal{-}QMC}$,
		in the D-QMC simulations. The values of the flow time $t$ are shown in the figure. The ratio of the failed samples, $R_\mathrm{fail}$,
		is also shown. The results are obtained for the eighteen-site Kitaev model with $K^x = K^y = K^z = 1/3$ and $h^x = h^y = h^z = 0.04$ at
		$\beta = 2.8125$. The statistical errorbars for the ALT-QMC results are calculated from ten independent samples, each consisting of
		$5 \times 10^3$ MC sweeps for thermalization and $4.5 \times 10^4$ MC sweeps for sampling. The statistical errorbars for the D-QMC
		[D-QMC ($N_\beta = 36$)] results are calculated from ten independent samples, each consisting of $10^5$ ($10^6$) MC sweeps for
		thermalization and $9 \times 10^5$ ($4 \times 10^6$) MC sweeps for sampling. All the errorbars are presented with $\sigma_\mathrm{SE}$.
	}
	\label{fig:alt_qmc_result}
\end{figure}

First of all, we show the behavior of the sign while varying $T$ and $h$ obtained by the D-QMC simulations. We find that the average sign
of the action, $S_\mathrm{D\textnormal{-}QMC}$, decreases rapidly as $T$ decreases, as plotted in Fig.~\ref{fig:field_temperature_sign}.
The decrease is more severe in the smaller field region for $h \lesssim 0.09$. The results suggest that the D-QMC simulations become
inefficient in the region below $T \simeq 0.3$ and $h \lesssim 0.09$, where $S_\mathrm{D\textnormal{-}QMC}$ becomes smaller than $10^{-3}$,
even for this relatively small cluster with eighteen sites.

Given the D-QMC results, we perform the ALT-QMC simulations at $T = 0.3\dot{5}$ ($\beta = 2.8125$) and $h = 0.04$, where
$S_\mathrm{D\textnormal{-}QMC}$ is small but still retains a reasonable value of $0.013(2)$. We show the results for $N_\beta = 2$, $4$,
$6$, and $8$ in Fig.~\ref{fig:alt_qmc_result}, in comparison with the D-QMC results which include the $N_\beta = 36$ data.
Figure~\ref{fig:alt_qmc_result}(a) shows the internal energy $E$, as a function of $1/N_\beta^2$. We find that the results by the D-QMC
and ALT-QMC methods agree with each other within the errorbars for all $N_\beta$. The data follows well a linear function of $1/N_\beta^2$
as expected from the discussion in Sec.~\ref{subsec:path_integral}. The fittings are shown for both results with the confidence interval of
the errors (hatched areas) estimated by a modified least-square method taking into account the errors of the data~\cite{taylor_1997}. We
find that the extrapolated values agree well with the result obtained by the D-QMC with sufficiently large $N_\beta = 36$.

Figure~\ref{fig:alt_qmc_result}(b) shows the total average sign $S_\mathrm{total}$ and the ratio of the failed samples, $R_\mathrm{fail}$,
obtained by the ALT-QMC technique. We also plot the average sign of the action, $S_\mathrm{D\textnormal{-}QMC}$, obtained by the D-QMC
technique for comparison. In the ALT-QMC simulations, we tune the flow time $t$ so that $S_\mathrm{total} \simeq 0.05$ for $N_\beta = 4$,
$6$, and $8$. The value of $t$ becomes larger and $R_\mathrm{fail}$ becomes higher for larger $N_\beta$, as shown in
Fig.~\ref{fig:alt_qmc_result}(b). Comparing $S_\mathrm{total}$ in the ALT-QMC simulations with $S_\mathrm{D\textnormal{-}QMC}$ in the D-QMC
simulations, we find the improvement of nearly four times for all $N_\beta$, except for nearly six times for the $N_\beta = 2$ case. Since
the number of measurements required to obtain a constant error is proportional to $(S_\mathrm{total})^{-2}$, our results suggest that the
performance in terms of the number of MC samples becomes nearly sixteen times higher in the ALT-QMC simulations. We note, however, that the
improvement is still not enough to replace the D-QMC method at this stage, when taking into account the computational cost; the calculation
complexity of the D-QMC method is $\mathcal{O}(N_\beta N_S)$, which is substantially better than that of the ALT-QMC,
$\mathcal{O}(N_\beta^3 N_S^3)$.

%
\section{Summary}
\label{sec:summary_and_outlook}
%

To summarize, we have developed a QMC method based on the asymptotic Lefschetz thimbles that is applicable to quantum spin models with
generic two-spin interactions and the Zeeman coupling to a magnetic field. The method is constructed by introducing the
Hubbard-Stratonovich transformation and the Popov-Fedotov transformation. As a demonstration, we applied the method dubbed the ALT-QMC
method to the Kitaev model in a magnetic field. For the four-site cluster, we visualized explicitly the time evolution of the asymptotic
Lefschetz thimbles in the complex space of the auxiliary variables in the Hubbard-Stratonovich transformation, and presented how the
ALT-QMC method alleviates the sign problem compared to the conventional determinant QMC method. We also showed the benchmark of the
eighteen-site cluster and demonstrated the potential of studying lower-temperature regions in a magnetic field compared to the conventional
method. Although the ALT-QMC method developed in the present study is still too costly, we emphasize that the average sign can be
considerably improved from that in the determinant QMC method. Our results may pave the way for constructing further efficient numerical
techniques.

\begin{acknowledgments}
This research was supported by Grant-in-Aid for Scientific Research Grants Numbers JP19H05822, JP20H00122, JP18K03447, and JST CREST
(JP-MJCR18T2). P.A.M. was supported by JSPS through a research fellowship for young scientists. Parts of the numerical calculations were
performed in the supercomputing systems in ISSP, the University of Tokyo.
\end{acknowledgments}

\bibliography{submit}

\providecommand{\noopsort}[1]{}\providecommand{\singleletter}[1]{#1}%
\begin{thebibliography}{105}%
\makeatletter
\providecommand \@ifxundefined [1]{%
 \@ifx{#1\undefined}
}%
\providecommand \@ifnum [1]{%
 \ifnum #1\expandafter \@firstoftwo
 \else \expandafter \@secondoftwo
 \fi
}%
\providecommand \@ifx [1]{%
 \ifx #1\expandafter \@firstoftwo
 \else \expandafter \@secondoftwo
 \fi
}%
\providecommand \natexlab [1]{#1}%
\providecommand \enquote  [1]{``#1''}%
\providecommand \bibnamefont  [1]{#1}%
\providecommand \bibfnamefont [1]{#1}%
\providecommand \citenamefont [1]{#1}%
\providecommand \href@noop [0]{\@secondoftwo}%
\providecommand \href [0]{\begingroup \@sanitize@url \@href}%
\providecommand \@href[1]{\@@startlink{#1}\@@href}%
\providecommand \@@href[1]{\endgroup#1\@@endlink}%
\providecommand \@sanitize@url [0]{\catcode `\\12\catcode `\$12\catcode
  `\&12\catcode `\#12\catcode `\^12\catcode `\_12\catcode `\%12\relax}%
\providecommand \@@startlink[1]{}%
\providecommand \@@endlink[0]{}%
\providecommand \url  [0]{\begingroup\@sanitize@url \@url }%
\providecommand \@url [1]{\endgroup\@href {#1}{\urlprefix }}%
\providecommand \urlprefix  [0]{URL }%
\providecommand \Eprint [0]{\href }%
\providecommand \doibase [0]{http://dx.doi.org/}%
\providecommand \selectlanguage [0]{\@gobble}%
\providecommand \bibinfo  [0]{\@secondoftwo}%
\providecommand \bibfield  [0]{\@secondoftwo}%
\providecommand \translation [1]{[#1]}%
\providecommand \BibitemOpen [0]{}%
\providecommand \bibitemStop [0]{}%
\providecommand \bibitemNoStop [0]{.\EOS\space}%
\providecommand \EOS [0]{\spacefactor3000\relax}%
\providecommand \BibitemShut  [1]{\csname bibitem#1\endcsname}%
\let\auto@bib@innerbib\@empty
\bibitem [{\citenamefont {Blankenbecler}\ \emph {et~al.}(1981)\citenamefont
  {Blankenbecler}, \citenamefont {Scalapino},\ and\ \citenamefont
  {Sugar}}]{blankenbecler_prd_24_1981}%
  \BibitemOpen
  \bibfield  {author} {\bibinfo {author} {\bibfnamefont {R.}~\bibnamefont
  {Blankenbecler}}, \bibinfo {author} {\bibfnamefont {D.~J.}\ \bibnamefont
  {Scalapino}}, \ and\ \bibinfo {author} {\bibfnamefont {R.~L.}\ \bibnamefont
  {Sugar}},\ }\bibfield  {title} {\enquote {\bibinfo {title} {{Monte Carlo
  calculations of coupled boson-fermion systems. I}},}\ }\href
  {https://doi.org/10.1103/PhysRevD.24.2278} {\bibfield  {journal} {\bibinfo
  {journal} {Phys. Rev. D}\ }\textbf {\bibinfo {volume} {24}},\ \bibinfo
  {pages} {2278} (\bibinfo {year} {1981})}\BibitemShut {NoStop}%
\bibitem [{\citenamefont {Scalapino}\ and\ \citenamefont
  {Sugar}(1981)}]{scalapino_prb_24_1981}%
  \BibitemOpen
  \bibfield  {author} {\bibinfo {author} {\bibfnamefont {D.~J.}\ \bibnamefont
  {Scalapino}}\ and\ \bibinfo {author} {\bibfnamefont {R.~L.}\ \bibnamefont
  {Sugar}},\ }\bibfield  {title} {\enquote {\bibinfo {title} {{Monte Carlo
  calculations of coupled boson-fermion systems. II}},}\ }\href
  {https://doi.org/10.1103/PhysRevB.24.4295} {\bibfield  {journal} {\bibinfo
  {journal} {Phys. Rev. B}\ }\textbf {\bibinfo {volume} {24}},\ \bibinfo
  {pages} {4295} (\bibinfo {year} {1981})}\BibitemShut {NoStop}%
\bibitem [{\citenamefont {von~der Linden}(1992)}]{linden_pys_rep_220_1992}%
  \BibitemOpen
  \bibfield  {author} {\bibinfo {author} {\bibfnamefont {W.}~\bibnamefont
  {von~der Linden}},\ }\bibfield  {title} {\enquote {\bibinfo {title} {{A
  quantum Monte Carlo approach to many-body physics}},}\ }\href
  {https://doi.org/10.1016/0370-1573(92)90029-Y} {\bibfield  {journal}
  {\bibinfo  {journal} {Phys. Rep.}\ }\textbf {\bibinfo {volume} {220}},\
  \bibinfo {pages} {53} (\bibinfo {year} {1992})}\BibitemShut {NoStop}%
\bibitem [{\citenamefont {Loh}\ and\ \citenamefont
  {Gubernatis}(1992)}]{loh_gubernatis_1992}%
  \BibitemOpen
  \bibfield  {author} {\bibinfo {author} {\bibfnamefont {E.~Y.}\ \bibnamefont
  {Loh}}\ and\ \bibinfo {author} {\bibfnamefont {J.~E.}\ \bibnamefont
  {Gubernatis}},\ }\href@noop {} {\emph {\bibinfo {title} {{"Stable Numerical
  Simulations of Models of Interacting Electrons in Condensed-Matter Physics"
  in Modern Problems in Condensed Matter Sciences, Electronic Phase
  Transitions}}}},\ Vol.~\bibinfo {volume} {32}\ (\bibinfo  {publisher}
  {Elsevier Science Publishers B.V.},\ \bibinfo {year} {1992})\ pp.\ \bibinfo
  {pages} {177--235}\BibitemShut {NoStop}%
\bibitem [{\citenamefont {dos Santos}(2003)}]{santos_bjp_33_2003}%
  \BibitemOpen
  \bibfield  {author} {\bibinfo {author} {\bibfnamefont {R.~R.}\ \bibnamefont
  {dos Santos}},\ }\bibfield  {title} {\enquote {\bibinfo {title}
  {{Introduction to quantum Monte Carlo simulations for fermionic systems}},}\
  }\href {http://dx.doi.org/10.1590/S0103-97332003000100003} {\bibfield
  {journal} {\bibinfo  {journal} {Braz. J. Phys.}\ }\textbf {\bibinfo {volume}
  {33}},\ \bibinfo {pages} {36} (\bibinfo {year} {2003})}\BibitemShut {NoStop}%
\bibitem [{\citenamefont {Bercx}\ \emph {et~al.}(2003)\citenamefont {Bercx},
  \citenamefont {Goth}, \citenamefont {Hofmann},\ and\ \citenamefont
  {Assaad}}]{bercx_scipost_phys_3_2017}%
  \BibitemOpen
  \bibfield  {author} {\bibinfo {author} {\bibfnamefont {M.}~\bibnamefont
  {Bercx}}, \bibinfo {author} {\bibfnamefont {F.}~\bibnamefont {Goth}},
  \bibinfo {author} {\bibfnamefont {J.~S.}\ \bibnamefont {Hofmann}}, \ and\
  \bibinfo {author} {\bibfnamefont {F.~F.}\ \bibnamefont {Assaad}},\ }\bibfield
   {title} {\enquote {\bibinfo {title} {{The ALF (Algorithms for Lattice
  Fermions) project release 1.0. Documentation for the auxiliary field quantum
  Monte Carlo code}},}\ }\href {https://arxiv.org/abs/1704.00131v2} {\bibfield
  {journal} {\bibinfo  {journal} {SciPost Phys.}\ }\textbf {\bibinfo {volume}
  {3}},\ \bibinfo {pages} {013} (\bibinfo {year} {2003})},\ \Eprint
  {http://arxiv.org/abs/arXiv:1704.00131v2} {arXiv:1704.00131v2} \BibitemShut
  {NoStop}%
\bibitem [{\citenamefont {Assaad}\ and\ \citenamefont
  {Evertz}(2008)}]{assaad_evertz}%
  \BibitemOpen
  \bibfield  {author} {\bibinfo {author} {\bibfnamefont {F.~F.}\ \bibnamefont
  {Assaad}}\ and\ \bibinfo {author} {\bibfnamefont {H.~G.}\ \bibnamefont
  {Evertz}},\ }\href@noop {} {\emph {\bibinfo {title} {{"World-line and
  determinant quantum monte carlo methods for spins, phonons and electrons" in
  Computational Many-Particle Physics, Lecture Notes in Physics}}}},\ Vol.\
  \bibinfo {volume} {739}\ (\bibinfo  {publisher} {Springer, Berlin,
  Heidelberg},\ \bibinfo {year} {2008})\ pp.\ \bibinfo {pages}
  {277--356}\BibitemShut {NoStop}%
\bibitem [{\citenamefont {Sato}\ and\ \citenamefont
  {Assaad}(2021)}]{sato_arxiv_2012.12283}%
  \BibitemOpen
  \bibfield  {author} {\bibinfo {author} {\bibfnamefont {T.}~\bibnamefont
  {Sato}}\ and\ \bibinfo {author} {\bibfnamefont {F.~F.}\ \bibnamefont
  {Assaad}},\ }\href {https://arxiv.org/abs/2012.12283v2} {\enquote {\bibinfo
  {title} {{Quantum Monte Carlo Simulation of Generalized Kitaev Models}},}\ }
  (\bibinfo {year} {2021}),\ \Eprint {http://arxiv.org/abs/arXiv:2012.12283v2}
  {arXiv:2012.12283v2} \BibitemShut {NoStop}%
\bibitem [{\citenamefont {Philipsen}(2009)}]{philipsen_2009}%
  \BibitemOpen
  \bibfield  {author} {\bibinfo {author} {\bibfnamefont {O.}~\bibnamefont
  {Philipsen}},\ }\href {https://arxiv.org/abs/1009.4089v2} {\emph {\bibinfo
  {title} {{"Lattice QCD at non-zero temperature and baryon density" in Modern
  perspectives in lattice QCD: Quantum field theory and high performance
  computing}}}}\ (\bibinfo  {publisher} {Oxford University Press},\ \bibinfo
  {address} {UK},\ \bibinfo {year} {2009})\ pp.\ \bibinfo {pages} {273--330},\
  \Eprint {http://arxiv.org/abs/arXiv:1009.4089v2} {arXiv:1009.4089v2}
  \BibitemShut {NoStop}%
\bibitem [{\citenamefont {Aarts}(2016)}]{aarts_jpcf_706_2016}%
  \BibitemOpen
  \bibfield  {author} {\bibinfo {author} {\bibfnamefont {G.}~\bibnamefont
  {Aarts}},\ }\bibfield  {title} {\enquote {\bibinfo {title} {{Introductory
  lectures on lattice QCD at nonzero baryon number}},}\ }\href
  {https://doi.org/10.1088/1742-6596/706/2/022004} {\bibfield  {journal}
  {\bibinfo  {journal} {J. Phys. Conf. Ser.}\ }\textbf {\bibinfo {volume}
  {706}},\ \bibinfo {pages} {022004} (\bibinfo {year} {2016})}\BibitemShut
  {NoStop}%
\bibitem [{\citenamefont {Inguscio}\ \emph {et~al.}(2007)\citenamefont
  {Inguscio}, \citenamefont {Ketterle},\ and\ \citenamefont
  {(editors)}}]{inguscio_pispef_164_2007}%
  \BibitemOpen
  \bibfield  {author} {\bibinfo {author} {\bibfnamefont {M.}~\bibnamefont
  {Inguscio}}, \bibinfo {author} {\bibfnamefont {W.}~\bibnamefont {Ketterle}},
  \ and\ \bibinfo {author} {\bibfnamefont {C.~Salomon}\ \bibnamefont
  {(editors)}},\ }\href@noop {} {\emph {\bibinfo {title} {{Ultra-cold Fermi
  Gases}}}},\ \bibinfo {series} {Proceedings of the International School of
  Physics ``Enrico Fermi''}, Vol.\ \bibinfo {volume} {164}\ (\bibinfo
  {publisher} {IOS Press},\ \bibinfo {address} {Amsterdam},\ \bibinfo {year}
  {2007})\BibitemShut {NoStop}%
\bibitem [{\citenamefont {Bloch}\ \emph {et~al.}(2008)\citenamefont {Bloch},
  \citenamefont {Dalibard},\ and\ \citenamefont {Zwerger}}]{bloch_rmp_80_2008}%
  \BibitemOpen
  \bibfield  {author} {\bibinfo {author} {\bibfnamefont {I.}~\bibnamefont
  {Bloch}}, \bibinfo {author} {\bibfnamefont {J.}~\bibnamefont {Dalibard}}, \
  and\ \bibinfo {author} {\bibfnamefont {W.}~\bibnamefont {Zwerger}},\
  }\bibfield  {title} {\enquote {\bibinfo {title} {{Many-body physics with
  ultracold gases}},}\ }\href {https://doi.org/10.1103/RevModPhys.80.885}
  {\bibfield  {journal} {\bibinfo  {journal} {Rev. Mod. Phys.}\ }\textbf
  {\bibinfo {volume} {80}},\ \bibinfo {pages} {885} (\bibinfo {year}
  {2008})}\BibitemShut {NoStop}%
\bibitem [{\citenamefont {Giorgini}\ \emph {et~al.}(2008)\citenamefont
  {Giorgini}, \citenamefont {Pitaevskii},\ and\ \citenamefont
  {Stringari}}]{giorgini_rmp_80_2008}%
  \BibitemOpen
  \bibfield  {author} {\bibinfo {author} {\bibfnamefont {S.}~\bibnamefont
  {Giorgini}}, \bibinfo {author} {\bibfnamefont {L.~P.}\ \bibnamefont
  {Pitaevskii}}, \ and\ \bibinfo {author} {\bibfnamefont {S.}~\bibnamefont
  {Stringari}},\ }\bibfield  {title} {\enquote {\bibinfo {title} {{Theory of
  ultracold atomic Fermi gases}},}\ }\href
  {https://doi.org/10.1103/RevModPhys.80.1215} {\bibfield  {journal} {\bibinfo
  {journal} {Rev. Mod. Phys.}\ }\textbf {\bibinfo {volume} {80}},\ \bibinfo
  {pages} {1215} (\bibinfo {year} {2008})}\BibitemShut {NoStop}%
\bibitem [{\citenamefont {Baeriswyl}\ \emph {et~al.}(1995)\citenamefont
  {Baeriswyl}, \citenamefont {Campbell}, \citenamefont {Carmelo}, \citenamefont
  {Guinea},\ and\ \citenamefont {Louis}}]{baeriswyl_1995}%
  \BibitemOpen
  \bibfield  {author} {\bibinfo {author} {\bibfnamefont {D.}~\bibnamefont
  {Baeriswyl}}, \bibinfo {author} {\bibfnamefont {D.~K.}\ \bibnamefont
  {Campbell}}, \bibinfo {author} {\bibfnamefont {J.~M.~P.}\ \bibnamefont
  {Carmelo}}, \bibinfo {author} {\bibfnamefont {F.}~\bibnamefont {Guinea}}, \
  and\ \bibinfo {author} {\bibfnamefont {E.}~\bibnamefont {Louis}},\
  }\href@noop {} {\emph {\bibinfo {title} {{The Hubbard Model, Its Physics and
  Mathematical Physics}}}},\ \bibinfo {edition} {1st}\ ed.\ (\bibinfo
  {publisher} {Springer},\ \bibinfo {address} {US},\ \bibinfo {year}
  {1995})\BibitemShut {NoStop}%
\bibitem [{\citenamefont {Lee}\ \emph {et~al.}(2006)\citenamefont {Lee},
  \citenamefont {Nagaosa},\ and\ \citenamefont {Wen}}]{lee_rmp_78_2006}%
  \BibitemOpen
  \bibfield  {author} {\bibinfo {author} {\bibfnamefont {P.~A.}\ \bibnamefont
  {Lee}}, \bibinfo {author} {\bibfnamefont {N.}~\bibnamefont {Nagaosa}}, \ and\
  \bibinfo {author} {\bibfnamefont {X.-G.}\ \bibnamefont {Wen}},\ }\bibfield
  {title} {\enquote {\bibinfo {title} {{Doping a Mott insulator: Physics of
  high-temperature superconductivity}},}\ }\href
  {https://doi.org/10.1103/RevModPhys.78.17} {\bibfield  {journal} {\bibinfo
  {journal} {Rev. Mod. Phys.}\ }\textbf {\bibinfo {volume} {78}},\ \bibinfo
  {pages} {17} (\bibinfo {year} {2006})}\BibitemShut {NoStop}%
\bibitem [{\citenamefont {Wen}(2007)}]{wen_2007}%
  \BibitemOpen
  \bibfield  {author} {\bibinfo {author} {\bibfnamefont {X.-G.}\ \bibnamefont
  {Wen}},\ }\href {https://doi.org/10.1093/acprof:oso/9780199227259.001.0001}
  {\emph {\bibinfo {title} {{Quantum Field Theory of Many-Body Systems: From
  the Origin of Sound to an Origin of Light and Electrons}}}}\ (\bibinfo
  {publisher} {Oxford University Press},\ \bibinfo {address} {UK},\ \bibinfo
  {year} {2007})\BibitemShut {NoStop}%
\bibitem [{\citenamefont {Balents}(2010)}]{balents_nature_464_2010}%
  \BibitemOpen
  \bibfield  {author} {\bibinfo {author} {\bibfnamefont {L.}~\bibnamefont
  {Balents}},\ }\bibfield  {title} {\enquote {\bibinfo {title} {{Spin liquids
  in frustrated magnets}},}\ }\href {https://doi.org/10.1038/nature08917}
  {\bibfield  {journal} {\bibinfo  {journal} {Nature}\ }\textbf {\bibinfo
  {volume} {464}},\ \bibinfo {pages} {199} (\bibinfo {year}
  {2010})}\BibitemShut {NoStop}%
\bibitem [{\citenamefont {Lacroix}\ \emph {et~al.}(2011)\citenamefont
  {Lacroix}, \citenamefont {Mendels},\ and\ \citenamefont
  {Mila}}]{lacroix_2011}%
  \BibitemOpen
  \bibfield  {author} {\bibinfo {author} {\bibfnamefont {C.}~\bibnamefont
  {Lacroix}}, \bibinfo {author} {\bibfnamefont {P.}~\bibnamefont {Mendels}}, \
  and\ \bibinfo {author} {\bibfnamefont {F.}~\bibnamefont {Mila}},\ }\href@noop
  {} {\emph {\bibinfo {title} {{Introduction to Frustrated Magnetism:
  Materials, Experiments, Theory}}}},\ \bibinfo {edition} {1st}\ ed.\ (\bibinfo
   {publisher} {Springer},\ \bibinfo {address} {Berlin, Heidelberg},\ \bibinfo
  {year} {2011})\BibitemShut {NoStop}%
\bibitem [{\citenamefont {Diep}(2013)}]{diep_2013}%
  \BibitemOpen
  \bibfield  {author} {\bibinfo {author} {\bibfnamefont {H.~T.}\ \bibnamefont
  {Diep}},\ }\href {https://doi.org/10.1142/8676} {\emph {\bibinfo {title}
  {{Frustrated Spin Systems}}}},\ \bibinfo {edition} {2nd}\ ed.\ (\bibinfo
  {publisher} {World Scientific},\ \bibinfo {address} {France},\ \bibinfo
  {year} {2013})\BibitemShut {NoStop}%
\bibitem [{\citenamefont {Savary}\ and\ \citenamefont
  {Balents}(2017)}]{savary_rpp_80_2017}%
  \BibitemOpen
  \bibfield  {author} {\bibinfo {author} {\bibfnamefont {L.}~\bibnamefont
  {Savary}}\ and\ \bibinfo {author} {\bibfnamefont {L.}~\bibnamefont
  {Balents}},\ }\bibfield  {title} {\enquote {\bibinfo {title} {{Quantum spin
  liquids: a review}},}\ }\href {https://doi.org/10.1088/0034-4885/80/1/016502}
  {\bibfield  {journal} {\bibinfo  {journal} {Rep. Prog. Phys.}\ }\textbf
  {\bibinfo {volume} {80}},\ \bibinfo {pages} {016502} (\bibinfo {year}
  {2017})}\BibitemShut {NoStop}%
\bibitem [{\citenamefont {Troyer}\ and\ \citenamefont
  {Wiese}(2005)}]{troyer_prl_94_2005}%
  \BibitemOpen
  \bibfield  {author} {\bibinfo {author} {\bibfnamefont {M.}~\bibnamefont
  {Troyer}}\ and\ \bibinfo {author} {\bibfnamefont {U.-J.}\ \bibnamefont
  {Wiese}},\ }\bibfield  {title} {\enquote {\bibinfo {title} {{Computational
  Complexity and Fundamental Limitations to Fermionic Quantum Monte Carlo
  Simulations}},}\ }\href {https://doi.org/10.1103/PhysRevLett.94.170201}
  {\bibfield  {journal} {\bibinfo  {journal} {Phys. Rev. Lett.}\ }\textbf
  {\bibinfo {volume} {94}},\ \bibinfo {pages} {170201} (\bibinfo {year}
  {2005})}\BibitemShut {NoStop}%
\bibitem [{\citenamefont
  {Witten}(2010{\natexlab{a}})}]{written_arxiv_1001.2933}%
  \BibitemOpen
  \bibfield  {author} {\bibinfo {author} {\bibfnamefont {E.}~\bibnamefont
  {Witten}},\ }\href {https://arxiv.org/abs/1001.2933v4} {\enquote {\bibinfo
  {title} {{Analytic Continuation Of Chern-Simons Theory}},}\ } (\bibinfo
  {year} {2010}{\natexlab{a}}),\ \Eprint
  {http://arxiv.org/abs/arXiv:1001.2933v4} {arXiv:1001.2933v4} \BibitemShut
  {NoStop}%
\bibitem [{\citenamefont
  {Witten}(2010{\natexlab{b}})}]{written_arxiv_1009.6032}%
  \BibitemOpen
  \bibfield  {author} {\bibinfo {author} {\bibfnamefont {E.}~\bibnamefont
  {Witten}},\ }\href {https://arxiv.org/abs/1009.6032} {\enquote {\bibinfo
  {title} {{A New Look At The Path Integral Of Quantum Mechanics}},}\ }
  (\bibinfo {year} {2010}{\natexlab{b}}),\ \Eprint
  {http://arxiv.org/abs/arXiv:1009.6032} {arXiv:1009.6032} \BibitemShut
  {NoStop}%
\bibitem [{\citenamefont {Cristoforetti}\ \emph {et~al.}(2012)\citenamefont
  {Cristoforetti}, \citenamefont {Renzo},\ and\ \citenamefont
  {Scorzato}}]{cristoforetti_prd_86_2012}%
  \BibitemOpen
  \bibfield  {author} {\bibinfo {author} {\bibfnamefont {M.}~\bibnamefont
  {Cristoforetti}}, \bibinfo {author} {\bibfnamefont {F.~Di}\ \bibnamefont
  {Renzo}}, \ and\ \bibinfo {author} {\bibfnamefont {L.}~\bibnamefont
  {Scorzato}},\ }\bibfield  {title} {\enquote {\bibinfo {title} {{New approach
  to the sign problem in quantum field theories: High density QCD on a
  Lefschetz thimble}},}\ }\href {https://doi.org/10.1103/PhysRevD.86.074506}
  {\bibfield  {journal} {\bibinfo  {journal} {Phys. Rev. D}\ }\textbf {\bibinfo
  {volume} {86}},\ \bibinfo {pages} {074506} (\bibinfo {year}
  {2012})}\BibitemShut {NoStop}%
\bibitem [{\citenamefont {Cristoforetti}\ \emph {et~al.}(2013)\citenamefont
  {Cristoforetti}, \citenamefont {Renzo}, \citenamefont {Mukherjee},\ and\
  \citenamefont {Scorzato}}]{cristoforetti_prd_88_2013}%
  \BibitemOpen
  \bibfield  {author} {\bibinfo {author} {\bibfnamefont {M.}~\bibnamefont
  {Cristoforetti}}, \bibinfo {author} {\bibfnamefont {F.~Di}\ \bibnamefont
  {Renzo}}, \bibinfo {author} {\bibfnamefont {A.}~\bibnamefont {Mukherjee}}, \
  and\ \bibinfo {author} {\bibfnamefont {L.}~\bibnamefont {Scorzato}},\
  }\bibfield  {title} {\enquote {\bibinfo {title} {{Monte Carlo simulations on
  the Lefschetz thimble: Taming the sign problem}},}\ }\href
  {https://doi.org/10.1103/PhysRevD.88.051501} {\bibfield  {journal} {\bibinfo
  {journal} {Phys. Rev. D}\ }\textbf {\bibinfo {volume} {88}},\ \bibinfo
  {pages} {051501(R)} (\bibinfo {year} {2013})}\BibitemShut {NoStop}%
\bibitem [{\citenamefont {Fujii}\ \emph {et~al.}(2013)\citenamefont {Fujii},
  \citenamefont {Honda}, \citenamefont {Kato}, \citenamefont {Kikukawa},
  \citenamefont {Komatsu},\ and\ \citenamefont {Sano}}]{fujii_jhep_10_2013}%
  \BibitemOpen
  \bibfield  {author} {\bibinfo {author} {\bibfnamefont {H.}~\bibnamefont
  {Fujii}}, \bibinfo {author} {\bibfnamefont {D.}~\bibnamefont {Honda}},
  \bibinfo {author} {\bibfnamefont {M.}~\bibnamefont {Kato}}, \bibinfo {author}
  {\bibfnamefont {Y.}~\bibnamefont {Kikukawa}}, \bibinfo {author}
  {\bibfnamefont {S.}~\bibnamefont {Komatsu}}, \ and\ \bibinfo {author}
  {\bibfnamefont {T.}~\bibnamefont {Sano}},\ }\bibfield  {title} {\enquote
  {\bibinfo {title} {{Hybrid Monte Carlo on Lefschetz thimbles -- A study of
  the residual sign problem}},}\ }\href
  {https://doi.org/10.1007/JHEP10(2013)147} {\bibfield  {journal} {\bibinfo
  {journal} {JHEP}\ }\textbf {\bibinfo {volume} {10}},\ \bibinfo {pages} {147}
  (\bibinfo {year} {2013})}\BibitemShut {NoStop}%
\bibitem [{\citenamefont {Mukherjee}\ \emph {et~al.}(2013)\citenamefont
  {Mukherjee}, \citenamefont {Cristoforetti},\ and\ \citenamefont
  {Scorzato}}]{mukherjee_prd_88_2013}%
  \BibitemOpen
  \bibfield  {author} {\bibinfo {author} {\bibfnamefont {A.}~\bibnamefont
  {Mukherjee}}, \bibinfo {author} {\bibfnamefont {M.}~\bibnamefont
  {Cristoforetti}}, \ and\ \bibinfo {author} {\bibfnamefont {L.}~\bibnamefont
  {Scorzato}},\ }\bibfield  {title} {\enquote {\bibinfo {title} {{Metropolis
  Monte Carlo integration on the Lefschetz thimble: Application to a
  one-plaquette model}},}\ }\href {https://doi.org/10.1103/PhysRevD.88.051502}
  {\bibfield  {journal} {\bibinfo  {journal} {Phys. Rev. D}\ }\textbf {\bibinfo
  {volume} {88}},\ \bibinfo {pages} {051502(R)} (\bibinfo {year}
  {2013})}\BibitemShut {NoStop}%
\bibitem [{\citenamefont {Cristoforetti}\ \emph
  {et~al.}(2014{\natexlab{a}})\citenamefont {Cristoforetti}, \citenamefont
  {Renzo}, \citenamefont {Mukherjee},\ and\ \citenamefont
  {Scorzato}}]{cristoforetti_lattice_2013}%
  \BibitemOpen
  \bibfield  {author} {\bibinfo {author} {\bibfnamefont {M.}~\bibnamefont
  {Cristoforetti}}, \bibinfo {author} {\bibfnamefont {F.~Di}\ \bibnamefont
  {Renzo}}, \bibinfo {author} {\bibfnamefont {A.}~\bibnamefont {Mukherjee}}, \
  and\ \bibinfo {author} {\bibfnamefont {L.}~\bibnamefont {Scorzato}},\
  }\bibfield  {title} {\enquote {\bibinfo {title} {{Quantum field theories on
  the Lefschetz thimble}},}\ }\href {https://arxiv.org/abs/1312.1052}
  {\bibfield  {journal} {\bibinfo  {journal} {PoS}\ }\textbf {\bibinfo {volume}
  {LATTICE2013}},\ \bibinfo {pages} {197} (\bibinfo {year}
  {2014}{\natexlab{a}})},\ \Eprint {http://arxiv.org/abs/arXiv:1312.1052}
  {arXiv:1312.1052} \BibitemShut {NoStop}%
\bibitem [{\citenamefont {Cristoforetti}\ \emph
  {et~al.}(2014{\natexlab{b}})\citenamefont {Cristoforetti}, \citenamefont
  {Renzo}, \citenamefont {Eruzzi}, \citenamefont {Mukherjee}, \citenamefont
  {Schmidt}, \citenamefont {Scorzato},\ and\ \citenamefont
  {Torrero}}]{cristoforetti_prd_89_2014}%
  \BibitemOpen
  \bibfield  {author} {\bibinfo {author} {\bibfnamefont {M.}~\bibnamefont
  {Cristoforetti}}, \bibinfo {author} {\bibfnamefont {F.~Di}\ \bibnamefont
  {Renzo}}, \bibinfo {author} {\bibfnamefont {G.}~\bibnamefont {Eruzzi}},
  \bibinfo {author} {\bibfnamefont {A.}~\bibnamefont {Mukherjee}}, \bibinfo
  {author} {\bibfnamefont {C.}~\bibnamefont {Schmidt}}, \bibinfo {author}
  {\bibfnamefont {L.}~\bibnamefont {Scorzato}}, \ and\ \bibinfo {author}
  {\bibfnamefont {C.}~\bibnamefont {Torrero}},\ }\bibfield  {title} {\enquote
  {\bibinfo {title} {{An efficient method to compute the residual phase on a
  Lefschetz thimble}},}\ }\href {https://doi.org/10.1103/PhysRevD.89.114505}
  {\bibfield  {journal} {\bibinfo  {journal} {Phys. Rev. D}\ }\textbf {\bibinfo
  {volume} {89}},\ \bibinfo {pages} {114505} (\bibinfo {year}
  {2014}{\natexlab{b}})}\BibitemShut {NoStop}%
\bibitem [{\citenamefont {Aarts}\ \emph {et~al.}(2014)\citenamefont {Aarts},
  \citenamefont {Bongiovanni}, \citenamefont {Seiler},\ and\ \citenamefont
  {Sexty}}]{aarts_jhep_10_2014}%
  \BibitemOpen
  \bibfield  {author} {\bibinfo {author} {\bibfnamefont {G.}~\bibnamefont
  {Aarts}}, \bibinfo {author} {\bibfnamefont {L.}~\bibnamefont {Bongiovanni}},
  \bibinfo {author} {\bibfnamefont {E.}~\bibnamefont {Seiler}}, \ and\ \bibinfo
  {author} {\bibfnamefont {D.}~\bibnamefont {Sexty}},\ }\bibfield  {title}
  {\enquote {\bibinfo {title} {{Some remarks on Lefschetz thimbles and complex
  Langevin dynamics}},}\ }\href {https://doi.org/10.1007/JHEP10(2014)159}
  {\bibfield  {journal} {\bibinfo  {journal} {JHEP}\ }\textbf {\bibinfo
  {volume} {10}},\ \bibinfo {pages} {159} (\bibinfo {year} {2014})}\BibitemShut
  {NoStop}%
\bibitem [{\citenamefont {Tanizaki}(2015)}]{tanizaki_prd_91_2015}%
  \BibitemOpen
  \bibfield  {author} {\bibinfo {author} {\bibfnamefont {Y.}~\bibnamefont
  {Tanizaki}},\ }\bibfield  {title} {\enquote {\bibinfo {title}
  {{Lefschetz-thimble techniques for path integral of zero-dimensional $O(n)$
  sigma models}},}\ }\href {https://doi.org/10.1103/PhysRevD.91.036002}
  {\bibfield  {journal} {\bibinfo  {journal} {Phys. Rev. D}\ }\textbf {\bibinfo
  {volume} {91}},\ \bibinfo {pages} {036002} (\bibinfo {year}
  {2015})}\BibitemShut {NoStop}%
\bibitem [{\citenamefont {Renzo}\ and\ \citenamefont
  {Eruzzi}(2015)}]{renzo_prd_92_2015}%
  \BibitemOpen
  \bibfield  {author} {\bibinfo {author} {\bibfnamefont {F.~Di}\ \bibnamefont
  {Renzo}}\ and\ \bibinfo {author} {\bibfnamefont {G.}~\bibnamefont {Eruzzi}},\
  }\bibfield  {title} {\enquote {\bibinfo {title} {{Thimble regularization at
  work: From toy models to chiral random matrix theories}},}\ }\href
  {https://doi.org/10.1103/PhysRevD.92.085030} {\bibfield  {journal} {\bibinfo
  {journal} {Phys. Rev. D}\ }\textbf {\bibinfo {volume} {92}},\ \bibinfo
  {pages} {085030} (\bibinfo {year} {2015})}\BibitemShut {NoStop}%
\bibitem [{\citenamefont {Fujii}\ \emph
  {et~al.}(2015{\natexlab{a}})\citenamefont {Fujii}, \citenamefont {Kamata},\
  and\ \citenamefont {Kikukawa}}]{fujii_jhep_11_2015}%
  \BibitemOpen
  \bibfield  {author} {\bibinfo {author} {\bibfnamefont {H.}~\bibnamefont
  {Fujii}}, \bibinfo {author} {\bibfnamefont {S.}~\bibnamefont {Kamata}}, \
  and\ \bibinfo {author} {\bibfnamefont {Y.}~\bibnamefont {Kikukawa}},\
  }\bibfield  {title} {\enquote {\bibinfo {title} {{Lefschetz thimble structure
  in one-dimensional lattice Thirring model at finite density}},}\ }\href
  {https://doi.org/10.1007/JHEP11(2015)078} {\bibfield  {journal} {\bibinfo
  {journal} {JHEP}\ }\textbf {\bibinfo {volume} {11}},\ \bibinfo {pages} {079}
  (\bibinfo {year} {2015}{\natexlab{a}})}\BibitemShut {NoStop}%
\bibitem [{\citenamefont {Kanazawa}\ and\ \citenamefont
  {Tanizaki}(2015)}]{kanazawa_jhep_03_2015}%
  \BibitemOpen
  \bibfield  {author} {\bibinfo {author} {\bibfnamefont {T.}~\bibnamefont
  {Kanazawa}}\ and\ \bibinfo {author} {\bibfnamefont {Y.}~\bibnamefont
  {Tanizaki}},\ }\bibfield  {title} {\enquote {\bibinfo {title} {{Structure of
  Lefschetz thimbles in simple fermionic systems}},}\ }\href
  {https://doi.org/10.1007/JHEP03(2015)044} {\bibfield  {journal} {\bibinfo
  {journal} {JHEP}\ }\textbf {\bibinfo {volume} {03}},\ \bibinfo {pages} {44}
  (\bibinfo {year} {2015})}\BibitemShut {NoStop}%
\bibitem [{\citenamefont {Fukushima}\ and\ \citenamefont
  {Tanizaki}(2015)}]{fukushima_ptep_2015_2015}%
  \BibitemOpen
  \bibfield  {author} {\bibinfo {author} {\bibfnamefont {K.}~\bibnamefont
  {Fukushima}}\ and\ \bibinfo {author} {\bibfnamefont {Y.}~\bibnamefont
  {Tanizaki}},\ }\bibfield  {title} {\enquote {\bibinfo {title} {{Hamilton
  dynamics for Lefschetz-thimble integration akin to the complex Langevin
  method}},}\ }\href {https://doi.org/10.1093/ptep/ptv152} {\bibfield
  {journal} {\bibinfo  {journal} {PTEP}\ }\textbf {\bibinfo {volume} {2015}},\
  \bibinfo {pages} {111A01} (\bibinfo {year} {2015})}\BibitemShut {NoStop}%
\bibitem [{\citenamefont {Fujii}\ \emph
  {et~al.}(2015{\natexlab{b}})\citenamefont {Fujii}, \citenamefont {Kamata},\
  and\ \citenamefont {Kikukawa}}]{fujii_jhep_12_2015}%
  \BibitemOpen
  \bibfield  {author} {\bibinfo {author} {\bibfnamefont {H.}~\bibnamefont
  {Fujii}}, \bibinfo {author} {\bibfnamefont {S.}~\bibnamefont {Kamata}}, \
  and\ \bibinfo {author} {\bibfnamefont {Y.}~\bibnamefont {Kikukawa}},\
  }\bibfield  {title} {\enquote {\bibinfo {title} {{Monte Carlo study of
  Lefschetz thimble structure in one-dimensional Thirring model at finite
  density}},}\ }\href {https://doi.org/10.1007/JHEP12(2015)125} {\bibfield
  {journal} {\bibinfo  {journal} {JHEP}\ }\textbf {\bibinfo {volume} {12}},\
  \bibinfo {pages} {125} (\bibinfo {year} {2015}{\natexlab{b}})}\BibitemShut
  {NoStop}%
\bibitem [{\citenamefont {Tanizaki}\ \emph {et~al.}(2016)\citenamefont
  {Tanizaki}, \citenamefont {Hidaka},\ and\ \citenamefont
  {Hayata}}]{tanizaki_njp_18_2016}%
  \BibitemOpen
  \bibfield  {author} {\bibinfo {author} {\bibfnamefont {Y.}~\bibnamefont
  {Tanizaki}}, \bibinfo {author} {\bibfnamefont {Y.}~\bibnamefont {Hidaka}}, \
  and\ \bibinfo {author} {\bibfnamefont {T.}~\bibnamefont {Hayata}},\
  }\bibfield  {title} {\enquote {\bibinfo {title} {{Lefschetz-thimble analysis
  of the sign problem in one-site fermion model}},}\ }\href
  {https://doi.org/10.1088/1367-2630/18/3/033002} {\bibfield  {journal}
  {\bibinfo  {journal} {New J. Phys.}\ }\textbf {\bibinfo {volume} {18}},\
  \bibinfo {pages} {033002} (\bibinfo {year} {2016})}\BibitemShut {NoStop}%
\bibitem [{\citenamefont {Tsutsui}\ and\ \citenamefont
  {Doi}(2016)}]{tsutsui_prd_94_2016}%
  \BibitemOpen
  \bibfield  {author} {\bibinfo {author} {\bibfnamefont {S.}~\bibnamefont
  {Tsutsui}}\ and\ \bibinfo {author} {\bibfnamefont {T.~M.}\ \bibnamefont
  {Doi}},\ }\bibfield  {title} {\enquote {\bibinfo {title} {{Improvement in
  complex Langevin dynamics from a view point of Lefschetz thimbles}},}\ }\href
  {https://doi.org/10.1103/PhysRevD.94.074009} {\bibfield  {journal} {\bibinfo
  {journal} {Phys. Rev. D}\ }\textbf {\bibinfo {volume} {94}},\ \bibinfo
  {pages} {074009} (\bibinfo {year} {2016})}\BibitemShut {NoStop}%
\bibitem [{\citenamefont {Alexandru}\ \emph
  {et~al.}(2016{\natexlab{a}})\citenamefont {Alexandru}, \citenamefont
  {Basar},\ and\ \citenamefont {Bedaque}}]{alexandru_prd_93_2016}%
  \BibitemOpen
  \bibfield  {author} {\bibinfo {author} {\bibfnamefont {A.}~\bibnamefont
  {Alexandru}}, \bibinfo {author} {\bibfnamefont {G.}~\bibnamefont {Basar}}, \
  and\ \bibinfo {author} {\bibfnamefont {P.}~\bibnamefont {Bedaque}},\
  }\bibfield  {title} {\enquote {\bibinfo {title} {{Monte Carlo algorithm for
  simulating fermions on Lefschetz thimbles}},}\ }\href
  {https://doi.org/10.1103/PhysRevD.93.014504} {\bibfield  {journal} {\bibinfo
  {journal} {Phys. Rev. D}\ }\textbf {\bibinfo {volume} {93}},\ \bibinfo
  {pages} {014504} (\bibinfo {year} {2016}{\natexlab{a}})}\BibitemShut
  {NoStop}%
\bibitem [{\citenamefont {Alexandru}\ \emph
  {et~al.}(2016{\natexlab{b}})\citenamefont {Alexandru}, \citenamefont {Basar},
  \citenamefont {Bedaque}, \citenamefont {Ridgway},\ and\ \citenamefont
  {Warrington}}]{alexandru_jhep_05_2016}%
  \BibitemOpen
  \bibfield  {author} {\bibinfo {author} {\bibfnamefont {A.}~\bibnamefont
  {Alexandru}}, \bibinfo {author} {\bibfnamefont {G.}~\bibnamefont {Basar}},
  \bibinfo {author} {\bibfnamefont {P.~F.}\ \bibnamefont {Bedaque}}, \bibinfo
  {author} {\bibfnamefont {G.~W.}\ \bibnamefont {Ridgway}}, \ and\ \bibinfo
  {author} {\bibfnamefont {N.~C.}\ \bibnamefont {Warrington}},\ }\bibfield
  {title} {\enquote {\bibinfo {title} {{Sign problem and Monte Carlo
  calculations beyond Lefschetz thimbles}},}\ }\href
  {https://doi.org/10.1007/JHEP05(2016)053} {\bibfield  {journal} {\bibinfo
  {journal} {JHEP}\ }\textbf {\bibinfo {volume} {05}},\ \bibinfo {pages} {53}
  (\bibinfo {year} {2016}{\natexlab{b}})}\BibitemShut {NoStop}%
\bibitem [{\citenamefont {Alexandru}\ \emph
  {et~al.}(2017{\natexlab{a}})\citenamefont {Alexandru}, \citenamefont {Basar},
  \citenamefont {Bedaque}, \citenamefont {Ridgway},\ and\ \citenamefont
  {Warrington}}]{alexandru_prd_93_2017}%
  \BibitemOpen
  \bibfield  {author} {\bibinfo {author} {\bibfnamefont {A.}~\bibnamefont
  {Alexandru}}, \bibinfo {author} {\bibfnamefont {G.}~\bibnamefont {Basar}},
  \bibinfo {author} {\bibfnamefont {P.~F.}\ \bibnamefont {Bedaque}}, \bibinfo
  {author} {\bibfnamefont {G.~W.}\ \bibnamefont {Ridgway}}, \ and\ \bibinfo
  {author} {\bibfnamefont {N.~C.}\ \bibnamefont {Warrington}},\ }\bibfield
  {title} {\enquote {\bibinfo {title} {{Monte Carlo calculations of the finite
  density Thirring model}},}\ }\href
  {https://doi.org/10.1103/PhysRevD.95.014502} {\bibfield  {journal} {\bibinfo
  {journal} {Phys. Rev. D}\ }\textbf {\bibinfo {volume} {95}},\ \bibinfo
  {pages} {014502} (\bibinfo {year} {2017}{\natexlab{a}})}\BibitemShut
  {NoStop}%
\bibitem [{\citenamefont {Alexandru}\ \emph
  {et~al.}(2017{\natexlab{b}})\citenamefont {Alexandru}, \citenamefont
  {Bedaque}, \citenamefont {Lamm},\ and\ \citenamefont
  {Lawrence}}]{alexandru_prd_96_2017_094505}%
  \BibitemOpen
  \bibfield  {author} {\bibinfo {author} {\bibfnamefont {A.}~\bibnamefont
  {Alexandru}}, \bibinfo {author} {\bibfnamefont {P.~F.}\ \bibnamefont
  {Bedaque}}, \bibinfo {author} {\bibfnamefont {H.}~\bibnamefont {Lamm}}, \
  and\ \bibinfo {author} {\bibfnamefont {S.}~\bibnamefont {Lawrence}},\
  }\bibfield  {title} {\enquote {\bibinfo {title} {{Deep learning beyond
  Lefschetz thimbles}},}\ }\href {https://doi.org/10.1103/PhysRevD.96.094505}
  {\bibfield  {journal} {\bibinfo  {journal} {Phys. Rev. D}\ }\textbf {\bibinfo
  {volume} {96}},\ \bibinfo {pages} {094505} (\bibinfo {year}
  {2017}{\natexlab{b}})}\BibitemShut {NoStop}%
\bibitem [{\citenamefont {Tanizaki}\ \emph {et~al.}(2017)\citenamefont
  {Tanizaki}, \citenamefont {Nishimura},\ and\ \citenamefont
  {Verbaarschot}}]{tanizaki_jhep_10_2017}%
  \BibitemOpen
  \bibfield  {author} {\bibinfo {author} {\bibfnamefont {Y.}~\bibnamefont
  {Tanizaki}}, \bibinfo {author} {\bibfnamefont {H.}~\bibnamefont {Nishimura}},
  \ and\ \bibinfo {author} {\bibfnamefont {J.~J.~M.}\ \bibnamefont
  {Verbaarschot}},\ }\bibfield  {title} {\enquote {\bibinfo {title} {{Gradient
  flows without blow-up for Lefschetz thimbles}},}\ }\href
  {https://doi.org/10.1007/JHEP10(2017)100} {\bibfield  {journal} {\bibinfo
  {journal} {JHEP}\ }\textbf {\bibinfo {volume} {10}},\ \bibinfo {pages} {100}
  (\bibinfo {year} {2017})}\BibitemShut {NoStop}%
\bibitem [{\citenamefont {Alexandru}\ \emph
  {et~al.}(2017{\natexlab{c}})\citenamefont {Alexandru}, \citenamefont {Basar},
  \citenamefont {Bedaque},\ and\ \citenamefont
  {Warrington}}]{alexandru_prd_96_2017}%
  \BibitemOpen
  \bibfield  {author} {\bibinfo {author} {\bibfnamefont {A.}~\bibnamefont
  {Alexandru}}, \bibinfo {author} {\bibfnamefont {G.}~\bibnamefont {Basar}},
  \bibinfo {author} {\bibfnamefont {P.~F.}\ \bibnamefont {Bedaque}}, \ and\
  \bibinfo {author} {\bibfnamefont {N.~C.}\ \bibnamefont {Warrington}},\
  }\bibfield  {title} {\enquote {\bibinfo {title} {{Tempered transitions
  between thimbles}},}\ }\href {https://doi.org/10.1103/PhysRevD.96.034513}
  {\bibfield  {journal} {\bibinfo  {journal} {Phys. Rev. D}\ }\textbf {\bibinfo
  {volume} {96}},\ \bibinfo {pages} {034513} (\bibinfo {year}
  {2017}{\natexlab{c}})}\BibitemShut {NoStop}%
\bibitem [{\citenamefont {M.~Fukuma}(2017)}]{fukuma_ptep_2017_2017}%
  \BibitemOpen
  \bibfield  {author} {\bibinfo {author} {\bibfnamefont {N.~Umeda}\
  \bibnamefont {M.~Fukuma}},\ }\bibfield  {title} {\enquote {\bibinfo {title}
  {{Parallel tempering algorithm for integration over Lefschetz thimbles}},}\
  }\href {https://doi.org/10.1093/ptep/ptx081} {\bibfield  {journal} {\bibinfo
  {journal} {PTEP}\ }\textbf {\bibinfo {volume} {2017}},\ \bibinfo {pages} {no.
  7, 073B01} (\bibinfo {year} {2017})}\BibitemShut {NoStop}%
\bibitem [{\citenamefont {Renzo}\ and\ \citenamefont
  {Eruzzi}(2018)}]{renzo_prd_97_2018}%
  \BibitemOpen
  \bibfield  {author} {\bibinfo {author} {\bibfnamefont {F.~Di}\ \bibnamefont
  {Renzo}}\ and\ \bibinfo {author} {\bibfnamefont {G.}~\bibnamefont {Eruzzi}},\
  }\bibfield  {title} {\enquote {\bibinfo {title} {{One-dimensional QCD in
  thimble regularization}},}\ }\href
  {https://doi.org/10.1103/PhysRevD.97.014503} {\bibfield  {journal} {\bibinfo
  {journal} {Phys. Rev. D}\ }\textbf {\bibinfo {volume} {97}},\ \bibinfo
  {pages} {014503} (\bibinfo {year} {2018})}\BibitemShut {NoStop}%
\bibitem [{\citenamefont {Bluecher}\ \emph {et~al.}(2018)\citenamefont
  {Bluecher}, \citenamefont {Pawlowski}, \citenamefont {Scherzer},
  \citenamefont {Schlosser}, \citenamefont {Stamatescu}, \citenamefont
  {Syrkowski},\ and\ \citenamefont {Ziegler}}]{bluecher_scipost_5_2018}%
  \BibitemOpen
  \bibfield  {author} {\bibinfo {author} {\bibfnamefont {S.}~\bibnamefont
  {Bluecher}}, \bibinfo {author} {\bibfnamefont {J.~M.}\ \bibnamefont
  {Pawlowski}}, \bibinfo {author} {\bibfnamefont {M.}~\bibnamefont {Scherzer}},
  \bibinfo {author} {\bibfnamefont {M.}~\bibnamefont {Schlosser}}, \bibinfo
  {author} {\bibfnamefont {I.-O.}\ \bibnamefont {Stamatescu}}, \bibinfo
  {author} {\bibfnamefont {S.}~\bibnamefont {Syrkowski}}, \ and\ \bibinfo
  {author} {\bibfnamefont {F.~P.~G.}\ \bibnamefont {Ziegler}},\ }\bibfield
  {title} {\enquote {\bibinfo {title} {{Reweighting Lefschetz Thimbles}},}\
  }\href {https://doi.org/10.21468/SciPostPhys.5.5.044} {\bibfield  {journal}
  {\bibinfo  {journal} {SciPost Phys.}\ }\textbf {\bibinfo {volume} {5}},\
  \bibinfo {pages} {44} (\bibinfo {year} {2018})}\BibitemShut {NoStop}%
\bibitem [{\citenamefont {Alexandru}\ \emph
  {et~al.}(2018{\natexlab{a}})\citenamefont {Alexandru}, \citenamefont
  {Bedaque}, \citenamefont {Lamm},\ and\ \citenamefont
  {Lawrence}}]{alexandru_prd_97_2018}%
  \BibitemOpen
  \bibfield  {author} {\bibinfo {author} {\bibfnamefont {A.}~\bibnamefont
  {Alexandru}}, \bibinfo {author} {\bibfnamefont {P.~F.}\ \bibnamefont
  {Bedaque}}, \bibinfo {author} {\bibfnamefont {H.}~\bibnamefont {Lamm}}, \
  and\ \bibinfo {author} {\bibfnamefont {S.}~\bibnamefont {Lawrence}},\
  }\bibfield  {title} {\enquote {\bibinfo {title} {{Finite-density Monte Carlo
  calculations on sign-optimized manifolds}},}\ }\href
  {https://doi.org/10.1103/PhysRevD.97.094510} {\bibfield  {journal} {\bibinfo
  {journal} {Phys. Rev. D}\ }\textbf {\bibinfo {volume} {97}},\ \bibinfo
  {pages} {094510} (\bibinfo {year} {2018}{\natexlab{a}})}\BibitemShut
  {NoStop}%
\bibitem [{\citenamefont {Alexandru}\ \emph
  {et~al.}(2018{\natexlab{b}})\citenamefont {Alexandru}, \citenamefont
  {Bedaque}, \citenamefont {Lamm}, \citenamefont {Lawrence},\ and\
  \citenamefont {Warrington}}]{alexandru_prl_121_2018}%
  \BibitemOpen
  \bibfield  {author} {\bibinfo {author} {\bibfnamefont {A.}~\bibnamefont
  {Alexandru}}, \bibinfo {author} {\bibfnamefont {P.~F.}\ \bibnamefont
  {Bedaque}}, \bibinfo {author} {\bibfnamefont {H.}~\bibnamefont {Lamm}},
  \bibinfo {author} {\bibfnamefont {S.}~\bibnamefont {Lawrence}}, \ and\
  \bibinfo {author} {\bibfnamefont {N.~C.}\ \bibnamefont {Warrington}},\
  }\bibfield  {title} {\enquote {\bibinfo {title} {{Fermions at Finite Density
  in $2 + 1$ Dimensions with Sign-Optimized Manifolds}},}\ }\href
  {https://doi.org/10.1103/PhysRevLett.121.191602} {\bibfield  {journal}
  {\bibinfo  {journal} {Phys. Rev. Lett.}\ }\textbf {\bibinfo {volume} {121}},\
  \bibinfo {pages} {191602} (\bibinfo {year} {2018}{\natexlab{b}})}\BibitemShut
  {NoStop}%
\bibitem [{\citenamefont {Alexandru}\ \emph
  {et~al.}(2018{\natexlab{c}})\citenamefont {Alexandru}, \citenamefont {Basar},
  \citenamefont {Bedaque}, \citenamefont {Lamm},\ and\ \citenamefont
  {Lawrence}}]{alexandru_prd_98_2018}%
  \BibitemOpen
  \bibfield  {author} {\bibinfo {author} {\bibfnamefont {A.}~\bibnamefont
  {Alexandru}}, \bibinfo {author} {\bibfnamefont {G.}~\bibnamefont {Basar}},
  \bibinfo {author} {\bibfnamefont {P.~F.}\ \bibnamefont {Bedaque}}, \bibinfo
  {author} {\bibfnamefont {H.}~\bibnamefont {Lamm}}, \ and\ \bibinfo {author}
  {\bibfnamefont {S.}~\bibnamefont {Lawrence}},\ }\bibfield  {title} {\enquote
  {\bibinfo {title} {{Finite density {\bf{QED}$_{1 + 1}$} near Lefschetz
  thimbles}},}\ }\href {https://doi.org/10.1103/PhysRevD.98.034506} {\bibfield
  {journal} {\bibinfo  {journal} {Phys. Rev. D}\ }\textbf {\bibinfo {volume}
  {98}},\ \bibinfo {pages} {034506} (\bibinfo {year}
  {2018}{\natexlab{c}})}\BibitemShut {NoStop}%
\bibitem [{\citenamefont {Mukherjee}\ and\ \citenamefont
  {Cristoforetti}(2014)}]{mukherjee_prb_90_2014}%
  \BibitemOpen
  \bibfield  {author} {\bibinfo {author} {\bibfnamefont {A.}~\bibnamefont
  {Mukherjee}}\ and\ \bibinfo {author} {\bibfnamefont {M.}~\bibnamefont
  {Cristoforetti}},\ }\bibfield  {title} {\enquote {\bibinfo {title}
  {{Lefschetz thimble Monte Carlo for many-body theories: A Hubbard model
  study}},}\ }\href {https://doi.org/10.1103/PhysRevB.90.035134} {\bibfield
  {journal} {\bibinfo  {journal} {Phys. Rev. B}\ }\textbf {\bibinfo {volume}
  {90}},\ \bibinfo {pages} {035134} (\bibinfo {year} {2014})}\BibitemShut
  {NoStop}%
\bibitem [{\citenamefont {Ulybyshev}\ and\ \citenamefont
  {Valgushev}(2017)}]{ulybyshev_arxiv_1712.02188}%
  \BibitemOpen
  \bibfield  {author} {\bibinfo {author} {\bibfnamefont {M.~V.}\ \bibnamefont
  {Ulybyshev}}\ and\ \bibinfo {author} {\bibfnamefont {S.~N.}\ \bibnamefont
  {Valgushev}},\ }\href {https://arxiv.org/abs/1712.02188} {\enquote {\bibinfo
  {title} {{Path integral representation for the Hubbard model with reduced
  number of Lefschetz thimbles}},}\ } (\bibinfo {year} {2017}),\ \Eprint
  {http://arxiv.org/abs/arXiv:1712.02188} {arXiv:1712.02188} \BibitemShut
  {NoStop}%
\bibitem [{\citenamefont {Ulybyshev}\ \emph {et~al.}(2019)\citenamefont
  {Ulybyshev}, \citenamefont {Winterowd},\ and\ \citenamefont
  {Zafeiropoulos}}]{ulybyshev_arxiv_1906.02726}%
  \BibitemOpen
  \bibfield  {author} {\bibinfo {author} {\bibfnamefont {M.~V.}\ \bibnamefont
  {Ulybyshev}}, \bibinfo {author} {\bibfnamefont {C.}~\bibnamefont
  {Winterowd}}, \ and\ \bibinfo {author} {\bibfnamefont {S.}~\bibnamefont
  {Zafeiropoulos}},\ }\href {https://arxiv.org/abs/1906.02726v2} {\enquote
  {\bibinfo {title} {{Taming the sign problem of the finite density Hubbard
  model via Lefschetz thimbles}},}\ } (\bibinfo {year} {2019}),\ \Eprint
  {http://arxiv.org/abs/arXiv:1906.02726v2} {arXiv:1906.02726v2} \BibitemShut
  {NoStop}%
\bibitem [{\citenamefont {Fukuma}\ \emph
  {et~al.}(2019{\natexlab{a}})\citenamefont {Fukuma}, \citenamefont
  {Matsumoto},\ and\ \citenamefont {Umeda}}]{fukuma_prd_100_2019}%
  \BibitemOpen
  \bibfield  {author} {\bibinfo {author} {\bibfnamefont {M.}~\bibnamefont
  {Fukuma}}, \bibinfo {author} {\bibfnamefont {N.}~\bibnamefont {Matsumoto}}, \
  and\ \bibinfo {author} {\bibfnamefont {N.}~\bibnamefont {Umeda}},\ }\bibfield
   {title} {\enquote {\bibinfo {title} {{Applying the tempered Lefschetz
  thimble method to the Hubbard model away from half filling}},}\ }\href
  {https://doi.org/10.1103/PhysRevD.100.114510} {\bibfield  {journal} {\bibinfo
   {journal} {Phys. Rev. D}\ }\textbf {\bibinfo {volume} {100}},\ \bibinfo
  {pages} {114510} (\bibinfo {year} {2019}{\natexlab{a}})}\BibitemShut
  {NoStop}%
\bibitem [{\citenamefont {Fukuma}\ \emph
  {et~al.}(2019{\natexlab{b}})\citenamefont {Fukuma}, \citenamefont
  {Matsumoto},\ and\ \citenamefont {Umeda}}]{fukuma_arxiv_1912.13303}%
  \BibitemOpen
  \bibfield  {author} {\bibinfo {author} {\bibfnamefont {M.}~\bibnamefont
  {Fukuma}}, \bibinfo {author} {\bibfnamefont {N.}~\bibnamefont {Matsumoto}}, \
  and\ \bibinfo {author} {\bibfnamefont {N.}~\bibnamefont {Umeda}},\ }\href
  {https://arxiv.org/abs/1912.13303v2} {\enquote {\bibinfo {title}
  {{Implementation of the HMC algorithm on the tempered Lefschetz thimble
  method}},}\ } (\bibinfo {year} {2019}{\natexlab{b}}),\ \Eprint
  {http://arxiv.org/abs/arXiv:1912.13303v2} {arXiv:1912.13303v2} \BibitemShut
  {NoStop}%
\bibitem [{\citenamefont {Ulybyshev}\ \emph
  {et~al.}(2020{\natexlab{a}})\citenamefont {Ulybyshev}, \citenamefont
  {Winterowd},\ and\ \citenamefont {Zafeiropoulos}}]{ulybyshev_prd_101_2020}%
  \BibitemOpen
  \bibfield  {author} {\bibinfo {author} {\bibfnamefont {M.}~\bibnamefont
  {Ulybyshev}}, \bibinfo {author} {\bibfnamefont {C.}~\bibnamefont
  {Winterowd}}, \ and\ \bibinfo {author} {\bibfnamefont {S.}~\bibnamefont
  {Zafeiropoulos}},\ }\bibfield  {title} {\enquote {\bibinfo {title}
  {{Lefschetz thimbles decomposition for the Hubbard model on the hexagonal
  lattice}},}\ }\href {https://doi.org/10.1103/PhysRevD.101.014508} {\bibfield
  {journal} {\bibinfo  {journal} {Phys. Rev. D}\ }\textbf {\bibinfo {volume}
  {101}},\ \bibinfo {pages} {014508} (\bibinfo {year}
  {2020}{\natexlab{a}})}\BibitemShut {NoStop}%
\bibitem [{\citenamefont {Ulybyshev}\ \emph
  {et~al.}(2020{\natexlab{b}})\citenamefont {Ulybyshev}, \citenamefont
  {Dorozhinskii},\ and\ \citenamefont {Pavlovskii}}]{ulybyshev_ppn_51_2020}%
  \BibitemOpen
  \bibfield  {author} {\bibinfo {author} {\bibfnamefont {M.~V.}\ \bibnamefont
  {Ulybyshev}}, \bibinfo {author} {\bibfnamefont {V.~I.}\ \bibnamefont
  {Dorozhinskii}}, \ and\ \bibinfo {author} {\bibfnamefont {O.~V.}\
  \bibnamefont {Pavlovskii}},\ }\bibfield  {title} {\enquote {\bibinfo {title}
  {{The Use of Neural Networks to Solve the Sign Problem in Physical
  Models}},}\ }\href {https://doi.org/10.1134/S1063779620030314} {\bibfield
  {journal} {\bibinfo  {journal} {Phys. Part. Nuclei}\ }\textbf {\bibinfo
  {volume} {51}},\ \bibinfo {pages} {363} (\bibinfo {year}
  {2020}{\natexlab{b}})}\BibitemShut {NoStop}%
\bibitem [{\citenamefont {Wynen}\ \emph {et~al.}(2021)\citenamefont {Wynen},
  \citenamefont {Berkowitz}, \citenamefont {Krieg}, \citenamefont {Luu},\ and\
  \citenamefont {Ostmeyer}}]{wynen_arxiv_2006.11221}%
  \BibitemOpen
  \bibfield  {author} {\bibinfo {author} {\bibfnamefont {J.-L.}\ \bibnamefont
  {Wynen}}, \bibinfo {author} {\bibfnamefont {E.}~\bibnamefont {Berkowitz}},
  \bibinfo {author} {\bibfnamefont {S.}~\bibnamefont {Krieg}}, \bibinfo
  {author} {\bibfnamefont {T.}~\bibnamefont {Luu}}, \ and\ \bibinfo {author}
  {\bibfnamefont {J.}~\bibnamefont {Ostmeyer}},\ }\bibfield  {title} {\enquote
  {\bibinfo {title} {{Machine learning to alleviate Hubbard-model sign
  problems}},}\ }\href {https://doi.org/10.1103/PhysRevB.103.125153} {\bibfield
   {journal} {\bibinfo  {journal} {Phys. Rev. B}\ }\textbf {\bibinfo {volume}
  {103}},\ \bibinfo {pages} {125153} (\bibinfo {year} {2021})}\BibitemShut
  {NoStop}%
\bibitem [{\citenamefont {Stratonovich}(1958)}]{stratonovich_spd_2_1958}%
  \BibitemOpen
  \bibfield  {author} {\bibinfo {author} {\bibfnamefont {R.~L.}\ \bibnamefont
  {Stratonovich}},\ }\bibfield  {title} {\enquote {\bibinfo {title} {{On a
  Method of Calculating Quantum Distribution Functions}},}\ }\href@noop {}
  {\bibfield  {journal} {\bibinfo  {journal} {Soviet Phys. Doklady}\ }\textbf
  {\bibinfo {volume} {2}},\ \bibinfo {pages} {416} (\bibinfo {year}
  {1958})}\BibitemShut {NoStop}%
\bibitem [{\citenamefont {Hubbard}(1959)}]{hubbard_prl_3_1959}%
  \BibitemOpen
  \bibfield  {author} {\bibinfo {author} {\bibfnamefont {J.}~\bibnamefont
  {Hubbard}},\ }\bibfield  {title} {\enquote {\bibinfo {title} {{Calculation of
  Partition Functions.{}}},}\ }\href {https://doi.org/10.1103/PhysRevLett.3.77}
  {\bibfield  {journal} {\bibinfo  {journal} {Phys. Rev, Lett.}\ }\textbf
  {\bibinfo {volume} {3}},\ \bibinfo {pages} {77} (\bibinfo {year}
  {1959})}\BibitemShut {NoStop}%
\bibitem [{\citenamefont {Popov}\ and\ \citenamefont
  {Fedotov}(1988)}]{popov_jetp_67_1988}%
  \BibitemOpen
  \bibfield  {author} {\bibinfo {author} {\bibfnamefont {V.~N.}\ \bibnamefont
  {Popov}}\ and\ \bibinfo {author} {\bibfnamefont {S.~A.}\ \bibnamefont
  {Fedotov}},\ }\bibfield  {title} {\enquote {\bibinfo {title} {{The
  functional-integration method and diagram technique for spin systems}},}\
  }\href@noop {} {\bibfield  {journal} {\bibinfo  {journal} {Sov. Phys. -
  JETP}\ }\textbf {\bibinfo {volume} {67}},\ \bibinfo {pages} {535} (\bibinfo
  {year} {1988})}\BibitemShut {NoStop}%
\bibitem [{pop(1991)}]{popov_psim_184_1991}%
  \BibitemOpen
  \href@noop {} {\bibfield  {journal} {\bibinfo  {journal} {Proc. Steklov Inst.
  Math.}\ }\textbf {\bibinfo {volume} {184}},\ \bibinfo {pages} {177} (\bibinfo
  {year} {1991})}\BibitemShut {NoStop}%
\bibitem [{\citenamefont {Gros}\ and\ \citenamefont
  {Johnson}(1990)}]{gros_physb_165-166_1990}%
  \BibitemOpen
  \bibfield  {author} {\bibinfo {author} {\bibfnamefont {C.}~\bibnamefont
  {Gros}}\ and\ \bibinfo {author} {\bibfnamefont {M.~D.}\ \bibnamefont
  {Johnson}},\ }\bibfield  {title} {\enquote {\bibinfo {title} {{An exact
  mapping of the $t$-$J$ model to the unrestricted Hilbert space}},}\ }\href
  {https://doi.org/10.1016/S0921-4526(09)80078-1} {\bibfield  {journal}
  {\bibinfo  {journal} {Physica B}\ }\textbf {\bibinfo {volume} {165-166}},\
  \bibinfo {pages} {985} (\bibinfo {year} {1990})}\BibitemShut {NoStop}%
\bibitem [{\citenamefont {Stein}\ and\ \citenamefont
  {Oppermann}(1991)}]{stain_zp_83_1991}%
  \BibitemOpen
  \bibfield  {author} {\bibinfo {author} {\bibfnamefont {J.}~\bibnamefont
  {Stein}}\ and\ \bibinfo {author} {\bibfnamefont {R.}~\bibnamefont
  {Oppermann}},\ }\bibfield  {title} {\enquote {\bibinfo {title} {{A
  spin-dependent Popov-Fedotov trick and a new loop expansion for the strong
  coupling negative $U$ Hubbard model}},}\ }\href
  {https://doi.org/10.1007/BF01313402} {\bibfield  {journal} {\bibinfo
  {journal} {Z. Phys. B}\ }\textbf {\bibinfo {volume} {83}},\ \bibinfo {pages}
  {333} (\bibinfo {year} {1991})}\BibitemShut {NoStop}%
\bibitem [{\citenamefont {Veits}\ \emph {et~al.}(1994)\citenamefont {Veits},
  \citenamefont {Oppermann}, \citenamefont {Binderberger},\ and\ \citenamefont
  {Stein}}]{veits_jsjp_4_1994}%
  \BibitemOpen
  \bibfield  {author} {\bibinfo {author} {\bibfnamefont {O.}~\bibnamefont
  {Veits}}, \bibinfo {author} {\bibfnamefont {R.}~\bibnamefont {Oppermann}},
  \bibinfo {author} {\bibfnamefont {M.}~\bibnamefont {Binderberger}}, \ and\
  \bibinfo {author} {\bibfnamefont {J.}~\bibnamefont {Stein}},\ }\bibfield
  {title} {\enquote {\bibinfo {title} {{Extension of the Popov-Fedotov method
  to arbitrary spin}},}\ }\href {https://doi.org/10.1051/jp1:1994154}
  {\bibfield  {journal} {\bibinfo  {journal} {J. Phys. I France}\ }\textbf
  {\bibinfo {volume} {4}},\ \bibinfo {pages} {493} (\bibinfo {year}
  {1994})}\BibitemShut {NoStop}%
\bibitem [{\citenamefont {Bouis}\ and\ \citenamefont
  {Kiselev}(1999)}]{bouis_physb_259-261_1999}%
  \BibitemOpen
  \bibfield  {author} {\bibinfo {author} {\bibfnamefont {F.}~\bibnamefont
  {Bouis}}\ and\ \bibinfo {author} {\bibfnamefont {M.~N.}\ \bibnamefont
  {Kiselev}},\ }\bibfield  {title} {\enquote {\bibinfo {title} {{Effective
  action for the Kondo lattice model. New approach for $S = 1/2$}},}\ }\href
  {https://doi.org/10.1016/S0921-4526(98)01171-5} {\bibfield  {journal}
  {\bibinfo  {journal} {Physica B}\ }\textbf {\bibinfo {volume} {259-261}},\
  \bibinfo {pages} {195} (\bibinfo {year} {1999})}\BibitemShut {NoStop}%
\bibitem [{\citenamefont {Azakov}\ \emph {et~al.}(2000)\citenamefont {Azakov},
  \citenamefont {Dilaver},\ and\ \citenamefont {Oztas}}]{azakov_ijmpb_14_2000}%
  \BibitemOpen
  \bibfield  {author} {\bibinfo {author} {\bibfnamefont {S.}~\bibnamefont
  {Azakov}}, \bibinfo {author} {\bibfnamefont {M.}~\bibnamefont {Dilaver}}, \
  and\ \bibinfo {author} {\bibfnamefont {A.~M.}\ \bibnamefont {Oztas}},\
  }\bibfield  {title} {\enquote {\bibinfo {title} {{The Low-temperature Phase
  of the Heisenberg Antiferromagnet in a Fermionic Representation}},}\ }\href
  {https://doi.org/10.1142/S0217979200000030} {\bibfield  {journal} {\bibinfo
  {journal} {Int. Journal of Modern Phys. B}\ }\textbf {\bibinfo {volume}
  {14}},\ \bibinfo {pages} {13} (\bibinfo {year} {2000})}\BibitemShut {NoStop}%
\bibitem [{\citenamefont {Kiselev}\ and\ \citenamefont
  {Oppermann}(2000{\natexlab{a}})}]{kiselev_prl_85_2000}%
  \BibitemOpen
  \bibfield  {author} {\bibinfo {author} {\bibfnamefont {M.~N.}\ \bibnamefont
  {Kiselev}}\ and\ \bibinfo {author} {\bibfnamefont {R.}~\bibnamefont
  {Oppermann}},\ }\bibfield  {title} {\enquote {\bibinfo {title}
  {{Schwinger-Keldysh Semionic Approach for Quantum Spin Systems}},}\ }\href
  {https://doi.org/10.1103/PhysRevLett.85.5631} {\bibfield  {journal} {\bibinfo
   {journal} {Phys. Rev. Lett}\ }\textbf {\bibinfo {volume} {85}},\ \bibinfo
  {pages} {5631} (\bibinfo {year} {2000}{\natexlab{a}})}\BibitemShut {NoStop}%
\bibitem [{\citenamefont {Kiselev}\ and\ \citenamefont
  {Oppermann}(2000{\natexlab{b}})}]{kiselev_jetp_lett_71_2000}%
  \BibitemOpen
  \bibfield  {author} {\bibinfo {author} {\bibfnamefont {M.~N.}\ \bibnamefont
  {Kiselev}}\ and\ \bibinfo {author} {\bibfnamefont {R.}~\bibnamefont
  {Oppermann}},\ }\bibfield  {title} {\enquote {\bibinfo {title} {{Spin-glass
  transition in a Kondo lattice with quenched disorder}},}\ }\href
  {https://doi.org/10.1134/1.568327} {\bibfield  {journal} {\bibinfo  {journal}
  {JETP Lett.}\ }\textbf {\bibinfo {volume} {71}},\ \bibinfo {pages} {250}
  (\bibinfo {year} {2000}{\natexlab{b}})}\BibitemShut {NoStop}%
\bibitem [{\citenamefont {Kiselev}\ \emph {et~al.}(2001)\citenamefont
  {Kiselev}, \citenamefont {Feldmann},\ and\ \citenamefont
  {Oppermann}}]{kiselev_epjb_22_2001}%
  \BibitemOpen
  \bibfield  {author} {\bibinfo {author} {\bibfnamefont {M.}~\bibnamefont
  {Kiselev}}, \bibinfo {author} {\bibfnamefont {H.}~\bibnamefont {Feldmann}}, \
  and\ \bibinfo {author} {\bibfnamefont {R.}~\bibnamefont {Oppermann}},\
  }\bibfield  {title} {\enquote {\bibinfo {title} {{Semi-fermionic
  representation of SU($\bm{\mathrm{N}}$) Hamiltonians}},}\ }\href
  {https://doi.org/10.1007/PL00011135} {\bibfield  {journal} {\bibinfo
  {journal} {Eur. Phys. J. B}\ }\textbf {\bibinfo {volume} {22}},\ \bibinfo
  {pages} {53} (\bibinfo {year} {2001})}\BibitemShut {NoStop}%
\bibitem [{\citenamefont {Kiselev}\ \emph {et~al.}(2002)\citenamefont
  {Kiselev}, \citenamefont {Kikoin},\ and\ \citenamefont
  {Oppermann}}]{kiselev_prb_65_2002}%
  \BibitemOpen
  \bibfield  {author} {\bibinfo {author} {\bibfnamefont {M.}~\bibnamefont
  {Kiselev}}, \bibinfo {author} {\bibfnamefont {K.}~\bibnamefont {Kikoin}}, \
  and\ \bibinfo {author} {\bibfnamefont {R.}~\bibnamefont {Oppermann}},\
  }\bibfield  {title} {\enquote {\bibinfo {title} {{Ginzburg-Landau functional
  for nearly antiferromagnetic perfect and disordered Kondo lattices}},}\
  }\href {https://doi.org/10.1103/PhysRevB.65.184410} {\bibfield  {journal}
  {\bibinfo  {journal} {Phys. Rev. B}\ }\textbf {\bibinfo {volume} {65}},\
  \bibinfo {pages} {184410} (\bibinfo {year} {2002})}\BibitemShut {NoStop}%
\bibitem [{\citenamefont {Kiselev}\ \emph {et~al.}(2003)\citenamefont
  {Kiselev}, \citenamefont {Kikoin},\ and\ \citenamefont
  {Molenkamp}}]{kiselev_prb_68_2003}%
  \BibitemOpen
  \bibfield  {author} {\bibinfo {author} {\bibfnamefont {M.~N.}\ \bibnamefont
  {Kiselev}}, \bibinfo {author} {\bibfnamefont {K.}~\bibnamefont {Kikoin}}, \
  and\ \bibinfo {author} {\bibfnamefont {L.~W.}\ \bibnamefont {Molenkamp}},\
  }\bibfield  {title} {\enquote {\bibinfo {title} {{Resonance Kondo tunneling
  through a double quantum dot at finite bias}},}\ }\href
  {https://doi.org/10.1103/PhysRevB.68.155323} {\bibfield  {journal} {\bibinfo
  {journal} {Phys. Rev. B}\ }\textbf {\bibinfo {volume} {68}},\ \bibinfo
  {pages} {155323} (\bibinfo {year} {2003})}\BibitemShut {NoStop}%
\bibitem [{\citenamefont {Coleman}\ and\ \citenamefont
  {Mao}(2004)}]{coleman_jpcm_16_2004}%
  \BibitemOpen
  \bibfield  {author} {\bibinfo {author} {\bibfnamefont {P.}~\bibnamefont
  {Coleman}}\ and\ \bibinfo {author} {\bibfnamefont {W.}~\bibnamefont {Mao}},\
  }\bibfield  {title} {\enquote {\bibinfo {title} {{Quantum reciprocity
  conjecture for the non-equilibrium steady state}},}\ }\href
  {https://doi.org/10.1088/0953-8984/16/20/l02} {\bibfield  {journal} {\bibinfo
   {journal} {J. Phys.: Condens. Matter}\ }\textbf {\bibinfo {volume} {16}},\
  \bibinfo {pages} {L263--L269} (\bibinfo {year} {2004})}\BibitemShut {NoStop}%
\bibitem [{\citenamefont {Dillenschneider}\ and\ \citenamefont
  {Richert}(2006)}]{dillenschneider_epjb_49_2006}%
  \BibitemOpen
  \bibfield  {author} {\bibinfo {author} {\bibfnamefont {R.}~\bibnamefont
  {Dillenschneider}}\ and\ \bibinfo {author} {\bibfnamefont {J.}~\bibnamefont
  {Richert}},\ }\bibfield  {title} {\enquote {\bibinfo {title} {{Magnetic
  properties of antiferromagnetic quantum Heisenberg spin systems with a strict
  single particle site occupation}},}\ }\href
  {https://doi.org/10.1140/epjb/e2006-00052-x} {\bibfield  {journal} {\bibinfo
  {journal} {Eur. Phys. J. B}\ }\textbf {\bibinfo {volume} {49}},\ \bibinfo
  {pages} {187} (\bibinfo {year} {2006})}\BibitemShut {NoStop}%
\bibitem [{\citenamefont {Prokof'ev}\ and\ \citenamefont
  {Svistunov}(2011)}]{prokofev_prb_84_2011}%
  \BibitemOpen
  \bibfield  {author} {\bibinfo {author} {\bibfnamefont {N.~V.}\ \bibnamefont
  {Prokof'ev}}\ and\ \bibinfo {author} {\bibfnamefont {B.~V.}\ \bibnamefont
  {Svistunov}},\ }\bibfield  {title} {\enquote {\bibinfo {title} {{From the
  Popov-Fedotov case to universal fermionization}},}\ }\href
  {https://doi.org/10.1103/PhysRevB.84.073102} {\bibfield  {journal} {\bibinfo
  {journal} {Phys. Rev. B}\ }\textbf {\bibinfo {volume} {84}},\ \bibinfo
  {pages} {073102} (\bibinfo {year} {2011})}\BibitemShut {NoStop}%
\bibitem [{\citenamefont {Sandvik}(2010)}]{sandvik_aip_1297_2010}%
  \BibitemOpen
  \bibfield  {author} {\bibinfo {author} {\bibfnamefont {A.~W.}\ \bibnamefont
  {Sandvik}},\ }\bibfield  {title} {\enquote {\bibinfo {title} {{Computational
  Studies of Quantum Spin Systems}},}\ }\href
  {https://doi.org/10.1063/1.3518900} {\bibfield  {journal} {\bibinfo
  {journal} {AIP Conference Proceedings}\ }\textbf {\bibinfo {volume} {1297}},\
  \bibinfo {pages} {135} (\bibinfo {year} {2010})}\BibitemShut {NoStop}%
\bibitem [{\citenamefont {Ulybyshev}\ \emph {et~al.}(2013)\citenamefont
  {Ulybyshev}, \citenamefont {Buividovich}, \citenamefont {Katsnelson},\ and\
  \citenamefont {Polikarpov}}]{ulybyshev_prl_111_2013}%
  \BibitemOpen
  \bibfield  {author} {\bibinfo {author} {\bibfnamefont {M.~V.}\ \bibnamefont
  {Ulybyshev}}, \bibinfo {author} {\bibfnamefont {P.~V.}\ \bibnamefont
  {Buividovich}}, \bibinfo {author} {\bibfnamefont {M.~I.}\ \bibnamefont
  {Katsnelson}}, \ and\ \bibinfo {author} {\bibfnamefont {M.~I.}\ \bibnamefont
  {Polikarpov}},\ }\bibfield  {title} {\enquote {\bibinfo {title} {{Monte Carlo
  Study of the Semimetal-Insulator Phase Transition in Monolayer Graphene with
  a Realistic Interelectron Interaction Potential}},}\ }\href
  {https://doi.org/10.1103/PhysRevLett.111.056801} {\bibfield  {journal}
  {\bibinfo  {journal} {Phys. Rev. Lett.}\ }\textbf {\bibinfo {volume} {111}},\
  \bibinfo {pages} {056801} (\bibinfo {year} {2013})}\BibitemShut {NoStop}%
\bibitem [{\citenamefont {Smith}\ and\ \citenamefont {von
  Smekal}(2014)}]{smith_prb_89_2014}%
  \BibitemOpen
  \bibfield  {author} {\bibinfo {author} {\bibfnamefont {D.}~\bibnamefont
  {Smith}}\ and\ \bibinfo {author} {\bibfnamefont {L.}~\bibnamefont {von
  Smekal}},\ }\bibfield  {title} {\enquote {\bibinfo {title} {{Monte Carlo
  simulation of the tight-binding model of graphene with partially screened
  Coulomb interactions}},}\ }\href {https://doi.org/10.1103/PhysRevB.89.195429}
  {\bibfield  {journal} {\bibinfo  {journal} {Phys. Rev. B}\ }\textbf {\bibinfo
  {volume} {89}},\ \bibinfo {pages} {195429} (\bibinfo {year}
  {2014})}\BibitemShut {NoStop}%
\bibitem [{\citenamefont {Buividovich}\ and\ \citenamefont
  {Polikarpov}(2012)}]{buividovich_prb_86_2012}%
  \BibitemOpen
  \bibfield  {author} {\bibinfo {author} {\bibfnamefont {P.~V.}\ \bibnamefont
  {Buividovich}}\ and\ \bibinfo {author} {\bibfnamefont {M.~I.}\ \bibnamefont
  {Polikarpov}},\ }\bibfield  {title} {\enquote {\bibinfo {title} {{Monte Carlo
  study of the electron transport properties of monolayer graphene within the
  tight-binding model}},}\ }\href {https://doi.org/10.1103/PhysRevB.86.245117}
  {\bibfield  {journal} {\bibinfo  {journal} {Phys. Rev. B}\ }\textbf {\bibinfo
  {volume} {86}},\ \bibinfo {pages} {245117} (\bibinfo {year}
  {2012})}\BibitemShut {NoStop}%
\bibitem [{\citenamefont {Altland}\ and\ \citenamefont
  {Simons}(2010)}]{altland}%
  \BibitemOpen
  \bibfield  {author} {\bibinfo {author} {\bibfnamefont {A.}~\bibnamefont
  {Altland}}\ and\ \bibinfo {author} {\bibfnamefont {B.~D.}\ \bibnamefont
  {Simons}},\ }\href@noop {} {\emph {\bibinfo {title} {{Condensed Matter Field
  Theory}}}},\ \bibinfo {edition} {2nd}\ ed.\ (\bibinfo  {publisher} {Cambridge
  University Press},\ \bibinfo {address} {New York},\ \bibinfo {year}
  {2010})\BibitemShut {NoStop}%
\bibitem [{\citenamefont {Cash}\ and\ \citenamefont
  {Karp}(1990)}]{cash_tms_16_1990}%
  \BibitemOpen
  \bibfield  {author} {\bibinfo {author} {\bibfnamefont {J.~R.}\ \bibnamefont
  {Cash}}\ and\ \bibinfo {author} {\bibfnamefont {A.~H.}\ \bibnamefont
  {Karp}},\ }\bibfield  {title} {\enquote {\bibinfo {title} {{A variable order
  Runge-Kutta method for initial value problems with rapidly varying right-hand
  sides}},}\ }\href {https://doi.org/10.1145/79505.79507} {\bibfield  {journal}
  {\bibinfo  {journal} {ACM Trans. Math. Softw.}\ }\textbf {\bibinfo {volume}
  {16}},\ \bibinfo {pages} {201} (\bibinfo {year} {1990})}\BibitemShut
  {NoStop}%
\bibitem [{\citenamefont {Press}\ \emph {et~al.}(1988)\citenamefont {Press},
  \citenamefont {Teukolsky}, \citenamefont {Vetterling},\ and\ \citenamefont
  {Flannery}}]{numerical_recipes_in_c}%
  \BibitemOpen
  \bibfield  {author} {\bibinfo {author} {\bibfnamefont {W.~H.}\ \bibnamefont
  {Press}}, \bibinfo {author} {\bibfnamefont {S.~A.}\ \bibnamefont
  {Teukolsky}}, \bibinfo {author} {\bibfnamefont {W.~T.}\ \bibnamefont
  {Vetterling}}, \ and\ \bibinfo {author} {\bibfnamefont {B.~P.}\ \bibnamefont
  {Flannery}},\ }\href@noop {} {\emph {\bibinfo {title} {{Numerical Recipes in
  C, The Art of Scientific Computing}}}},\ \bibinfo {edition} {2nd}\ ed.\
  (\bibinfo  {publisher} {Cambridge University Press},\ \bibinfo {address}
  {Cambridge, New York, Port Chester, Melbourne, Sydney},\ \bibinfo {year}
  {1988})\BibitemShut {NoStop}%
\bibitem [{\citenamefont {Kitaev}(2006)}]{kitaev_ann_phys_321_2006}%
  \BibitemOpen
  \bibfield  {author} {\bibinfo {author} {\bibfnamefont {A.}~\bibnamefont
  {Kitaev}},\ }\bibfield  {title} {\enquote {\bibinfo {title} {{Anyons in an
  exactly solved model and beyond}},}\ }\href
  {https://doi.org/10.1016/j.aop.2005.10.005} {\bibfield  {journal} {\bibinfo
  {journal} {Ann. Phys.}\ }\textbf {\bibinfo {volume} {321}},\ \bibinfo {pages}
  {2} (\bibinfo {year} {2006})}\BibitemShut {NoStop}%
\bibitem [{\citenamefont {Baskaran}\ \emph {et~al.}(2007)\citenamefont
  {Baskaran}, \citenamefont {Mandal},\ and\ \citenamefont
  {Shankar}}]{baskaran_prl_98_2007}%
  \BibitemOpen
  \bibfield  {author} {\bibinfo {author} {\bibfnamefont {G.}~\bibnamefont
  {Baskaran}}, \bibinfo {author} {\bibfnamefont {S.}~\bibnamefont {Mandal}}, \
  and\ \bibinfo {author} {\bibfnamefont {R.}~\bibnamefont {Shankar}},\
  }\bibfield  {title} {\enquote {\bibinfo {title} {{Exact Results for Spin
  Dynamics and Fractionalization in the Kitaev Model}},}\ }\href
  {https://doi.org/10.1103/PhysRevLett.98.247201} {\bibfield  {journal}
  {\bibinfo  {journal} {Phys. Rev. Lett.}\ }\textbf {\bibinfo {volume} {98}},\
  \bibinfo {pages} {247201} (\bibinfo {year} {2007})}\BibitemShut {NoStop}%
\bibitem [{\citenamefont {Baek}\ \emph {et~al.}(2017)\citenamefont {Baek},
  \citenamefont {Do}, \citenamefont {Choi}, \citenamefont {Kwon}, \citenamefont
  {Wolter}, \citenamefont {Nishimoto}, \citenamefont {van~den Brink},\ and\
  \citenamefont {Buchner}}]{baek_prl_119_2017}%
  \BibitemOpen
  \bibfield  {author} {\bibinfo {author} {\bibfnamefont {S.-H.}\ \bibnamefont
  {Baek}}, \bibinfo {author} {\bibfnamefont {S.-H.}\ \bibnamefont {Do}},
  \bibinfo {author} {\bibfnamefont {K.-Y.}\ \bibnamefont {Choi}}, \bibinfo
  {author} {\bibfnamefont {Y.~S.}\ \bibnamefont {Kwon}}, \bibinfo {author}
  {\bibfnamefont {A.~U.~B.}\ \bibnamefont {Wolter}}, \bibinfo {author}
  {\bibfnamefont {S.}~\bibnamefont {Nishimoto}}, \bibinfo {author}
  {\bibfnamefont {J.}~\bibnamefont {van~den Brink}}, \ and\ \bibinfo {author}
  {\bibfnamefont {B.}~\bibnamefont {Buchner}},\ }\bibfield  {title} {\enquote
  {\bibinfo {title} {{Evidence for a Field-Induced Quantum Spin Liquid in
  $\alpha$-RuCl$_3$}},}\ }\href
  {https://doi.org/10.1103/PhysRevLett.119.037201} {\bibfield  {journal}
  {\bibinfo  {journal} {Phys. Rev. Lett.}\ }\textbf {\bibinfo {volume} {119}},\
  \bibinfo {pages} {037201} (\bibinfo {year} {2017})}\BibitemShut {NoStop}%
\bibitem [{\citenamefont {Zheng}\ \emph {et~al.}(2017)\citenamefont {Zheng},
  \citenamefont {Ran}, \citenamefont {Li}, \citenamefont {Wang}, \citenamefont
  {Wang}, \citenamefont {Liu}, \citenamefont {Liu}, \citenamefont {Normand},
  \citenamefont {Wen},\ and\ \citenamefont {Yu}}]{zheng_prl_119_2017}%
  \BibitemOpen
  \bibfield  {author} {\bibinfo {author} {\bibfnamefont {J.}~\bibnamefont
  {Zheng}}, \bibinfo {author} {\bibfnamefont {K.}~\bibnamefont {Ran}}, \bibinfo
  {author} {\bibfnamefont {T.}~\bibnamefont {Li}}, \bibinfo {author}
  {\bibfnamefont {J.}~\bibnamefont {Wang}}, \bibinfo {author} {\bibfnamefont
  {P.}~\bibnamefont {Wang}}, \bibinfo {author} {\bibfnamefont {B.}~\bibnamefont
  {Liu}}, \bibinfo {author} {\bibfnamefont {Z.-X.}\ \bibnamefont {Liu}},
  \bibinfo {author} {\bibfnamefont {B.}~\bibnamefont {Normand}}, \bibinfo
  {author} {\bibfnamefont {J.}~\bibnamefont {Wen}}, \ and\ \bibinfo {author}
  {\bibfnamefont {W.}~\bibnamefont {Yu}},\ }\bibfield  {title} {\enquote
  {\bibinfo {title} {{Gapless Spin Excitations in the Field-Induced Quantum
  Spin Liquid Phase of $\alpha$-RuCl$_3$}},}\ }\href
  {https://doi.org/10.1103/PhysRevLett.119.227208} {\bibfield  {journal}
  {\bibinfo  {journal} {Phys. Rev. Lett.}\ }\textbf {\bibinfo {volume} {119}},\
  \bibinfo {pages} {227208} (\bibinfo {year} {2017})}\BibitemShut {NoStop}%
\bibitem [{\citenamefont {Wang}\ \emph {et~al.}(2017)\citenamefont {Wang},
  \citenamefont {Reschke}, \citenamefont {Huvonen}, \citenamefont {Do},
  \citenamefont {Choi}, \citenamefont {Gensch}, \citenamefont {Nagel},
  \citenamefont {Room},\ and\ \citenamefont {Loidl}}]{wang_prl_119_2017}%
  \BibitemOpen
  \bibfield  {author} {\bibinfo {author} {\bibfnamefont {Z.}~\bibnamefont
  {Wang}}, \bibinfo {author} {\bibfnamefont {S.}~\bibnamefont {Reschke}},
  \bibinfo {author} {\bibfnamefont {D.}~\bibnamefont {Huvonen}}, \bibinfo
  {author} {\bibfnamefont {S.-H.}\ \bibnamefont {Do}}, \bibinfo {author}
  {\bibfnamefont {K.-Y.}\ \bibnamefont {Choi}}, \bibinfo {author}
  {\bibfnamefont {M.}~\bibnamefont {Gensch}}, \bibinfo {author} {\bibfnamefont
  {U.}~\bibnamefont {Nagel}}, \bibinfo {author} {\bibfnamefont
  {T.}~\bibnamefont {Room}}, \ and\ \bibinfo {author} {\bibfnamefont
  {A.}~\bibnamefont {Loidl}},\ }\bibfield  {title} {\enquote {\bibinfo {title}
  {{Magnetic Excitations and Continuum of a Possibly Field-Induced Quantum Spin
  Liquid in $\alpha$-RuCl$_3$}},}\ }\href
  {https://doi.org/10.1103/PhysRevLett.119.227202} {\bibfield  {journal}
  {\bibinfo  {journal} {Phys. Rev. Lett.}\ }\textbf {\bibinfo {volume} {119}},\
  \bibinfo {pages} {227202} (\bibinfo {year} {2017})}\BibitemShut {NoStop}%
\bibitem [{\citenamefont {Ponomaryov}\ \emph {et~al.}(2017)\citenamefont
  {Ponomaryov}, \citenamefont {Schulze}, \citenamefont {Wosnitza},
  \citenamefont {Lampen-Kelley}, \citenamefont {Banerjee}, \citenamefont {Yan},
  \citenamefont {Bridges}, \citenamefont {Mandrus}, \citenamefont {Nagler},
  \citenamefont {Kolezhuk},\ and\ \citenamefont
  {Zvyagin}}]{ponomaryov_prb_96_2017}%
  \BibitemOpen
  \bibfield  {author} {\bibinfo {author} {\bibfnamefont {A.~N.}\ \bibnamefont
  {Ponomaryov}}, \bibinfo {author} {\bibfnamefont {E.}~\bibnamefont {Schulze}},
  \bibinfo {author} {\bibfnamefont {J.}~\bibnamefont {Wosnitza}}, \bibinfo
  {author} {\bibfnamefont {P.}~\bibnamefont {Lampen-Kelley}}, \bibinfo {author}
  {\bibfnamefont {A.}~\bibnamefont {Banerjee}}, \bibinfo {author}
  {\bibfnamefont {J.-Q.}\ \bibnamefont {Yan}}, \bibinfo {author} {\bibfnamefont
  {C.~A.}\ \bibnamefont {Bridges}}, \bibinfo {author} {\bibfnamefont {D.~G.}\
  \bibnamefont {Mandrus}}, \bibinfo {author} {\bibfnamefont {S.~E.}\
  \bibnamefont {Nagler}}, \bibinfo {author} {\bibfnamefont {A.~K.}\
  \bibnamefont {Kolezhuk}}, \ and\ \bibinfo {author} {\bibfnamefont {S.~A.}\
  \bibnamefont {Zvyagin}},\ }\bibfield  {title} {\enquote {\bibinfo {title}
  {{Unconventional spin dynamics in the honeycomb-lattice material
  $\alpha$-RuCl$_3$: High-field electron spin resonance studies}},}\ }\href
  {https://doi.org/10.1103/PhysRevB.96.241107} {\bibfield  {journal} {\bibinfo
  {journal} {Phys. Rev. B}\ }\textbf {\bibinfo {volume} {96}},\ \bibinfo
  {pages} {241107(R)} (\bibinfo {year} {2017})}\BibitemShut {NoStop}%
\bibitem [{\citenamefont {Banerjee}\ \emph {et~al.}(2018)\citenamefont
  {Banerjee}, \citenamefont {Lampen-Kelley}, \citenamefont {Knolle},
  \citenamefont {Balz}, \citenamefont {Aczel}, \citenamefont {Winn},
  \citenamefont {Liu}, \citenamefont {Pajerowski}, \citenamefont {Yan},
  \citenamefont {Bridges}, \citenamefont {Savici}, \citenamefont {Chakoumakos},
  \citenamefont {Lumsden}, \citenamefont {Tennant}, \citenamefont {Moessner},
  \citenamefont {Mandrus},\ and\ \citenamefont {Nagler}}]{banerjee_nqm_3_2018}%
  \BibitemOpen
  \bibfield  {author} {\bibinfo {author} {\bibfnamefont {A.}~\bibnamefont
  {Banerjee}}, \bibinfo {author} {\bibfnamefont {P.}~\bibnamefont
  {Lampen-Kelley}}, \bibinfo {author} {\bibfnamefont {J.}~\bibnamefont
  {Knolle}}, \bibinfo {author} {\bibfnamefont {C.}~\bibnamefont {Balz}},
  \bibinfo {author} {\bibfnamefont {A.~A.}\ \bibnamefont {Aczel}}, \bibinfo
  {author} {\bibfnamefont {B.}~\bibnamefont {Winn}}, \bibinfo {author}
  {\bibfnamefont {Y.}~\bibnamefont {Liu}}, \bibinfo {author} {\bibfnamefont
  {D.}~\bibnamefont {Pajerowski}}, \bibinfo {author} {\bibfnamefont
  {J.}~\bibnamefont {Yan}}, \bibinfo {author} {\bibfnamefont {C.~A.}\
  \bibnamefont {Bridges}}, \bibinfo {author} {\bibfnamefont {A.~T.}\
  \bibnamefont {Savici}}, \bibinfo {author} {\bibfnamefont {B.~C.}\
  \bibnamefont {Chakoumakos}}, \bibinfo {author} {\bibfnamefont {M.~D.}\
  \bibnamefont {Lumsden}}, \bibinfo {author} {\bibfnamefont {D.~A.}\
  \bibnamefont {Tennant}}, \bibinfo {author} {\bibfnamefont {R.}~\bibnamefont
  {Moessner}}, \bibinfo {author} {\bibfnamefont {D.~G.}\ \bibnamefont
  {Mandrus}}, \ and\ \bibinfo {author} {\bibfnamefont {S.~E.}\ \bibnamefont
  {Nagler}},\ }\bibfield  {title} {\enquote {\bibinfo {title} {{Excitations in
  the field-induced quantum spin liquid state of $\alpha$-RuCl$_3$}},}\ }\href
  {https://doi.org/10.1038/s41535-018-0079-2} {\bibfield  {journal} {\bibinfo
  {journal} {npj Quant. Mater.}\ }\textbf {\bibinfo {volume} {3}},\ \bibinfo
  {pages} {8} (\bibinfo {year} {2018})}\BibitemShut {NoStop}%
\bibitem [{\citenamefont {Jansa}\ \emph {et~al.}(2018)\citenamefont {Jansa},
  \citenamefont {Zorko}, \citenamefont {Gomilsek}, \citenamefont {Pregelj},
  \citenamefont {Kramer}, \citenamefont {Biner}, \citenamefont {Biffin},
  \citenamefont {Ruegg},\ and\ \citenamefont {Klanjsek}}]{jansa_nphys_14_2018}%
  \BibitemOpen
  \bibfield  {author} {\bibinfo {author} {\bibfnamefont {N.}~\bibnamefont
  {Jansa}}, \bibinfo {author} {\bibfnamefont {A.}~\bibnamefont {Zorko}},
  \bibinfo {author} {\bibfnamefont {M.}~\bibnamefont {Gomilsek}}, \bibinfo
  {author} {\bibfnamefont {M.}~\bibnamefont {Pregelj}}, \bibinfo {author}
  {\bibfnamefont {K.~W.}\ \bibnamefont {Kramer}}, \bibinfo {author}
  {\bibfnamefont {D.}~\bibnamefont {Biner}}, \bibinfo {author} {\bibfnamefont
  {A.}~\bibnamefont {Biffin}}, \bibinfo {author} {\bibfnamefont
  {C.}~\bibnamefont {Ruegg}}, \ and\ \bibinfo {author} {\bibfnamefont
  {M.}~\bibnamefont {Klanjsek}},\ }\bibfield  {title} {\enquote {\bibinfo
  {title} {{Observation of two types of fractional excitation in the Kitaev
  honeycomb magnet}},}\ }\href {https://doi.org/10.1038/s41567-018-0129-5}
  {\bibfield  {journal} {\bibinfo  {journal} {Nat. Phys.}\ }\textbf {\bibinfo
  {volume} {14}},\ \bibinfo {pages} {786} (\bibinfo {year} {2018})}\BibitemShut
  {NoStop}%
\bibitem [{\citenamefont {Wellm}\ \emph {et~al.}(2018)\citenamefont {Wellm},
  \citenamefont {Zeisner}, \citenamefont {Alfonsov}, \citenamefont {Wolter},
  \citenamefont {Roslova}, \citenamefont {Isaeva}, \citenamefont {Doert},
  \citenamefont {Vojta}, \citenamefont {Buchner},\ and\ \citenamefont
  {Kataev}}]{wellm_prb_98_2018}%
  \BibitemOpen
  \bibfield  {author} {\bibinfo {author} {\bibfnamefont {C.}~\bibnamefont
  {Wellm}}, \bibinfo {author} {\bibfnamefont {J.}~\bibnamefont {Zeisner}},
  \bibinfo {author} {\bibfnamefont {A.}~\bibnamefont {Alfonsov}}, \bibinfo
  {author} {\bibfnamefont {A.~U.~B.}\ \bibnamefont {Wolter}}, \bibinfo {author}
  {\bibfnamefont {M.}~\bibnamefont {Roslova}}, \bibinfo {author} {\bibfnamefont
  {A.}~\bibnamefont {Isaeva}}, \bibinfo {author} {\bibfnamefont
  {T.}~\bibnamefont {Doert}}, \bibinfo {author} {\bibfnamefont
  {M.}~\bibnamefont {Vojta}}, \bibinfo {author} {\bibfnamefont
  {B.}~\bibnamefont {Buchner}}, \ and\ \bibinfo {author} {\bibfnamefont
  {V.}~\bibnamefont {Kataev}},\ }\bibfield  {title} {\enquote {\bibinfo {title}
  {{Signatures of low-energy fractionalized excitations in $\alpha$-RuCl$_3$
  from field-dependent microwave absorption}},}\ }\href
  {https://doi.org/10.1103/PhysRevB.98.184408} {\bibfield  {journal} {\bibinfo
  {journal} {Phys. Rev. B}\ }\textbf {\bibinfo {volume} {98}},\ \bibinfo
  {pages} {184408} (\bibinfo {year} {2018})}\BibitemShut {NoStop}%
\bibitem [{\citenamefont {Nagai}\ \emph {et~al.}(2020)\citenamefont {Nagai},
  \citenamefont {Jinno}, \citenamefont {Yoshitake}, \citenamefont {Nasu},
  \citenamefont {Motome}, \citenamefont {Itoh},\ and\ \citenamefont
  {Shimizu}}]{nagai_prb_101_2020}%
  \BibitemOpen
  \bibfield  {author} {\bibinfo {author} {\bibfnamefont {Y.}~\bibnamefont
  {Nagai}}, \bibinfo {author} {\bibfnamefont {T.}~\bibnamefont {Jinno}},
  \bibinfo {author} {\bibfnamefont {Y.}~\bibnamefont {Yoshitake}}, \bibinfo
  {author} {\bibfnamefont {J.}~\bibnamefont {Nasu}}, \bibinfo {author}
  {\bibfnamefont {Y.}~\bibnamefont {Motome}}, \bibinfo {author} {\bibfnamefont
  {M.}~\bibnamefont {Itoh}}, \ and\ \bibinfo {author} {\bibfnamefont
  {Y.}~\bibnamefont {Shimizu}},\ }\bibfield  {title} {\enquote {\bibinfo
  {title} {{Two-step gap opening across the quantum critical point in the
  Kitaev honeycomb magnet $\alpha$-RuCl$_3$}},}\ }\href
  {https://doi.org/10.1103/PhysRevB.101.020414} {\bibfield  {journal} {\bibinfo
   {journal} {Phys. Rev. B}\ }\textbf {\bibinfo {volume} {101}},\ \bibinfo
  {pages} {020414(R)} (\bibinfo {year} {2020})}\BibitemShut {NoStop}%
\bibitem [{\citenamefont {Motome}\ and\ \citenamefont
  {Nasu}(2020)}]{motome_jpsj_89_2020}%
  \BibitemOpen
  \bibfield  {author} {\bibinfo {author} {\bibfnamefont {Y.}~\bibnamefont
  {Motome}}\ and\ \bibinfo {author} {\bibfnamefont {J.}~\bibnamefont {Nasu}},\
  }\bibfield  {title} {\enquote {\bibinfo {title} {{Hunting Majorana Fermions
  in Kitaev Magnets}},}\ }\href {https://doi.org/10.7566/JPSJ.89.012002}
  {\bibfield  {journal} {\bibinfo  {journal} {J. Phys. Soc. Jpn.}\ }\textbf
  {\bibinfo {volume} {89}},\ \bibinfo {pages} {012002} (\bibinfo {year}
  {2020})}\BibitemShut {NoStop}%
\bibitem [{\citenamefont {Takagi}\ \emph {et~al.}(2019)\citenamefont {Takagi},
  \citenamefont {Takayama}, \citenamefont {Jackeli},\ and\ \citenamefont
  {Khaliullin}}]{takagi_nrp_1_2019}%
  \BibitemOpen
  \bibfield  {author} {\bibinfo {author} {\bibfnamefont {H.}~\bibnamefont
  {Takagi}}, \bibinfo {author} {\bibfnamefont {T.}~\bibnamefont {Takayama}},
  \bibinfo {author} {\bibfnamefont {G.}~\bibnamefont {Jackeli}}, \ and\
  \bibinfo {author} {\bibfnamefont {G.}~\bibnamefont {Khaliullin}},\ }\bibfield
   {title} {\enquote {\bibinfo {title} {{Concept and realization of Kitaev
  quantum spin liquids}},}\ }\href {https://doi.org/10.1038/s42254-019-0038-2}
  {\bibfield  {journal} {\bibinfo  {journal} {Nat. Rev. Phys.}\ }\textbf
  {\bibinfo {volume} {1}},\ \bibinfo {pages} {264} (\bibinfo {year}
  {2019})}\BibitemShut {NoStop}%
\bibitem [{\citenamefont {Kasahara}\ \emph {et~al.}(2018)\citenamefont
  {Kasahara}, \citenamefont {Ohnishi}, \citenamefont {Mizukami}, \citenamefont
  {Tanaka}, \citenamefont {Ma}, \citenamefont {Sugii}, \citenamefont {Kurita},
  \citenamefont {Tanaka}, \citenamefont {Nasu}, \citenamefont {Motome},
  \citenamefont {Shibauchi},\ and\ \citenamefont
  {Matsuda}}]{kasahra_nature_559_2018}%
  \BibitemOpen
  \bibfield  {author} {\bibinfo {author} {\bibfnamefont {Y.}~\bibnamefont
  {Kasahara}}, \bibinfo {author} {\bibfnamefont {T.}~\bibnamefont {Ohnishi}},
  \bibinfo {author} {\bibfnamefont {Y.}~\bibnamefont {Mizukami}}, \bibinfo
  {author} {\bibfnamefont {O.}~\bibnamefont {Tanaka}}, \bibinfo {author}
  {\bibfnamefont {S.}~\bibnamefont {Ma}}, \bibinfo {author} {\bibfnamefont
  {K.}~\bibnamefont {Sugii}}, \bibinfo {author} {\bibfnamefont
  {N.}~\bibnamefont {Kurita}}, \bibinfo {author} {\bibfnamefont
  {H.}~\bibnamefont {Tanaka}}, \bibinfo {author} {\bibfnamefont
  {J.}~\bibnamefont {Nasu}}, \bibinfo {author} {\bibfnamefont {Y.}~\bibnamefont
  {Motome}}, \bibinfo {author} {\bibfnamefont {T.}~\bibnamefont {Shibauchi}}, \
  and\ \bibinfo {author} {\bibfnamefont {Y.}~\bibnamefont {Matsuda}},\
  }\bibfield  {title} {\enquote {\bibinfo {title} {{Majorana quantization and
  half-integer thermal quantum Hall effect in a Kitaev spin liquid}},}\ }\href
  {https://doi.org/10.1038/s41586-018-0274-0} {\bibfield  {journal} {\bibinfo
  {journal} {Nature}\ }\textbf {\bibinfo {volume} {559}},\ \bibinfo {pages}
  {227} (\bibinfo {year} {2018})}\BibitemShut {NoStop}%
\bibitem [{\citenamefont {Nasu}\ \emph {et~al.}(2014)\citenamefont {Nasu},
  \citenamefont {Udagawa},\ and\ \citenamefont {Motome}}]{nasu_prl_113_2014}%
  \BibitemOpen
  \bibfield  {author} {\bibinfo {author} {\bibfnamefont {J.}~\bibnamefont
  {Nasu}}, \bibinfo {author} {\bibfnamefont {M.}~\bibnamefont {Udagawa}}, \
  and\ \bibinfo {author} {\bibfnamefont {Y.}~\bibnamefont {Motome}},\
  }\bibfield  {title} {\enquote {\bibinfo {title} {{Vaporization of Kitaev Spin
  Liquids}},}\ }\href {https://doi.org/10.1103/PhysRevLett.113.197205}
  {\bibfield  {journal} {\bibinfo  {journal} {Phys. Rev. Lett.}\ }\textbf
  {\bibinfo {volume} {113}},\ \bibinfo {pages} {197205} (\bibinfo {year}
  {2014})}\BibitemShut {NoStop}%
\bibitem [{\citenamefont {Nasu}\ \emph {et~al.}(2015)\citenamefont {Nasu},
  \citenamefont {Udagawa},\ and\ \citenamefont {Motome}}]{nasu_prb_92_2015}%
  \BibitemOpen
  \bibfield  {author} {\bibinfo {author} {\bibfnamefont {J.}~\bibnamefont
  {Nasu}}, \bibinfo {author} {\bibfnamefont {M.}~\bibnamefont {Udagawa}}, \
  and\ \bibinfo {author} {\bibfnamefont {Y.}~\bibnamefont {Motome}},\
  }\bibfield  {title} {\enquote {\bibinfo {title} {{Thermal fractionalization
  of quantum spins in a Kitaev model: Temperature-linear specific heat and
  coherent transport of Majorana fermions}},}\ }\href {\doibase
  10.1103/PhysRevB.92.115122} {\bibfield  {journal} {\bibinfo  {journal} {Phys.
  Rev. B}\ }\textbf {\bibinfo {volume} {92}},\ \bibinfo {pages} {115122}
  (\bibinfo {year} {2015})}\BibitemShut {NoStop}%
\bibitem [{\citenamefont {Nasu}\ and\ \citenamefont
  {Motome}(2015)}]{nasu_prl_115_2015}%
  \BibitemOpen
  \bibfield  {author} {\bibinfo {author} {\bibfnamefont {J.}~\bibnamefont
  {Nasu}}\ and\ \bibinfo {author} {\bibfnamefont {Y.}~\bibnamefont {Motome}},\
  }\bibfield  {title} {\enquote {\bibinfo {title} {{Thermodynamics of Chiral
  Spin Liquids with Abelian and Non-Abelian Anyons}},}\ }\href
  {https://doi.org/10.1103/PhysRevLett.115.087203} {\bibfield  {journal}
  {\bibinfo  {journal} {Phys. Rev. Lett.}\ }\textbf {\bibinfo {volume} {115}},\
  \bibinfo {pages} {087203} (\bibinfo {year} {2015})}\BibitemShut {NoStop}%
\bibitem [{\citenamefont {Nasu}\ \emph {et~al.}(2016)\citenamefont {Nasu},
  \citenamefont {Knolle}, \citenamefont {Kovrizhin}, \citenamefont {Motome},\
  and\ \citenamefont {Moessner}}]{nasu_nphys_12_2016}%
  \BibitemOpen
  \bibfield  {author} {\bibinfo {author} {\bibfnamefont {J.}~\bibnamefont
  {Nasu}}, \bibinfo {author} {\bibfnamefont {J.}~\bibnamefont {Knolle}},
  \bibinfo {author} {\bibfnamefont {D.~L.}\ \bibnamefont {Kovrizhin}}, \bibinfo
  {author} {\bibfnamefont {Y.}~\bibnamefont {Motome}}, \ and\ \bibinfo {author}
  {\bibfnamefont {R.}~\bibnamefont {Moessner}},\ }\bibfield  {title} {\enquote
  {\bibinfo {title} {{Fermionic response from fractionalization in an
  insulating two-dimensional magnet}},}\ }\href
  {https://doi.org/doi:10.1038/nphys3809} {\bibfield  {journal} {\bibinfo
  {journal} {Nature Physics}\ }\textbf {\bibinfo {volume} {12}},\ \bibinfo
  {pages} {912} (\bibinfo {year} {2016})}\BibitemShut {NoStop}%
\bibitem [{\citenamefont {Mishchenko}\ \emph {et~al.}(2017)\citenamefont
  {Mishchenko}, \citenamefont {Kato},\ and\ \citenamefont
  {Motome}}]{mishchenko_prb_96_2017}%
  \BibitemOpen
  \bibfield  {author} {\bibinfo {author} {\bibfnamefont {P.~A.}\ \bibnamefont
  {Mishchenko}}, \bibinfo {author} {\bibfnamefont {Y.}~\bibnamefont {Kato}}, \
  and\ \bibinfo {author} {\bibfnamefont {Y.}~\bibnamefont {Motome}},\
  }\bibfield  {title} {\enquote {\bibinfo {title} {{Finite-temperature phase
  transition to a Kitaev spin liquid phase on a hyperoctagon lattice: A
  large-scale quantum Monte Carlo study}},}\ }\href
  {https://doi.org/10.1103/PhysRevB.96.125124} {\bibfield  {journal} {\bibinfo
  {journal} {Phys. Rev. B}\ }\textbf {\bibinfo {volume} {96}},\ \bibinfo
  {pages} {125124} (\bibinfo {year} {2017})}\BibitemShut {NoStop}%
\bibitem [{\citenamefont {Eschmann}\ \emph {et~al.}(2019)\citenamefont
  {Eschmann}, \citenamefont {Mishchenko}, \citenamefont {Bojesen},
  \citenamefont {Kato}, \citenamefont {Hermanns}, \citenamefont {Motome},\ and\
  \citenamefont {Trebst}}]{eschmann_prr_1_2019}%
  \BibitemOpen
  \bibfield  {author} {\bibinfo {author} {\bibfnamefont {T.}~\bibnamefont
  {Eschmann}}, \bibinfo {author} {\bibfnamefont {P.~A.}\ \bibnamefont
  {Mishchenko}}, \bibinfo {author} {\bibfnamefont {T.~A.}\ \bibnamefont
  {Bojesen}}, \bibinfo {author} {\bibfnamefont {Y.}~\bibnamefont {Kato}},
  \bibinfo {author} {\bibfnamefont {M.}~\bibnamefont {Hermanns}}, \bibinfo
  {author} {\bibfnamefont {Y.}~\bibnamefont {Motome}}, \ and\ \bibinfo {author}
  {\bibfnamefont {S.}~\bibnamefont {Trebst}},\ }\bibfield  {title} {\enquote
  {\bibinfo {title} {{Thermodynamics of a gauge-frustrated Kitaev spin
  liquid}},}\ }\href {https://doi.org/10.1103/PhysRevResearch.1.032011}
  {\bibfield  {journal} {\bibinfo  {journal} {Phys. Rev. Research}\ }\textbf
  {\bibinfo {volume} {1}},\ \bibinfo {pages} {032011(R)} (\bibinfo {year}
  {2019})}\BibitemShut {NoStop}%
\bibitem [{\citenamefont {Mishchenko}\ \emph {et~al.}(2020)\citenamefont
  {Mishchenko}, \citenamefont {Kato}, \citenamefont {O'Brien}, \citenamefont
  {Bojesen}, \citenamefont {Eschmann}, \citenamefont {Hermanns}, \citenamefont
  {Trebst},\ and\ \citenamefont {Motome}}]{mishchenko_prb_101_2020}%
  \BibitemOpen
  \bibfield  {author} {\bibinfo {author} {\bibfnamefont {P.~A.}\ \bibnamefont
  {Mishchenko}}, \bibinfo {author} {\bibfnamefont {Y.}~\bibnamefont {Kato}},
  \bibinfo {author} {\bibfnamefont {K.}~\bibnamefont {O'Brien}}, \bibinfo
  {author} {\bibfnamefont {T.~A.}\ \bibnamefont {Bojesen}}, \bibinfo {author}
  {\bibfnamefont {T.}~\bibnamefont {Eschmann}}, \bibinfo {author}
  {\bibfnamefont {M.}~\bibnamefont {Hermanns}}, \bibinfo {author}
  {\bibfnamefont {S.}~\bibnamefont {Trebst}}, \ and\ \bibinfo {author}
  {\bibfnamefont {Y.}~\bibnamefont {Motome}},\ }\bibfield  {title} {\enquote
  {\bibinfo {title} {{Chiral spin liquids with crystalline $\mathbb{Z}_2$ gauge
  order in a three-dimensional Kitaev model}},}\ }\href
  {https://doi.org/10.1103/PhysRevB.101.045118} {\bibfield  {journal} {\bibinfo
   {journal} {Phys. Rev. B}\ }\textbf {\bibinfo {volume} {101}},\ \bibinfo
  {pages} {045118} (\bibinfo {year} {2020})}\BibitemShut {NoStop}%
\bibitem [{\citenamefont {Eschmann}\ \emph {et~al.}(2020)\citenamefont
  {Eschmann}, \citenamefont {Mishchenko}, \citenamefont {O'Brien},
  \citenamefont {Bojesen}, \citenamefont {Kato}, \citenamefont {Hermanns},
  \citenamefont {Motome},\ and\ \citenamefont
  {Trebst}}]{eschmann_prb_102_2020}%
  \BibitemOpen
  \bibfield  {author} {\bibinfo {author} {\bibfnamefont {T.}~\bibnamefont
  {Eschmann}}, \bibinfo {author} {\bibfnamefont {P.~A.}\ \bibnamefont
  {Mishchenko}}, \bibinfo {author} {\bibfnamefont {K.}~\bibnamefont {O'Brien}},
  \bibinfo {author} {\bibfnamefont {T.~A.}\ \bibnamefont {Bojesen}}, \bibinfo
  {author} {\bibfnamefont {Y.}~\bibnamefont {Kato}}, \bibinfo {author}
  {\bibfnamefont {M.}~\bibnamefont {Hermanns}}, \bibinfo {author}
  {\bibfnamefont {Y.}~\bibnamefont {Motome}}, \ and\ \bibinfo {author}
  {\bibfnamefont {S.}~\bibnamefont {Trebst}},\ }\bibfield  {title} {\enquote
  {\bibinfo {title} {{Thermodynamic classification of three-dimensional Kitaev
  spin liquids}},}\ }\href {https://doi.org/10.1103/PhysRevB.102.075125}
  {\bibfield  {journal} {\bibinfo  {journal} {Phys. Rev. B}\ }\textbf {\bibinfo
  {volume} {102}},\ \bibinfo {pages} {075125} (\bibinfo {year}
  {2020})}\BibitemShut {NoStop}%
\bibitem [{\citenamefont {Nasu}\ \emph {et~al.}(2017)\citenamefont {Nasu},
  \citenamefont {Yoshitake},\ and\ \citenamefont {Motome}}]{nasu_prl_119_2017}%
  \BibitemOpen
  \bibfield  {author} {\bibinfo {author} {\bibfnamefont {J.}~\bibnamefont
  {Nasu}}, \bibinfo {author} {\bibfnamefont {J.}~\bibnamefont {Yoshitake}}, \
  and\ \bibinfo {author} {\bibfnamefont {Y.}~\bibnamefont {Motome}},\
  }\bibfield  {title} {\enquote {\bibinfo {title} {{Thermal Transport in the
  Kitaev Model}},}\ }\href {https://doi.org/10.1103/PhysRevLett.119.127204}
  {\bibfield  {journal} {\bibinfo  {journal} {Phys. Rev. Lett.}\ }\textbf
  {\bibinfo {volume} {119}},\ \bibinfo {pages} {127204} (\bibinfo {year}
  {2017})}\BibitemShut {NoStop}%
\bibitem [{\citenamefont {Taylor}(1997)}]{taylor_1997}%
  \BibitemOpen
  \bibfield  {author} {\bibinfo {author} {\bibfnamefont {J.~R.}\ \bibnamefont
  {Taylor}},\ }\href@noop {} {\emph {\bibinfo {title} {{An Introduction to
  Error Analysis: The Study of Uncertainties in Physical Measurements}}}},\
  \bibinfo {edition} {2nd}\ ed.\ (\bibinfo  {publisher} {University Science
  Books},\ \bibinfo {address} {Sausalito, California},\ \bibinfo {year}
  {1997})\BibitemShut {NoStop}%
\end{thebibliography}%

\widetext
\appendix

%
\section{Explicit form of the action derived for the four site Kitaev model}
\label{sec:appendixA}
%

The explicit form of the action $\mathcal{S}(\bm{\varphi})$ in Eq.~(\ref{action}) derived for the model in Eq.~(\ref{4_sites_kitaev_h})
with $h^y = h^z = 0$ is expressed as
\begin{align*}
&
\mathcal{S}(\varphi_1, \varphi_2, \varphi_3) = \frac{\Delta}{2} \left( \varphi_1^2 + \varphi_2^2 + \varphi_3^2 \right)
- \log \left[ 2 \left( 1 + i \right) \right] \nonumber \\
&
- \log \left( 1 + e^{-2\Delta\sqrt{K^z}\varphi_3} \right)
- 2 \log \cosh \left( \Delta h^x \right)
- \log \cosh \left[ \Delta \left( h^x - \sqrt{K^x}\varphi_1 \right) \right]
- \log \cosh \left( \Delta              \sqrt{K^y}\varphi_2 \right)         \nonumber \\
&
- \log
\left\{
\left( 1 - ie^{2\Delta\sqrt{K^z}\varphi_3} \right)
\cosh \left[ \Delta \left( h^x - \sqrt{K^x} \varphi_1 - \sqrt{K^y} \varphi_2 \right) \right]
+
\left( -i + e^{2\Delta\sqrt{K^z}\varphi_3} \right)
\cosh \left[ \Delta \left( h^x - \sqrt{K^x} \varphi_1 + \sqrt{K^y} \varphi_2 \right) \right]
\right\}
\nonumber.
\end{align*}

\end{document}